\documentclass[onecolumn,nolinenumbers]{aastex63}
\usepackage{aas_macros}

\newcommand{\mydeg}{{$^{\circ}$}}

\newcommand\nh{{\em New Horizons}}
\newcommand\gaia{{\em Gaia}}
\newcommand\hst{{\em HST}}

\newcommand{\deltav}{{$\Delta$V}}

\submitjournal{PSJ}
\shortauthors{Buie et al.}
\shorttitle{KEM target search}

\begin{document}

\title{The \textit{New Horizons} Extended Mission Target: Arrokoth Search and Discovery}

\correspondingauthor{Marc Buie}
\email{buie@boulder.swri.edu}

\AuthorCollaborationLimit=4

\author[0000-0003-0854-745X]{Marc W. Buie}
\affiliation{Southwest Research Institute, 1301 Walnut St., Suite 400, Boulder, CO 80302}

\author[0000-0003-4452-8109]{John R. Spencer}
\affiliation{Southwest Research Institute, 1301 Walnut St., Suite 400, Boulder, CO 80302}

\author[0000-0003-0333-6055]{Simon B. Porter}
\affiliation{Southwest Research Institute, 1301 Walnut St., Suite 400, Boulder, CO 80302}

\author[0000-0001-8821-5927]{Susan D. Benecchi}	
\affiliation{Planetary Science Institute, 1700 East Fort Lowell, Suite 106, Tucson, AZ 85719}

\author[0000-0002-6722-0994]{Alex H. Parker}
\affiliation{Southwest Research Institute, 1301 Walnut St., Suite 400, Boulder, CO 80302}

\author[0000-0001-5018-7537]{S. Alan Stern}
\affiliation{Southwest Research Institute, 1301 Walnut St., Suite 400, Boulder, CO 80302}
		
\author[0000-0002-1542-1903]{Michael Belton}
\affiliation{Belton Space Exploration Initiatives, 430 S. Randolph Way, Tucson, AZ 85716; Emeritus Astronomer, Kitt Peak National Observatory, Tucson, AZ 85719}

\author[0000-0002-9995-7341]{Richard P. Binzel}
\affiliation{Department of Earth, Atmospheric, and Planetary Sciences, Massachusetts Institute of Technology, 77 Massachusetts Avenue, Cambridge, MA 02139}

\author[0000-0002-8003-7115]{David Borncamp}
\affiliation{Space Telescope Science Institute, INS, Baltimore, MD, 21218}

\author[0000-0002-8397-4219]{Francesca DeMeo}
\affiliation{Department of Earth, Atmospheric, and Planetary Sciences, Massachusetts Institute of Technology, 77 Massachusetts Avenue, Cambridge, MA 02139}

\author[0000-0003-2239-7988]{S. Fabbro}
\affiliation{National Research Council of Canada, Herzberg Astronomy \& Astrophysics Program, 5071 West Saanich Road, Victoria, BC, V9E 2E7, Canada}

\author[0000-0002-5211-0020]{Cesar Fuentes}
\affiliation{Northern Arizona University, Flagstaff, AZ}

\author[0000-0002-6174-8165]{Hisanori Furusawa}
\affiliation{Astronomy Data Center, National Astronomical Observatory of Japan, Osawa 2--21--1, Mitaka, Tokyo 181--8588, Japan}

\author[0000-0002-6174-8165]{Tetsuharu Fuse}
\affiliation{Public Relations Center, National Astronomical Observatory of Japan, Osawa 2--21--1, Mitaka, Tokyo 181--8588, Japan}

\author{Pamela L. Gay}
\affiliation{Southern Illinois University}

\author{Stephen Gwyn}
\affiliation{National Research Council of Canada, Herzberg Astronomy \& Astrophysics Research Centre, 5071 West Saanich Road, Victoria, BC V9E 2E7, Canada}

\author[0000-0002-1139-4880]{Matthew J. Holman}
\affiliation{Center for Astrophysics, Harvard \& Smithsonian, 60 Garden Street, Cambridge, MA 02138}

\author{H. Karoji}
\affiliation{National Astronomical Observatory of Japan, Osawa 2--21--1, Mitaka, Tokyo 181--8588, Japan}

\author[0000-0001-7032-5255]{J. J. Kavelaars}
\affiliation{National Research Council of Canada, Herzberg Astronomy \& Astrophysics Research Centre, 5071 West Saanich Road, Victoria, BC V9E 2E7, Canada}

\author{Daisuke Kinoshita}
\affiliation{Institute of Astronomy, National Central University, Jhongli, Taiwan}

\author[0000-0002-1962-904X]{Satoshi Miyazaki}
\affiliation{Subaru Telescope, National Astronomical Observatory of Japan, 650 N Aohoku Place Hilo Hawaii 96720 USA}

\author{Matt Mountain}
\affiliation{Space Telescope Science Institute, INS, Baltimore, MD, 21218}

\author[0000-0002-6013-9384]{Keith S. Noll}
\affiliation{NASA Goddard Spaceflight Center, 8800 Greenbelt Rd., Code 693.0, 20771, Greenbelt, MD}

\author[0000-0003-0412-9664]{David J. Osip}
\affiliation{Carnegie Observatories, Las Campanas Observatory (Chile)}

\author[0000-0003-0407-2266]{Jean-Marc Petit}
\affiliation{Institut UTINAM UMR6213, CNRS, OSU Theta F-25000 Besancon, France}

\author{Neill I. Reid}
\affiliation{Space Telescope Science Institute, INS, Baltimore, MD, 21218}

\author[0000-0003-3145-8682]{Scott S. Sheppard}
\affiliation{Earth and Planets Laboratory, Carnegie Institution for Science, 5241 Broad Branch Rd. NW, Washington, DC 20015}

\author[0000-0002-8580-4053]{Mark Showalter}
\affiliation{SETI Institute, 189 Bernardo Ave., Mountain View, CA 94043}

\author[0000-0002-5358-392X]{Andrew J. Steffl}
\affiliation{Southwest Research Institute, 1301 Walnut St., Suite 400, Boulder, CO 80302}

\author{Ray E. Sterner}
\affiliation{Johns Hopkins University Applied Physics Laboratory, 11100 Johns Hopkins Road, Laurel, MD 20723}

\author[0000-0001-8813-9338]{Akito Tajitsu}
\affiliation{Subaru Telescope Okayama Branch, National Astronomical Observatory of Japan, 3037-5, Honjou, Kamogata, Asakuchi, Okayama 719-0232, Japan}

\author[0000-0003-0773-1888]{David J. Tholen}
\affiliation{Institute for Astronomy, 2680 Woodlawn Drive, Honolulu, HI 96822}

\author[0000-0003-4580-3790]{David E. Trilling}
\affiliation{Department of Astronomy and Planetary Science, Northern Arizona University, P.O. Box 6010 Flagstaff, AZ, 86011; School of Informatics, Computing, and Cyber Systems, Northern Arizona University, P.O. Box 5693, Flagstaff, AZ, 86011}

\author[0000-0003-0951-7762]{Harold A. Weaver}
\affiliation{Johns Hopkins University Applied Physics Laboratory, 11100 Johns Hopkins Road, Laurel, MD 20723}

\author[0000-0002-3323-9304]{Anne J. Verbiscer}
\affiliation{Department of Astronomy, University of Virginia, Charlottesville, VA}

\author{Lawrence H. Wasserman}
\affiliation{Lowell Observatory, 1400 W Mars Hill Road, Flagstaff, AZ 86001}

\author{Takuji Yamashita}
\affiliation{Thirty Meter Telescope (TMT) Project, National Astronomical Observatory of Japan, Osawa 2--21--1, Mitaka, Tokyo 181--8588, Japan}

\author{Toshifumi Yanagisawa} 
\affiliation{Chofu headquarters, Japan Aerospace Exploration Agency, Jindaiji Higashimachi 7--44--1, Chofu, Tokyo 182--0012, Japan}

\author[0000-0002-3286-911X]{Fumi Yoshida}
\affiliation{School of Medicine, Department of Basic Sciences, University of Occupational and Environmental Health, 1-1 Iseigaoka, Yahata, Kitakyusyu, Fukuoka 807-8555, Japan; Planetary Exploration Research Center, Chiba Institute of Technology, 2-17-1 Tsudanuma, Narashino, Chiba 275-0016, Japan}

\author{Amanda M. Zangari}
\affiliation{Southwest Research Institute, 1301 Walnut St., Suite 400, Boulder, CO 80302}

\begin{abstract}

Following the Pluto fly-by of the \nh\ spacecraft, the mission provided a unique opportunity to explore the Kuiper Belt in-situ. The possibility existed to fly-by a Kuiper Belt object (KBO) as well as to observe additional objects at distances closer than are feasible from earth-orbit facilities. However, at the time of launch no KBOs were known about that were accessible by the spacecraft. In this paper we present the results of 10 years of observations and three uniquely dedicated efforts -- two ground-based using the Subaru Suprime Camera, the Magellan MegaCam and IMACS Cameras, and one with the {\em Hubble Space Telescope} -- to find such KBOs for study. In this paper we overview the search criteria and strategies employed in our work and detail the analysis efforts to locate and track faint objects in the galactic plane. We also present a summary of all of the KBOs that were discovered as part of our efforts and how spacecraft targetability was assessed, including a detailed description of our astrometric analysis which included development of an extensive secondary calibration network. Overall, these efforts resulted in the discovery of 89 KBOs including 11 which became objects for distant observation by \nh\ and (486958) Arrokoth which became the first post-Pluto fly-by destination. 

\end{abstract}

\section{Introduction}

NASA's \nh\ mission \citep{Stern2008} flew by Pluto in July 2015 and then deeper into the Kuiper Belt, providing the opportunity of close encounters with other objects.  This is the only opportunity for close-up observations of Kuiper Belt objects (KBOs) for the foreseeable future.  Prior to the launch of \nh, no KBOs were known that were accessible within the propulsion limits of the spacecraft. Finding suitable targets became a critical need for the extended mission. We estimated that objects numerous enough to be targetable with available \deltav\ were likely to be in the 50 km size range, with apparent R magnitudes near 26 \citep{Spencer2003}.  The search area lay in the Milky Way, in Sagittarius, because of the location of Pluto and the trajectory of the spacecraft, requiring KBOs to be found against a very high density of background stars.

\begin{center}
\includegraphics[scale=0.20]{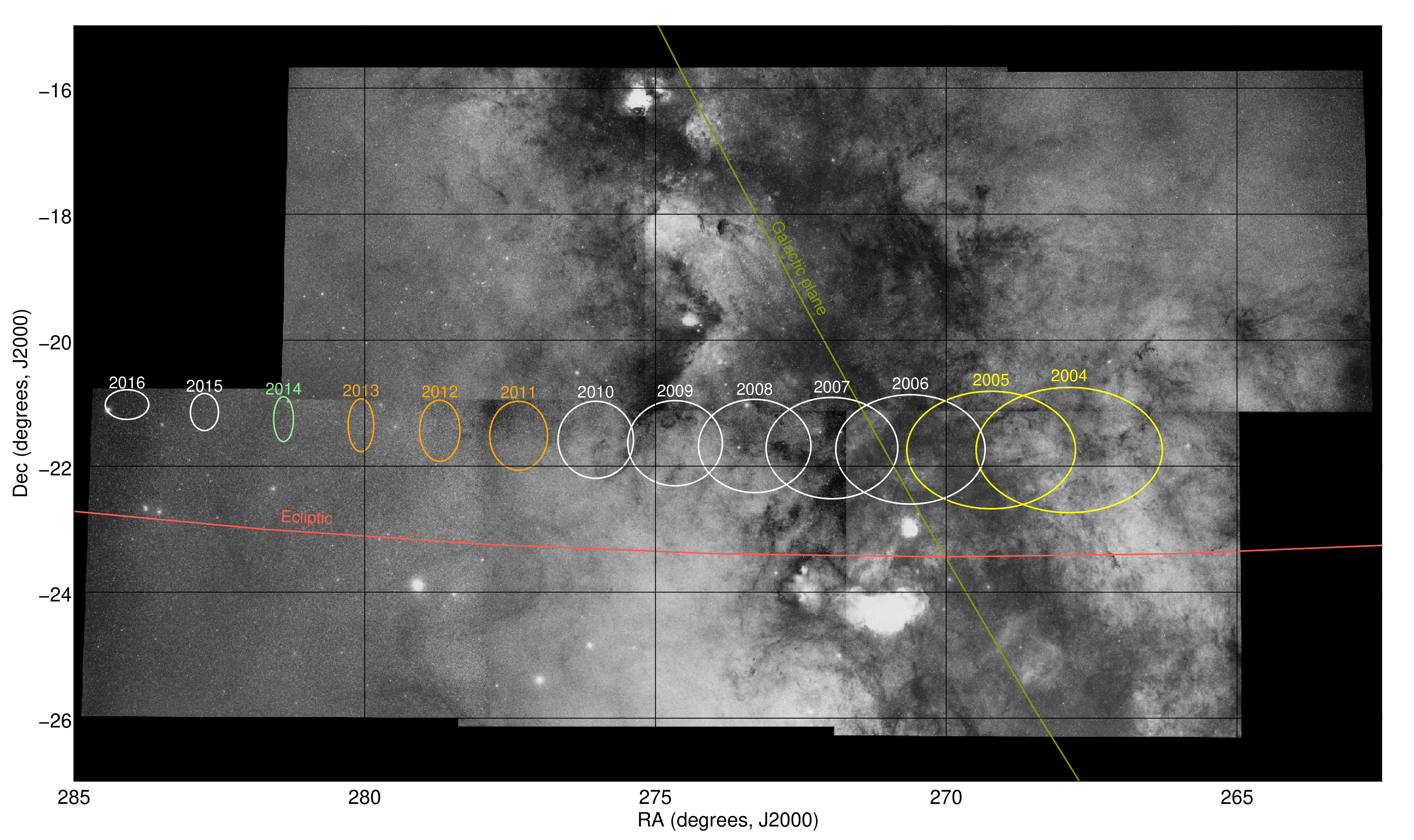}
\figcaption[figoverview.eps]{\label{fig-overview}\scriptsize
Map showing the search areas on the sky during the timespan of this project.  The sky is shown with an overlay of the equatorial sky-plane coordinates.  Superimposed are ellipses for the area that includes 67\% of the encounterable cold-classical KBO population at opposition for the indicated year.  The different colors correspond to the three distinct search epochs discussed in this paper (color coded by common effort).  No searches were conducted in years with white ellipses.  Also shown are curves for the ecliptic and the galactic plane.
}
\end{center}

A key element affecting our search efforts was the size and orbit distribution of the KBO population.  Based on the orbit distribution known even in the early 2000's, it was clear that the spacecraft trajectory would be passing through the cold-classical KBOs (CCKBOs) that are characterized by low inclination and low eccentricity orbits and represent a dynamically cold and presumably undisturbed primordial population.  The nature of this component of the KBO population was thus very important to the outcome of the search.  The hot classical component provided an additional small chance for an encounterable option.  No other dynamical types of object were considered when designing our search program due to the extremely large search area required for even a miniscule probability of detecting an encounterable object from the non-classical populations.

The size distribution also strongly affected the outcome of our search efforts.  Our search needed to reach to fainter brightness and thus smaller size than was fully characterized at the time.  There were already hints of a break in the slope of the size-frequency distribution when we started the project, but a firm consensus had not yet been reached \citep[eg.,][]{Bernstein2004,Fraser2008}. Ongoing work has provided much better characterization of this important attribute of the CCKBOs \citep{Fraser2014, Kavelaars2021}.  Our earliest estimate of the sky-density of flyby targets used a single-slope power law based on objects in the 20-23 mag brightness range.  Now we know that this provided a very optimistic prediction for the outcome of the search.

We will not review all the work and improvements to the characteristics of the general KBO population that was happening during this project.  But, we were well aware of such work going on and folded such new results into our search efforts as they came available.  Equally important to our work and the orbit estimation needed to support \nh\ mission operations was the astrometric catalog produced by the \gaia\ mission \citep{Gaia2016a,Gaia2016b,Lindegren2016}. By the end of this project, the astrometric support catalog was a solved problem but along the way this was a major challenge to overcome prior to the \gaia\ DR2 catalog release \citep{Gaia2018,Lindegren2018}.

In this paper we describe the full extent of our efforts to locate objects for both a  fly-by visit by the spacecraft as well as observations of objects at distances that provide higher resolution than the {\em Hubble Space Telescope} (\hst). Fig.~\ref{fig-overview} shows the search area by year as well as the background field. We first describe our initial wide and shallow search for objects which began in 2004 and set the stage for determining what actually needed to be done to ensure finding a fly-by target. We then describe our dedicated ground-based searches from 2011-2014 and our \hst\ large program which ran during the summer of 2014 and resulted in the discovery of (486958) Arrokoth which became the \nh\ fly-by target. We describe our various reduction techniques for each of the different observing configurations, ground- and space-based, and our efforts to assess and determine the targetability of potential objects both in our search area and then of the discovered objects themselves. Discussion details include the astrometric catalogs used in our work and the development of this process over time. An appendix provides extensive documentation of the mathematics, library routines, and other support tools developed for this project as well.

\section{Wide and Shallow Search -- 2004-2005}

Our initial search, intended as a pilot study to assess the effort required for a future deeper search, was carried out in 2004 and 2005 with the then newly commissioned SuprimeCam prime focus imager at the Subaru telescope on Mauna Kea.  The search area to cover was then about 7~deg$^2$.  The field-of-view (FOV) of SuprimeCam was 0.3 deg$^2$, then the largest FOV on any of the sufficiently large telescopes.  The readout time was about 20 seconds, crucial for part of our observing strategy.  Additional details about this camera system can be found in \citet{Miyazaki2002}.  As expected, the biggest problem for this search was the large number of stars in the background.  Our intent from the beginning was to use optimal image subtraction as described by \citet{Alard1998} to remove the stellar background, leaving the solar system objects to be identified and measured.  Our observing strategy for this project was to collect a series of relatively short exposures (either 90 or 120 seconds depending on sky conditions) at multiple distinct epochs, or visits.  Using short exposures minimized saturated stars and by stacking many images we built up longer integration times to reach a fainter limit.  At this stage in our efforts we did not attempt to shift the images to track target motions.  As a result, the length of time for any set of frames to be stacked was kept below the threshold where trailing losses could begin to affect our sensitivity.  The limiting magnitude targeted for this search was $R=25$ and was largely dictated by the relatively large search area and the amount of telescope time we could obtain.  In total, we were allocated 16 nights of time in 2004 and 6 in 2005 to this project and realized 67 hours of integration time on 58 distinct search fields.  Table~\ref{tbl04obs-new} gives an overview of the these runs and Table~\ref{tbl-04obs} provides the pointing details and the estimated relative probability of finding a targetable KBO on that field. In a given run a field was imaged on at least 2 nights with two visits on one night and one visit on an adjacent night. 

\begin{center}
\includegraphics[scale=0.50]{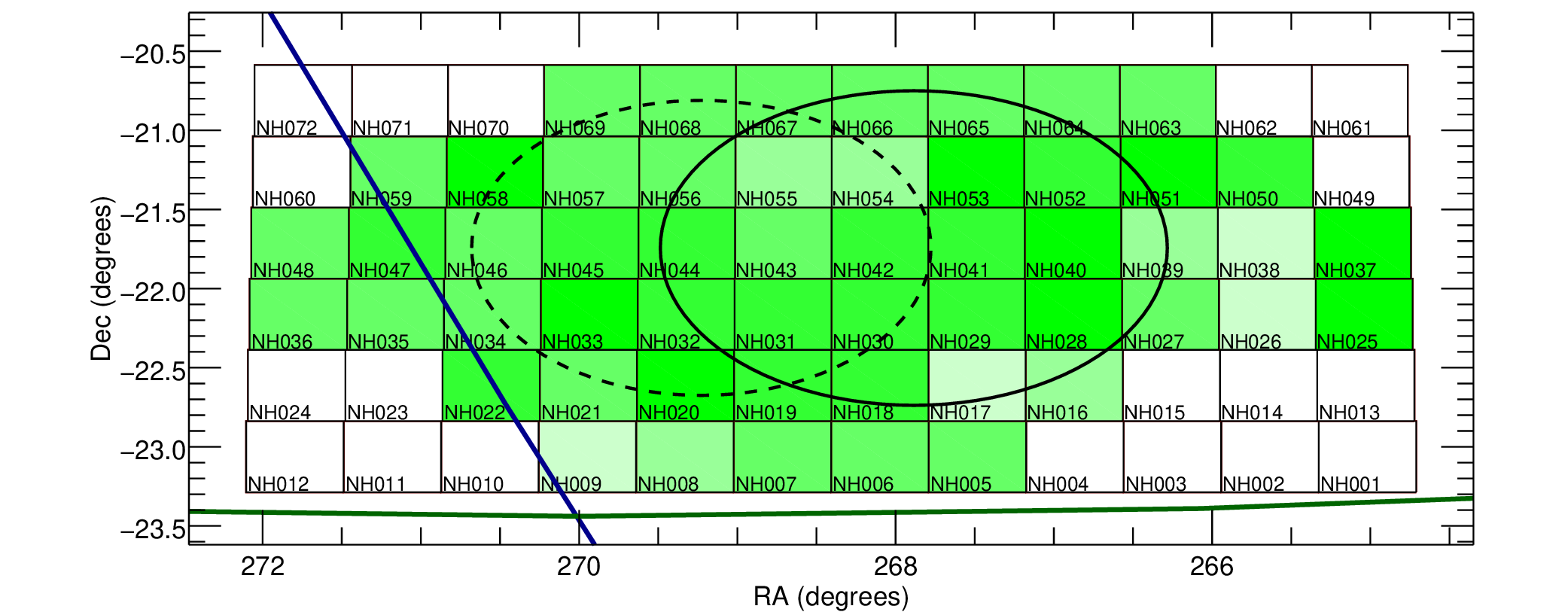}
\figcaption[fig2004.eps]{\label{fig-2004}\scriptsize
Map showing the field pointings for the 2004 search plotted on the plane of the sky in equatorial coordinates.  The solid blue line is the galactic equator and the center of the displayed segment is just 10\mydeg\ from the galactic center.  The solid green line at the bottom is the ecliptic.  The green boxes are the size of the SuprimeCam FOV and show the positions from Table~\ref{tbl-04obs}. The solid ellipse shows the 1-$\sigma$ search area in 2004 and the dashed ellipse shows the search area in 2005.
}
\end{center}

The search pattern for this phase of the project is shown in Fig.~\ref{fig-2004}.  Each tile in this pattern represents a single FOV on the sky for a single pointing of the telescope (Table~\ref{tbl-04obs}).  The gridded area covers 87\% of the distribution of targetable objects on the sky. The highest probability tile contained 4.2\% of the distribution. The goal for each surveyed field was to get five visits over a small number of days to serve as the discovery dataset.  This choice was dictated by the software requirements of the Japanese search team members who already had software for the search based on this required number of visits. Additional observations in 2004 and 2005, consisting of pairs or triplets of visits were used for followup observations.  The coverage of fields in the search area was such that covering the search area also served for followup without needing to specifically target the newly discovered objects.  This part of our strategy was driven by the realized expectation that data analysis would take a very long time to complete compared to the observational phase of the project.

\begin{deluxetable}{cccccc}
\tablecaption{Summary of 2004 Search Runs\label{tbl04obs-new}}
\tablewidth{0pt}
\tablehead{
\colhead{UT Start Date}&
\colhead{N nights}&
\colhead{N fields}&
\colhead{N images}&
\colhead{Total time [h]}
}
\startdata
2004-06-09& 3&  24& 486& 13.9\\
2004-06-16& 2&  11&  96&  3.2\\
2004-06-22& 3&  22& 290&  9.7\\
2004-07-14& 4&  36& 459& 15.3\\
2004-08-09& 4&  35& 307& 10.2\\
2005-04-02& 4&  23& 265&  7.8\\
2005-08-02& 2&  30& 208&  6.9\\
\enddata
\tablecomments{\scriptsize
Total time is the aggregate on-target exposure time in hours.
}
\end{deluxetable}

%

\begin{deluxetable}{cccccccccc}
\tablecaption{Summary of 2004 Search Fields\label{tbl-04obs}}
\tablewidth{0pt}
\tablehead{
\colhead{Name}&
\colhead{R.A.$^1$}&
\colhead{Dec.$^1$}&
\colhead{Prob$^2$}&
\colhead{}&
\colhead{Name}&
\colhead{R.A.$^1$}&
\colhead{Dec.$^1$}&
\colhead{Prob$^2$}&
\colhead{}
}
\startdata
NH001& 17:40:04.4& -23:03:45& 0.001& & NH037& 17:40:12.3& -21:42:45& 0.005& \\
NH002& 17:42:32.3& -23:03:45& 0.002& & NH038& 17:42:38.8& -21:42:45& 0.009& \\
NH003& 17:45:00.2& -23:03:45& 0.005& & NH039& 17:45:05.2& -21:42:45& 0.012& \\
NH004& 17:47:28.1& -23:03:45& 0.007& & NH040& 17:47:31.7& -21:42:45& 0.028& \\
NH005& 17:49:56.0& -23:03:45& 0.009& & NH041& 17:49:58.1& -21:42:45& 0.029& \\
NH006& 17:52:23.8& -23:03:45& 0.011& & NH042& 17:52:24.5& -21:42:45& 0.033& \\
NH007& 17:54:51.7& -23:03:45& 0.007& & NH043& 17:54:51.0& -21:42:45& 0.042& \\
NH008& 17:57:19.6& -23:03:45& 0.006& & NH044& 17:57:17.4& -21:42:45& 0.026& \\
NH009& 17:59:47.5& -23:03:45& 0.006& & NH045& 17:59:43.9& -21:42:45& 0.021& \\
NH010& 18:02:15.4& -23:03:45& 0.004& & NH046& 18:02:10.3& -21:42:45& 0.014& \\
NH011& 18:04:43.2& -23:03:45& 0.001& & NH047& 18:04:36.8& -21:42:45& 0.012& \\
NH012& 18:07:11.1& -23:03:45& 0.002& & NH048& 18:07:03.2& -21:42:45& 0.006& \\
NH013& 17:40:07.1& -22:36:45& 0.001& & NH049& 17:40:14.8& -21:15:45& 0.001& \\
NH014& 17:42:34.5& -22:36:45& 0.003& & NH050& 17:42:40.8& -21:15:45& 0.004& \\
NH015& 17:45:01.9& -22:36:45& 0.009& & NH051& 17:45:06.8& -21:15:45& 0.005& \\
NH016& 17:47:29.3& -22:36:45& 0.017& & NH052& 17:47:32.8& -21:15:45& 0.017& \\
NH017& 17:49:56.7& -22:36:45& 0.021& & NH053& 17:49:58.8& -21:15:45& 0.023& \\
NH018& 17:52:24.1& -22:36:45& 0.021& & NH054& 17:52:24.8& -21:15:45& 0.022& \\
NH019& 17:54:51.5& -22:36:45& 0.025& & NH055& 17:54:50.8& -21:15:45& 0.022& \\
NH020& 17:57:18.9& -22:36:45& 0.020& & NH056& 17:57:16.8& -21:15:45& 0.012& \\
NH021& 17:59:46.3& -22:36:45& 0.011& & NH057& 17:59:42.8& -21:15:45& 0.014& \\
NH022& 18:02:13.6& -22:36:45& 0.006& & NH058& 18:02:08.8& -21:15:45& 0.007& \\
NH023& 18:04:41.0& -22:36:45& 0.005& & NH059& 18:04:34.8& -21:15:45& 0.005& \\
NH024& 18:07:08.4& -22:36:45& 0.004& & NH060& 18:07:00.8& -21:15:45& 0.004& \\
NH025& 17:40:09.7& -22:09:45& 0.002& & NH061& 17:40:17.2& -20:48:45& 0.001& \\
NH026& 17:42:36.7& -22:09:45& 0.010& & NH062& 17:42:42.8& -20:48:45& 0.002& \\
NH027& 17:45:03.6& -22:09:45& 0.014& & NH063& 17:45:08.3& -20:48:45& 0.004& \\
NH028& 17:47:30.5& -22:09:45& 0.024& & NH064& 17:47:33.9& -20:48:45& 0.007& \\
NH029& 17:49:57.4& -22:09:45& 0.036& & NH065& 17:49:59.4& -20:48:45& 0.009& \\
NH030& 17:52:24.3& -22:09:45& 0.035& & NH066& 17:52:25.0& -20:48:45& 0.008& \\
NH031& 17:54:51.2& -22:09:45& 0.035& & NH067& 17:54:50.6& -20:48:45& 0.010& \\
NH032& 17:57:18.1& -22:09:45& 0.023& & NH068& 17:57:16.1& -20:48:45& 0.006& \\
NH033& 17:59:45.1& -22:09:45& 0.020& & NH069& 17:59:41.7& -20:48:45& 0.006& \\
NH034& 18:02:12.0& -22:09:45& 0.014& & NH070& 18:02:07.2& -20:48:45& 0.003& \\
NH035& 18:04:38.9& -22:09:45& 0.011& & NH071& 18:04:32.8& -20:48:45& 0.002& \\
NH036& 18:07:05.8& -22:09:45& 0.009& & NH072& 18:06:58.3& -20:48:45& 0.001& \\
\enddata
\tablecomments{\scriptsize
(1) The coordinates are all equinox J2000.  
(2) ``Prob'' is the estimated relative probability of finding a targetable KBO.
}
\end{deluxetable}

\subsection{Image calibration and characterization}

The raw images were processed using standard techniques prior to analysis. These steps include overscan corrections, bias-frame corrections, and flat-fielding corrections.  Our flat fields were generated from the imaging data directly with a robust stack of all data.  Building flats and other image stacking operations required a good characterization of the sky signal in each image. We developed a new routine ({\tt skyfit.pro}\footnote{All references to software with a name ending in ``.pro'' are routines from the Buie IDL Library, \url{http://www.boulder.swri.edu/~buie/idl} and further details about these programs are covered in Appendix \ref{app-idl}.}), that includes robust estimation and simultaneous fitting of a low-order 2-D polynomial to the sky background that uses incomplete information on the distribution of sky signal due to the crowded fields to reconstruct the full noise distribution. See Appendix \ref{app-idl} for more details about this process. The spatially variable sky signal was subtracted from the images, and the mean value near the center of the array was added back to roughly preserve the original detected signal level.

The required data processing was facilitated by collecting key derived quantities for each image and saving all of this in a relational database to prevent having to recompute these values in each step of the processing.  Information collected in this way was the image quality, photometric depth, sky signal properties, detector gain.  Some derived information was too voluminous to be practical for storage in the database, such as lists of detected sources and the full astrometric solution for each image at every step in that process.  This derived information was very valuable for examining data from a given night as a whole to help detect those images where the processing failed for some reason.

\subsection{Astrometric support catalogs}

Accurate astrometric solutions for the images is essential to the stacking and subtraction process.  SuprimeCam exhibits a large amount of spatial distortion in the focal plane between the center and corners of the mosaic.  The positions near the corner are offset compared to what a linear focal plane would provide.  It was not practical to control the pointing for the search images to the sub-pixel accuracy needed for stacking so the distortion had to be directly addressed.

We were well aware of the need to avoid saturation, hence the plan to take multiple short exposures and stack.  Under good sky conditions, the limiting magnitude was 23.9 mag.  More importantly, stars brighter than 18.2 were saturated and of no use for astrometric calibration.  This brightness limit meant that the best data had an insufficient number of astrometric reference stars to permit determining an astrometric solution of the distorted field.

We used the USNO-B1.0 catalog (\citet{Monet2003}) for this phase of the project. This catalog includes proper motions, but we used the form of the catalog extracted for use with the Deep Ecliptic Survey \citep{Elliot2005} where the format of the earlier USNO-A1.0 catalog files was used with all stars saved at an epoch of 2005.  Our first attempt was to use this catalog on a chip-by-chip reduction, but the combination of strong distortion and minimal overlap in brightness between the catalog and images made this simple approach insufficient.  Obtaining good astrometric solutions required two adjustments in procedure.  First, we had to create a secondary astrometric catalog that went to a fainter limiting magnitude. Second, we developed tools to treat the entire mosaic as a single astrometric device.

To create the secondary catalog we identified the data taken with the worst transparency from each field.  Every field had at least a few images taken with several magnitudes of extinction.  These low-throughput images could then be processed using the USNO-B catalog since most of the catalog stars were now not saturated.  More importantly, even though these images were photometrically shallower, they still pushed significantly deeper than the USNO-B catalog and we could extract the fainter stars for a deeper secondary catalog.  Note that the secondary catalog was constructed from the average of all data.  No attempt was made to correct or otherwise adjust these positions for proper motion.

Obtaining an astrometric solution was also complicated.  The large number of frames pushed us to automatic pipeline processing rather than a supervised manual operation.  The automatic tools available then were effective on only about half of the data.  To overcome this problem, we used the focal plane distortion map provided by NAOJ \citep{Miyazaki2002} for an initial mapping of pixel coordinates onto the sky plane. This map was provided in the form of software rather than as formulas and coefficients.

For each image, we used the NAOJ tool to convert the pixel positions of all measured sources in the image to RA and Dec given the knowledge of the telescope pointing.  The result of this step served to linearize the positions.  From here, the RA and Dec estimates were converted to tangent plane coordinates using the header position. The overlapping secondary star catalog positions were also converted to tangent plane coordinates using the same tangent point.  These two lists were close to each other in position, but always had offsets to be determined as well as occasionally needing to be rotated up to a couple of degrees.

Once the lists were correlated, a new astrometric solution was relatively easy to derive.  To support a full-mosaic coordinate system, we developed our own method for describing the focal plane distortion. This mathematical description and its mapping onto a set of FITS keywords is described more fully in Appendix \ref{app-ast}.  Deriving a full solution required an iterative approach for our fourth order fit.  Given a solution, we calculated the new location of the center of the mosaic, thus changing the tangent point.  This required a new solution, leading to a new tangent point.  This process converged within 3 iterations to provide a self-consistent tangent point and astrometric fit.

Another component of the astrometric fit is the relative offset and orientation of all the detectors.  One chip (\#2) was selected as the reference device and its coordinate system is extended to cover the full focal plane.  In principle, the choice of reference chip is arbitrary, but we found it convenient to use a device in the center of the mosaic where the distortion is lower.  The original distortion map for SuprimeCam was used to initially determine these chip offsets and orientations.  For all images, the fit is held constant and the offset and orientation is adjusted to find the best fitting solution.

In principle, this method assumes that the astrometric solution can be different on each image.  The constant and linear terms are always seen to vary with time.  The quadratic and higher-order terms were never seen to exhibit statistically significant variations within a run.  We found the results to be more internally consistent and stable when the mean values of the high-order terms were imposed on the fit, leaving only the constant and linear terms to be fitted for each image.

Additionally, the relative locations of the chips was not expected to vary.  This expectation seems like a reasonable assumption, but we chose a more conservative assumption that the devices don't perceptibly move during an observing run.  Finding the offsets is not entirely straightforward.  In general, there is a coupling between the astrometric solution for the entire mosaic and the values of the offsets and rotations used for each chip.  We did not have the computing resources and tools at that time to attempt a simultaneous determination of the solution and the chip positions.  Instead, we iteratively determined one while fixing the other and repeating until the answer converged. This proved to be a very effective method if somewhat tedious.  Doing this required a starting guess for the offsets and rotations which we had from the NAOJ documentation.  Then, we fixed these chip locations and solved for the chip-by-chip astrometric solution.  These solutions where then averaged to get mean high-order coefficients and the linear and constant terms were refit.  From this point, the solutions were fixed and the chip offset and rotation relative to the reference was scanned to find the values most consistent with the fixed solution.  This is done for all images in a run.  Given these values, a mean offset and rotation is computed using a robust mean estimator ({\tt robomean.pro}). These mean values were then adopted, and the solutions were iterated again.  This process was repeated two or three times for each run.

The measure of success in all of this intense astrometric calibration was in looking at multiple views of the same patch of sky.  A high-speed animation loop looking at the same region registered by their astrometric solutions should only reveal the usual variations in transparency and image quality induced by the atmosphere, but there should be no apparent motion in the positions of the sources.  In the early stages when chip-by-chip independent solutions were used, there was a lot of residual motion -- sometimes many pixels in the corner chips.  After this more involved calibration process, the images were seen to be internally consistent.  Given that this process is referenced to an astrometric catalog, the final absolute astrometric positions should be as good as the underlying catalog.

\subsection{Photometric calibration}

The usual process of flat-fielding an image preserves surface brightness. This is useful for removing pixel-to-pixel variations, over small and large spatial scales.  In the presence of optical distortions, flat fielding can introduce systematic shifts to point-source photometry. The reason for this is that the area of a pixel, on the plane of the sky, varies with such distortion.  To correct for the effect of this distortion on fixed-aperture photometry, we compute the area of each pixel in arcsec$^2$.  The map is normalized by the largest area.  Then, the flux of any source is divided by the value from the normalized map at the location it was measured before converting to an instrumental magnitude.  For SuprimeCam, this correction is as large as 30\% towards the more heavily distorted corners of the mosaic.  The computation of this map is time-consuming but straightfoward given a good astrometric description of an image.  We did not apply this correction to the images themselves, but used it to correct the extracted photometry.

\subsection{Image registration}

Given an accurate astrometric solution, image registration is straightforward.  In the early phase of the project we tried to determine a polynomial expansion approximation for the mapping from one image to another image.  The goal in this case was to pick the best image of a set and then resample the other images to match.  Operationally, this method should work but the high-order differences between two distorted images required too many terms and that fit was numerically unstable for a significant number of images.  A better methodology turned out to be to resample all images to a linear grid defined by the mean position and orientation of all images of the same field within a run.  Image scale for the output images was set to the scale of the image at the center of the mosaic.  The astrometric fit already determined was sufficient to map the observed images onto the idealized grid.  The output grid was sized to bound all of the input images, meaning that there could be pixels that map onto only a subset of the input images.

The images were all re-sampled with a flux conserving interpolation that is functionally equivalent to sinc interpolation.  This process used the Buie library routine {\tt dewarp.pro} that itself uses the built-in IDL image interpolation tool, {\tt congrid}.

\subsection{Image subtraction}

The power of image subtraction for picking out changes in images cannot be denied. What sets this project apart is its time-dependent nature. We routinely needed to work on data with a much larger range of image quality than is typically tolerated by other investigations of a largely static sky.  The quality of subtraction was seen to degrade as the difference in seeing between two images increased.

The first step in the subtraction process is to take all of the images for a given field that are consecutive in time and average them together. This averaging process rejects outliers such as cosmic ray strikes and fast moving objects in the field but preserves stars and KBOs. Prior to the stacking process, each image had its spatially variable sky level subtracted and then flux scaled to match the flux of the best image in the set. The constant term of the sky fit from the best image was then added to the average to return the image to pixel values closely approximating the original to facilitate tracking the noise properties of the images. All images in a set were used to build the stack.

Any pair of stacked images can be used for image subtraction at this point, but not all pairs were equally effective.  For each image, we computed the difference against the best quality stack.  The final difference pairs were computed and saved for the next step of searching.

\subsection{Searching for objects}

After the visit difference images were completed, we visually examined the data to  look for slow-moving objects.  This process was essentially the same as used by the Deep Ecliptic Survey in using color composites for comparing two epochs of data at a time.  Traditional image blinking could be used if desired.  When an object was noted, its position was measured in the difference image.  Since the astrometric grid is known, it then becomes a simple matter to convert from pixel coordinates to sky coordinates. This tool used a low-order polynomial short-arc predictor to project the position of a measured object onto each image not yet measured. This system allowed finding an object across the chip gaps.

Much later in the project, we implemented a parallel system for finding objects as a citizen-science project.  A broad discussion of what citizen science entails can be found at \url{https://www.citizenscience.gov/}.  In our case, we engaged our participants by providing image data to search for transient sources that included variable stars and moving objects with distinctions between main-belt asteroids and KBOs. This system was called ``Ice Hunters'' and hosted as part of the ``Zoo'' collection of citizen science arenas \url{https://www.zooniverse.org/}.  After analyzing data from a number of observing runs we found no difference in the detection list between our scanning and Ice Hunters. In the end we found that technology constraints at the time served to make the citizen-science approach much slower and more effort to manage than just using a local tool with a couple of people working to scan the data.  More details about this experiment can be found in Appendix~\ref{ap-icehunt}.

\subsection{Wide and shallow search results}

Our search effort was able to find 24 objects.  Table~\ref{tbl-04obj} summarizes our results. The faintest object seen was at R=24.9, exactly as expected from our initial estimates prior to the search.  Of those objects detected, 14 (58\%) were likely cold-classical and the lowest \deltav\ for \nh\ was 1.1 km/s, or about 10 times bigger than the available amount.  The closest candidate is one of two objects that are currently designated (though clearly there are others with good enough orbits for designations).  NI09 (or 2004~LW$_{31}$) now has a much better orbit than was available from just our search data.  It now has a 15-year arc with 39 unique observations (including ours).  The \deltav\ estimate shown in Table~\ref{tbl-04obj} for this object is good to 10\% showing that even short-arc data are sufficient to strongly constrain the list of potential targets.  The orbit estimates for these objects as well as those in following efforts evolved a great deal during this project.  The initial tools used were based on the approach of \citet{Bernstein2000}.  Our tools evolved to include a more direct probabilistic estimate meant to improve on the uncertainties and be more ameable to addressing encounterability.  By the end of this project, we moved to a full Markov-Chain Monte-Carlo (MCMC) approach that is described elsewhere \citep{Porter2018,Porter2022}.  The orbital elements tabulated in this work should not be taken as the final word on these objects.  They were vetted only enough to assess the encounter probability.


\begin{deluxetable}{ccccccccccccc}
\tablecaption{Summary of 2004 Search Discoveries\label{tbl-04obj}}
\tablewidth{0pt}
\tablehead{
\colhead{Object}&
\colhead{$a$}&
\colhead{$e$}&
\colhead{$i$}&
\colhead{$H_V$}&
\colhead{$R_d$}&
\colhead{arc}&
\colhead{discovery}&
\colhead{Nobs}&
\colhead{Desig}&
\colhead{CA Date}&
\colhead{CA}&
\colhead{$\Delta$V}\\
& (AU)& & (deg)& & & (days)& UT& & & & (AU)& (km/sec)
}
\startdata
NI01& 42$\pm$22&    0.03$\pm$0.57&  4.1$\pm$1.7&   8.1& 24.3&   1.0& 2004/06/16&  3& \nodata& 2017-07-17& 3.5& 7.9\\
NI02& 40$\pm$ 7&    0.18$\pm$0.30&  9.1$\pm$0.7&   7.9& 23.4&  61.8& 2004/06/09&  7& \nodata& 2015-01-07& 5.2& \nodata\\
NI03& 49$\pm$ 6&    0.30$\pm$0.17&  3.16$\pm$0.00& 6.8& 24.8& 419.9& 2004/06/09& 12& \nodata& 2019-11-24& 2.0& 2.3\\
NI04& 59$\pm$26&    0.33$\pm$0.31& 13.0$\pm$2.4&   7.0& 24.4&  35.9& 2004/06/09&  7& \nodata& 2018-02-02& 4.0& 8.2\\
NI05& 48$\pm$25&    0.03$\pm$0.57& 34$\pm$15&      8.0& 24.9&   2.1& 2004/06/09&  5& \nodata& 2019-06-03& 8.7& 11.3\\
NI06& 40$\pm$21&    0.02$\pm$0.57&  3.9$\pm$1.3&   8.6& 24.2&   2.1& 2004/06/09&  5& \nodata& 2017-08-17& 1.2& 3.2\\
NI07& 59$\pm$33&    0.47$\pm$0.40&  2.97$\pm$0.27& 8.1& 24.0&  37.0& 2004/06/09&  8& \nodata& 2017-01-19& 0.9& 3.3\\
NI09& 46.2$\pm$0.1& 0.003$\pm$0.004&  3.062$\pm$0.002&  6.9& 24.8& 420.0& 2004/06/09& 12& 2004 LW$_{31}$& 2019-11-26& 1.0& 1.1\\
NI10& 41$\pm$19&    0.35$\pm$0.72&  7.1$\pm$0.6&   8.3& 24.6&  50.9& 2004/07/14&  6& \nodata& 2019-01-18& 2.5& 3.6\\
NI11& 41$\pm$19&    0.42$\pm$0.90&  4.1$\pm$1.6&   8.3& 24.9&  36.9& 2004/07/16&  7& \nodata& 2020-07-14& 2.6& 2.6\\
NI12& 42$\pm$221&   0.02$\pm$0.57&  3.7$\pm$1.3&   8.3& 24.4&   1.0& 2004/06/22&  3& \nodata& 2018-02-23& 2.4& 4.9\\
NI13& 52$\pm$9&     0.29$\pm$0.33&  8.5$\pm$0.5&   8.2& 24.3&  51.0& 2004/07/15&  8& \nodata& 2017-12-26& 2.4& 5.2\\
NI14& 39$\pm$21&    0.03$\pm$0.57&  5.0$\pm$1.8&   8.8& 24.0&   2.1& 2004/06/16&  4& \nodata& 2017-02-08& 1.6& 5.9\\
NI16& 39$\pm$19&    0.23$\pm$0.76&  0.96$\pm$0.06& 7.2& 23.3&  21.7& 2004/06/23&  7& \nodata& 2015-01-25& 3.2& \nodata\\
NI17& 46$\pm$5&     0.17$\pm$0.30&  2.27$\pm$0.00& 8.6& 24.9& 419.9& 2004/06/09&  7& \nodata& 2020-01-10& 2.2& 2.4\\
NI18& 42$\pm$6&     0.15$\pm$0.77&  6.2$\pm$0.7&   8.0& 24.1&  54.9& 2004/06/16&  6& \nodata& 2017-03-08& 4.8& 16.2\\
NI19& 40$\pm$13&    0.11$\pm$0.70& 18.7$\pm$3.1&   8.9& 23.9&  37.0& 2004/06/09&  8& \nodata& 2016-08-24& 4.4& 24.3\\
NI20& 51$\pm$7&     0.39$\pm$0.15&  6.63$\pm$0.01& 8.7& 24.2& 420.0& 2004/06/09& 10& 2004 LV$_{31}$& 2017-09-18& 2.1& 5.1\\
NI21& 46$\pm$24&    0.03$\pm$0.56&  3.0$\pm$1.6&   8.7& 24.7&   1.0& 2004/06/10&  3& \nodata& 2019-04-27& 1.8& 2.4\\
NI22& 47$\pm$24&    0.02$\pm$0.56& 13.1$\pm$4.9&   8.6& 24.8&   2.1& 2004/06/09&  4& \nodata& 2019-10-31& 3.0& 3.5\\
NI25& 36$\pm$19&    0.03$\pm$0.58&  0.77$\pm$0.23& 7.7& 23.6&   1.0& 2004/07/16&  3& \nodata& 2015-11-08& 2.9& $>$\\
NI30& 38$\pm$19&    0.03$\pm$0.56&  6.5$\pm$2.3&   8.6& 23.9&  37.9& 2004/06/09&  5& \nodata& 2016-08-07& 4.0& 23.1\\
NI34& 48$\pm$25&    0.44$\pm$0.46&  2.54$\pm$0.30& 7.8& 24.4&  46.9& 2004/06/24&  5& \nodata& 2018-03-06& 1.8& 3.6\\
NI35& 45$\pm$23&    0.02$\pm$0.56&  2.7$\pm$1.2&   8.3& 24.7&   1.1& 2004/06/09&  4& \nodata& 2019-03-06& 2.4& 3.3\\
\enddata
\tablecomments{\scriptsize
The astrometry for these objects from this search is provided in the supplemental data file for this table.  All values given here are from the discovery data only.  Blank $\Delta$V values indicate that close approach conflicted with the Pluto encounter (on or before).  $\Delta$V values listed as `$>$' indicate a non-physical value.
}
\end{deluxetable}

Even though this early search work did not result in the detection of a mission target, important lessons were learned that were carried forward to the later search efforts.  Here we summarize the more important of these lessons.  Also, one of these objects (2004 LW$_{31}$) made it to the list of bodies to be observed with long-range encounters by \nh\ before and after its encounter with Arrokoth \citep{Porter2022}.

The search area was variegated enough to include some areas of ``dark lanes'' where foreground dust in our galaxy obscures the stars that would normally have been seen.  We did not specifically target these areas, but recognized that they would enhance our probability of finding objects due to a reduced amount of confusion.  Unfortunately, nothing useful was found in those limited areas.

\subsection{Lessons learned}

The results of this effort showed us that we clearly needed a much deeper search to have a reasonable chance to find an extended mission target.  This effort also clearly showed the critical need for a secondary reference star catalog that could be directly tied to the search data.  It likewise demonstrated the influence of high quality seeing ($<0.7$ arcsec) on our object discovery rate; quality seeing with lower transparency was better than poor seeing with clear skies.  All of these lessons provided essential information that guided the implementation of the next search effort.

Finally, nearly all the software tools and the lessons learned therein were then available as a starting point for follow-on searches that were characterized by ever shortening timescales for data processing.  This initial search effort did not result in finding a targetable object, but was an absolutely essential step along the way.

\section{Deep Ground-based Search -- 2011-2014}

A deeper search was undertaken starting in 2011.  By this time, the search area for 80\% of the candidate pool had dropped down to $\sim$2~deg$^2$ and we needed only 7 discrete SuprimeCam fields to cover the search region.  The reduction in the area made it feasible to consider much longer integrations and thus deeper image stacks to reach a fainter limiting magnitude while still covering most, if not all, of the search area.  Once again we used the Subaru SuprimeCam imager, but there was a second option available with Megacam (\citet{McLeod2015})and IMACS (\citet{Dressler2006}) on the Magellan telescopes.  A list of observing runs for this search effort which includes a run identification, the instrument used, the UT start and end time for each run, the number of distinct fields imaged and the number of individual frames collected can be found in Table~\ref{tbl-11obs}.

\subsection{Astrometric support catalog}

Driven by the lessons learned in the initial 2004-2005 pilot search, we developed a special astrometric support catalog for the calibration of the new search data.  Targeted observations with the Canada-France-Hawaii Telescope (CFHT) were obtained in queue mode.  The goal each year was to cover the sky for the following year so that the catalog was ready for each new apparition of search data.  All observations were taken with MegaCam \citep{Boulade2003} and an $r'$ filter with 30 second exposures. Good image quality was essential for these crowded fields and the queue observing system ensured we got the necessary quality data.  Most data were taken under photometric conditions even though this wasn't absolutely required.  The data were processed at CADC using their standard tools for the MegaCam mosaic \citep{Gwyn2014}.  With its 1~deg$^2$ FOV camera it was able to easily cover the search region. Toward the end of the search phase this catalog was extended to cover the entire field through the expected target position at the time of the \nh\ encounter.  The catalog reaches a useful depth of r=23-24 which provided ample overlap with the brightness range of the ground-based data and the available astrometric star catalogs. Initially, the UCAC4 catalog (\citet{Zacharias2012}) was used as the underlying reference upon which the support catalog was based.  As the project progressed we moved to hybrid catalogs, URAT1 from \citet{Zacharias2015}, and the Gaia DR1 catalog \citep{Gaia2016a,Gaia2016b,Lindegren2016}, and then ultimately to the Gaia DR2 catalog \citep{Gaia2018,Lindegren2018}. Mean proper motion corrections were applied so that the support catalog had a mean epoch contemporaneous with the search.  This catalog ensured that there were at least a couple of hundred reference stars for each detector in the search cameras.

\begin{center}
\includegraphics[scale=0.4]{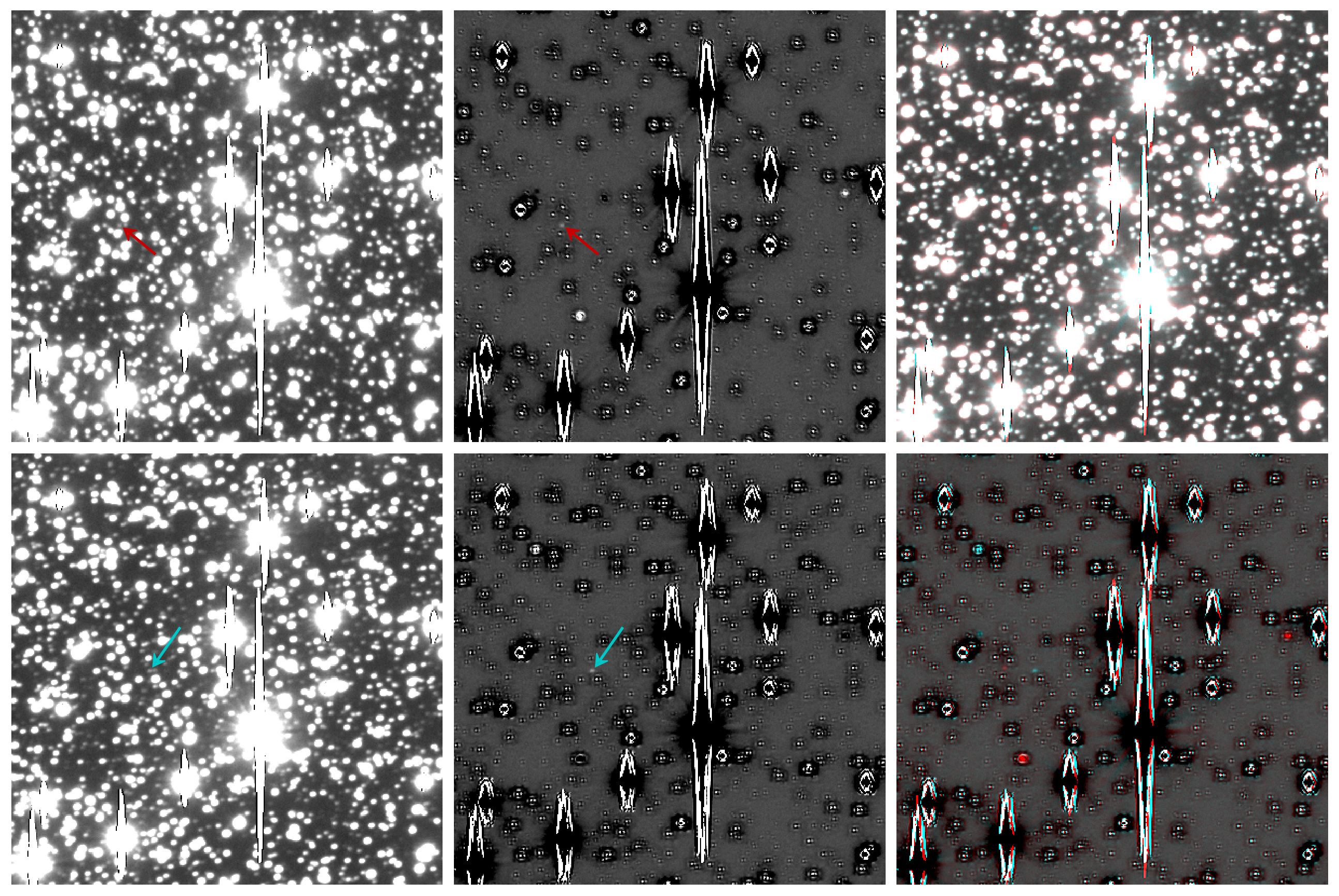}
\figcaption{\label{fig-image}
Sample SuprimeCam image data from 2011.  Two epochs are shown of the discovery images for 2011~HL$_{103}$.  The red arrow (or image) shows the location at the first epoch and the cyan arrow (or image) shows the location at the second epoch.  The stacked images are in the left column.  Template subtracted images are in the middle column.  The right column shows the color stack of the two epochs.  See text for more details.
}
\end{center}

\subsection{Data Processing}

The data analysis for this phase of the search was very similar to that used for the prior search, but there were some improvements and other changes needed.  First, we had multiple independent pipelines for searching the data.  Algorithmically, these pipelines performed similar operations and are not separately described.  The Subaru data were again processed as a single multi-chip astrometric device.  The Magellan data were processed with each image in the detector mosaics as independent systems.

\begin{deluxetable}{ccccccc}
\tablecaption{Summary of 2011-2014 Search Runs\label{tbl-11obs}}
\tablewidth{0pt}
\tablehead{
\colhead{RunID}&
\colhead{Observatory}&
\colhead{Instr}&
\colhead{UT start.}&
\colhead{UT end}&
\colhead{Nf}&
\colhead{Ni}
}
\startdata
 m1104&Magellan&    MegaCam& 2011-04-25 09:28:55& 2011-05-06 09:00:11&  9&   589\\
 m1105&Magellan&    MegaCam& 2011-05-27 06:00:37& 2011-06-04 10:31:09&  5&   399\\
  1105&Subaru& SuprimeCam& 2011-05-29 09:41:57& 2011-06-01 14:39:17& 12&  6090\\
m1107a&Magellan&      IMACS& 2011-07-01 01:02:04& 2011-07-02 08:42:21&  9&  2752\\
  1107&Subaru& SuprimeCam& 2011-07-02 06:35:18& 2011-07-02 13:32:49& 12&  1730\\
m1107b&Magellan&      IMACS& 2011-07-22 23:34:14& 2011-07-24 03:31:46&  4&  1256\\
 m1108&Magellan&      IMACS& 2011-08-27 23:37:14& 2011-08-30 04:36:53&  7&  2400\\
 m1204&Magellan&    MegaCam& 2012-04-17 06:48:17& 2012-04-22 07:33:43&  3&   485\\
 m1205&Magellan&      IMACS& 2012-05-24 03:51:35& 2012-05-24 07:24:44&  4&   232\\
 m1206&Magellan&      IMACS& 2012-06-13 23:43:04& 2012-06-15 07:51:39&  1&   400\\
 m1207&Magellan&      IMACS& 2012-07-11 05:01:52& 2012-07-23 04:29:45&  9&  3512\\
  1207&Subaru& SuprimeCam& 2012-07-16 10:07:35& 2012-07-18 12:09:21&  2&  2320\\
  1208&Subaru& SuprimeCam& 2012-08-13 05:52:22& 2012-08-16 11:00:24&  2&  4780\\
 m1306&Magellan&    MegaCam& 2013-06-03 04:37:07& 2013-06-13 09:05:56&  3&   753\\
 m1307&Magellan&    MegaCam& 2013-07-02 03:42:05& 2013-07-04 06:56:15&  2&   192\\
  1307&Subaru& SuprimeCam& 2013-07-04 07:15:26& 2013-07-14 12:42:26&  4& 10079\\
 m1308&Magellan&    MegaCam& 2013-08-28 00:35:57& 2013-09-06 03:12:50&  1&   431\\
 m1404&Magellan&    MegaCam& 2014-04-25 05:58:51& 2014-05-06 10:04:02&  3&   608\\
\enddata
\tablecomments{\scriptsize
Nf is the number of distinct fields imaged and Ni is the number of
individual images collected.  The UT times are the start times of the
first and last images collected in that run.
}
\end{deluxetable}


\begin{deluxetable}{cccccccccc}
\tablecaption{Logs Notes from 2011-2014 Observing Runs (Sample{$^1$})\label{tbl-11search-short}}
\tablewidth{0pt}
\tablehead{
\colhead{Field}&
\colhead{UT Date}&
\colhead{UT Start Time}&
\colhead{FWHM ($''$)}&
\colhead{Quality}&
\colhead{Comments}&
}
\startdata
\
MA05&110428&04:56&0.7&1& 0.85$''$ at start, some elongation, transparency good\\
MA04&110428&05:40&0.6&1& Good image quality\\
MA07&110428&06:12&0.6&1& Up to 0.7$''$ for some exp.\\
MA01&110428&06:48&0.55&1& Moonrise 06:57 UT\\
MA05&110428&07:26&0.5&1&  \\
MA04&110428&07:59&0.45&1& \\
MA07&110428&08:36&0.55&1& Cross meridian so more rotation\\
MA01&110428&09:19&0.55&1& \\
MA04&110428&09:51&0.55&3& Elevated bkg due to twilight\\
MA06&110429&04:52&1.0&1& 0.9-1.1``  Photometric all night\\
MA03&110429&05:29&0.8&1& $<$1.0`` down to 0.7``\\
MA02&110429&06:04&0.7&1& $<$0.7``\\
\enddata
\tablecomments{\scriptsize
(1) This table is available in its entirety in machine-readable and Virtual Observatory (VO) forms in the online journal. A portion is shown here for guidance regarding its form and content.
}
\end{deluxetable}

This search used much longer dwell times per visit to extend the search to fainter objects, but still took short exposures to be stacked. An important change was in the creation of template images for each pointing.  This time, we used all available images to build a master template image optimized for each pointing for each telescope.  Given the range of image quality in the data, we could not use all images for the template but needed to ensure a sufficient range of time so that any KBOs were filtered out of the template images.

To make a template image, all of the candidate images were sorted by photometric depth.  We tried using a list sorted by image quality, but that does not de-weight images that might be taken through some clouds.  The photometric depth is dependent on both transparency and image quality and worked best as the sorting metric.  A reference image was defined based on the mean pointing of all images, the mean orientation on the sky, and a linear image scale that matched the nominal scale of the images in the center of the mosaics.  All images were resampled to this reference.

As the list of images is traversed from good to bad, each new image was assessed for its ability to improve the stack.  We used a simple noise model based on the sky signal, sky noise, and photometric depth of the image to estimate the change in the SNR of the proposed new stack from adding each new image.  If the new image would improve the SNR, it was then added to the stack and the SNR was updated.  Any image not improving the SNR was dropped from inclusion in the template.  In all cases, every search field had enough data and enough distinct epochs to ensure a good template where moving objects and cosmic ray strikes were filtered out. Once the templates were complete, all individual short-exposure images were differenced against the appropriate template image.

The consequence of moving to longer dwell times was that trailing losses were no longer negligible, even for KBOs.  A simple stack as done for the previous search would leave a smeared image of a KBO that would adversely affect our limiting magnitude.  To generate the shift needed we used a population of synthetic KBOs that matched the orbital element distribution of cold-classical KBOs.  The mean velocity of this population was computed for each visit, field, and device within the detector mosaic.  This single shift value was applied to the subtracted images requiring another resampling of the images.  Once these images were shifted, they could then be stacked.  At this stage, cosmic ray strikes and some of the residuals of the stellar background were removed. The resulting image products could be searched using our standard color-composite and blinking tools.  

A very small sample image that is typical of our dataset is shown in Fig.~\ref{fig-image}.  This sample image shows two of the twenty-two epochs of data on object 2011~HL$_{103}$.  Each of these images is a stack of seven 90-second images for a total of 630 seconds of integration.  The first epoch had a FWHM of 3.4 pixels and the second epoch was 3.5 pixels.  All displayed images are stretched identically from $-3\sigma$ to $+8\sigma$ about the sky mean where $\sigma$ is the measured noise in the sky at the first epoch.  In the figure, the left column shows the stacked images from the same region in the two images with the first epoch on the top and the second epoch on the bottom.  The arrows point to the position of the object and the red arrow is for the first epoch and the cyan arrow is for the second epoch.  The middle column shows the same data but after the template image is subtracted.  The object is clearly visible at both epochs in the each type of presentation, but is clearly easier to see once the background stars are subtracted.  

The images in the right column show what we examined to discover the objects.  On the top is a color composite of both epochs where the red channel contains the first epoch and the green and blue channels contain the second epoch.  The static sky shows as shades of grey to white.  The transient component shows as either red or cyan depending on its location or properties in the two epochs.  Once again the object can be seen in both epochs, but is much easier to spot in the lower right where the background template has been subtracted.  It was this background subtracted color image composite that was used for all object detections.  Note that additional transient features are apparent in this image product.  Most notable are three apparent variable stars, two of which were brighter in the first epoch and thus show as red, and one that was brighter in the second epoch and shows as cyan.  These transients are discarded during the search when there is no pairing at a sensible KBOs sky-plane velocity.

In addition to the manual search process described above, the 2011 data were also searched by automatic computer searching methods as well as by the ``Ice Hunters.''  As with the earlier results, Ice Hunters did no better than our internal team search efforts and was not pursued for later epochs of search data.  The computer search tools did provide some complementary coverage to the visual searching that increased the aggregate completeness of the search effort.  From 2012 onward, the search continued with just in-house processing and search efforts by manual and computer
searches.

\subsection{Results}


\startlongtable
\begin{deluxetable}{chhccccccccccccc}
\tabletypesize{\scriptsize}
\tablecaption{Summary of 2011-2014 Search Discoveries\label{tbl-11obj}}
\tablewidth{0pt}
\tablehead{
\colhead{Object}&
\nocolhead{AHP-ID}&
\nocolhead{MWB-ID}&
\colhead{$a$}&
\colhead{$e$}&
\colhead{$i$}&
\colhead{$H_V$}&
\colhead{$R^1$}&
\colhead{arc}&
\colhead{discovery}&
\colhead{Nobs}&
\colhead{Desig}&
\colhead{type}&
\colhead{CA Date}&
\colhead{CA}&
\colhead{$\Delta$V}\\
&
&
&
(AU)&
&
(deg)&
&
&
(days)&
(UT)&
&
&
&
(UT)&
(AU)&
m/sec
}
\startdata
VNH0008& MA01c15a& NI112& 44.00$\pm$0.01&  0.04&   2.60&    7.32&  23.22&  358& 2011-04-28&  36& 2011JY31&        C& 2018-10-07&    0.15&     245\\
VNH0010& MA04c28a& NI113& 43.15$\pm$0.03&  0.03&   2.39&    8.03&  24.03&  359& 2011-04-28&  33& 2011HZ102&       C& 2018-11-09&    0.16&     250\\
VNH0007& MA01c12a& NI124&    46$\pm$1&     0.14&   1.87&    7.95&   23.85&  357& 2011-04-29&  20& 2011JW31&        C& 2018-08-22&    0.16&     259\\
VNH0006& MA04c06a& NI115&    51$\pm$26&    0.02&   2.51&    7.76&   24.56&    3& 2011-04-30&   6& \nodata&   \nodata& 2021-07-01&    0.38&     318\\
VNH0009& MA04c14d& NI104&  45.2$\pm$0.6&   0.13&   3.31&    6.92&   23.22&  359& 2011-04-28&  54& 2011JX31&        C& 2020-07-15&    0.36&     358\\
VNH0050&         & NI194&    45$\pm$23&    0.02&   2.90&    9.99&   26.09& 0.08& 2013-06-07&   3& \nodata&   \nodata& 2019-05-21&    0.3&      375\\
VNH0036& ML02c12a& NI199&    43$\pm$22&    0.03&   1.88&   10.54&   26.54&    3& 2013-06-04&   9& \nodata&   \nodata& 2018-09-18&    0.26&     421\\
VNH0044&         & NI187&    48$\pm$21&    0.33&   3.40&   11.23&   27.43&   33& 2013-06-03&  14& \nodata&   \nodata& 2020-09-25&    0.45&     435\\
VNH0002& MC07c02a& NI111& 53.34$\pm$0.05&  0.31&   6.42&    9.73&   25.73&  862& 2011-04-28&  55& 2011HK103&       S& 2019-09-01&    0.27&     441\\
VNH0005& MC06c03a& NI105&  43.0$\pm$0.2&   0.04&   3.05&    9.46&   25.46&  446& 2011-04-29&  24& 2011HF103&       C& 2018-12-12&    0.30&     445\\
VNH0034& ML03c03a& NI180&    44$\pm$1&     0.06&   2.99&   10.62&   26.72&   94& 2013-06-05&  18& 2013LU35&        C& 2019-04-20&    0.33&     450\\
VNH0029&         & NI156&    67$\pm$31&    0.04&   2.77&    8.09&   26.09&    5& 2011-04-28&   5& \nodata&   \nodata& 2027-06-01&    1.1&      457\\
VNH0054&         & NI179&    47$\pm$47&    0.03&   2.65&   11.21&   27.61&    1& 2012-04-17&   6& \nodata&   \nodata& 2019-11-16&    0.42&     492\\
VNH0019& MF01c02a&      &    46$\pm$10&    0.02&   2.68&    7.00&   23.40&    2& 2012-04-17&   6& \nodata&   \nodata& 2019-12-24&    0.44&     505\\
VNH0052&         & NI196&    44$\pm$23&    0.25&   2.57&   11.34&   27.34&    5& 2013-06-06&   7& \nodata&   \nodata& 2019-06-07&    0.43&     562\\
VNH0021& MF01c16a& NI174& 44.42$\pm$0.03&  0.10&   3.03&   11.11&   26.81&  506& 2012-04-18&  42& \nodata&       9:5& 2017-12-09&    0.32&     716\\
VNH0014& MA04c21c& NI163&    56$\pm$17&    0.44&   3.22&   10.53&   26.63&  355& 2011-05-02&  19& 2011JA32&        C& 2018-02-10&    0.38&     767\\
VNH0058&         & NI190&    42$\pm$22&    0.02&   2.19&   10.75&   26.45&    7& 2013-07-07&   6& \nodata&   \nodata& 2018-04-20&    0.41&     771\\
VNH0017&         &      &    47$\pm$24&    0.26&   6.72& \nodata& \nodata&   33& 2011-04-28&  10& \nodata&   \nodata& 2021-02-02&    0.93&     833\\
VNH0040&         & NI169&    48$\pm$25&    0.02&   3.17&   10.05&   26.45&    1& 2011-07-23&   3& \nodata&   \nodata& 2020-05-05&    0.85&     886\\
VNH0003& MC07c02b& NI107&    47$\pm$2&     0.19&   5.36&    9.31&   25.01&   65& 2011-04-28&  22& 2011HJ103&       C& 2017-10-13&    0.40&     949\\
VNH0012& MA01c25a& NI162&  43.7$\pm$0.2&   0.08&   6.83&   10.84&   27.34&  357& 2011-04-28&  24& 2011HE103& \nodata& 2020-04-18&    0.92&     975\\
VNH0022& MF01c23a& NI177&  46.2$\pm$0.4&   0.21&   5.44&   11.35&   27.25&  505& 2012-04-17&  35& 516977&          S& 2017-12-10&    0.49&    1.1k\\
VNH0033& ML01c31a& NI193&    58$\pm$30&    0.58&   5.93&    8.89&   26.09&   10& 2013-06-03&  16& \nodata&   \nodata& 2022-05-10&    1.5&     1.1k\\
VNH0025&         & NI101&    46$\pm$24&    0.02&   2.92&    7.28&   23.58&    1& 2011-05-31&   5& \nodata&   \nodata& 2019-09-25&    1.0&     1.2k\\
VNH0016& MA04c35a& NI114&    53$\pm$27&    0.49&   5.76&    9.55&   26.45&  358& 2011-04-28&  24& 2011HD103& \nodata& 2020-11-11&    1.4&     1.3k\\
VNH0047& ML02c03b&      &    47$\pm$24&    0.02&  14.55&    9.45&   26.45& 0.12& 2013-06-06&   3& \nodata&   \nodata& 2020-01-10&    1.4&     1.5k\\
VNH0046&         & NI189&    45$\pm$23&    0.02&   7.43&   10.45&   26.45& 0.17& 2013-07-07&   3& \nodata&   \nodata& 2019-05-22&    1.2&     1.6k\\
VNH0027& ML01c14a& NI110&    54$\pm$29&    0.76&   6.95&    8.86&   25.46&  703& 2011-07-02&   8& \nodata&   \nodata& 2023-05-11&    2.5&     1.6k\\
VNH0037&         &      &    58$\pm$30&    0.02&  13.83&    8.98&   26.18& 0.08& 2011-07-01&   2& \nodata&   \nodata& 2023-10-23&    2.7&     1.6k\\
VNH0035& ML01c23a&      &    42$\pm$22&    0.02&  12.75&    7.79&   23.49&    2& 2013-06-11&   4& \nodata&   \nodata& 2018-03-05&    0.84&    1.7k\\
VNH0055&         & NI182&    48$\pm$18&    0.23&   6.96&   11.56&   27.16&   33& 2013-06-03&  14& \nodata&   \nodata& 2018-03-17&    0.86&    1.7k\\
VNH0032& ML01c08a&      &    41$\pm$21&    0.02&   7.62&    8.77&   24.47&    4& 2013-06-04&   6& \nodata&   \nodata& 2018-01-28&    0.90&    1.9k\\
VNH0020& MF01c08a& NI178&    58$\pm$28&    0.24&  10.20&   11.05&   27.25&    1& 2012-04-18&  15& \nodata&   \nodata& 2019-03-30&    1.4&     2.0k\\
VNH0026&         & NI109&    45$\pm$23&    0.02&   9.08&    7.57&   23.67& 0.19& 2011-07-02&   3& \nodata&   \nodata& 2019-04-19&    1.7&     2.3k\\
VNH0061& ML02c03a&      &    45$\pm$23&    0.02&  16.06&    8.28&   24.38& 0.12& 2013-06-05&   3& \nodata&   \nodata& 2019-04-26&    1.8&     2.4k\\
VNH0059&         & NI192&    87$\pm$3&     0.57&   9.91&   10.76&   26.36&   88& 2013-06-08&  19& \nodata&         S& 2017-09-29&    1.1&     2.8k\\
VNH0051&         & NI195&    43$\pm$18&    0.06&  12.01&   10.91&   26.81&    6& 2013-06-05&   9& \nodata&   \nodata& 2018-06-28&    1.6&     2.8k\\
VNH0024& MF01c31a& NI175&  41.7$\pm$0.9&   0.15&   4.56&   10.62&   25.82&    3& 2012-04-17&  23& \nodata&   \nodata& 2016-05-01&    0.34&    3.0k\\
VNH0018&         &      &    45$\pm$23&    0.02&  16.61& \nodata& \nodata& 0.06& 2011-06-01&   8& \nodata&   \nodata& 2019-04-01&    2.3&     3.2k\\
VNH0060&         & NI198&    40$\pm$12&    0.09&   7.26&   11.57&   27.07&    8& 2013-06-05&   9& \nodata&   \nodata& 2017-03-29&    1.0&     3.4k\\
VNH0042&         & NI183&    40$\pm$20&    0.03&  19.92&   11.02&   26.72&    2& 2013-07-04&   4& \nodata&   \nodata& 2018-01-29&    1.9&     4.0k\\
VNH0045&         & NI188&    39$\pm$20&    0.02&  18.89&   11.67&   27.07&    7& 2013-07-07&   6& \nodata&   \nodata& 2017-03-15&    1.6&     5.3k\\
VNH0023& MF01c26a& NI176&    39$\pm$3&     0.02&  19.28&   11.12&   26.72&    2& 2012-04-17&  18& \nodata&   \nodata& 2017-03-29&    1.6&     5.3k\\
VNH0057&         & NI186&    41$\pm$22&    0.02&  31.74&   10.75&   26.45&    1& 2013-07-05&   5& \nodata&   \nodata& 2018-01-03&    2.5&     5.4k\\
VNH0011& MA01c17a& NI100&  42.3$\pm$0.8&   0.15&  14.44&    6.53&   24.03&   65& 2011-04-28&  25& 2011HL103&     5:3& 2016-08-29&    1.4&     7.8k\\
VNH0041& MM01c26a& NI181&    45$\pm$19&    0.62&  12.20&   10.33&   25.73&   23& 2013-06-13&  15& \nodata&   \nodata& 2016-07-11&    1.3&     8.3k\\
VNH0013& MA04c16a& NI103&    54$\pm$8&     0.41&  12.94&    9.99&   25.19&   63& 2011-04-30&  39& 2011JD32&  \nodata& 2015-12-26&    1.1&    25.9k\\
VNH0001& MC04c01b& NI108& 30.12$\pm$0.01&  0.08&  29.38&    7.49&   21.69&  354& 2011-04-29& 131& 2011HM102&     1:1& 2013-11-04&    1.2&  \nodata\\
VNH0031&         & NI164&    23$\pm$23&    0.32&   6.94&   12.40&   26.90&    5& 2011-04-28&   5& \nodata&   \nodata& 2014-06-29&    0.73& \nodata\\
VNH0048&         & NI191&    37$\pm$9&     0.25&  22.34&   12.41&   26.81&    9& 2013-06-04&  12& \nodata&   \nodata& 2014-07-02&    1.3&  \nodata\\
VNH0004& MC02c05a& NI102&    37$\pm$14&    0.30&   3.81&    9.40&   24.20&   34& 2011-05-29&  12& 2011KW48&  \nodata& 2015-02-12&    0.43& \nodata\\
VNH0053&         & NI197&    34$\pm$18&    0.25&  25.09&   12.02&   26.72&    4& 2013-06-09&   9& \nodata&   \nodata& 2015-06-12&    0.95& \nodata\\
VNH0056&         & NI184&    42$\pm$21&    0.61&  16.00&   11.36&   26.36&   23& 2013-06-13&   9& \nodata&   \nodata& 2015-10-15&    0.90& \nodata\\
VNH0015& MA04c34a&      &    35$\pm$21&    0.03&  18.95&    9.01&   24.11&    5& 2011-04-28&   6& \nodata&   \nodata& 2015-10-21&    1.8&      $>$\\
VNH0039&         &      &  2000$\pm$1200&  0.84& 171.95& \nodata&   25.28& 0.11& 2011-07-02&   2& \nodata&   \nodata&    \nodata& \nodata& \nodata\\
\enddata 
\tablecomments{\scriptsize 
(1) $R$ magnitude is the apparent brightness at the time of discovery.
}
\end{deluxetable}

The search depth of the survey was quite variable due to weather and site conditions. The best dataset from the entire effort was at Magellan where we reached a 50\% efficiency detection limit of R=26.0 in multi-hour stares with exceptional seeing of $\sim0.45$ arcsec.  While seeing on individual images ranging from 0.4 greater than 2.0 arcsec, the average was 0.84, 1.06, 0.75 and 0.71 arcsec per year respectively, from 2011-2014 (Table~\ref{tbl-11search-short}). Likewise, each run was given a unique set of field names, but with significant overlapping sky area between the runs in different months. This allowed for orbit extensions on any new objects to determine their observability/accessability by the spacecraft.  In effect, all of the project search area was observed each year with multi-month overlap of the fields as conditions allowed. 2011 was our most productive year accumulating 12.5 degrees$^2$ of sky area (covering the search area multiple times at different epochs). The following years accumulated 4.4, 2.17 and 0.48 degrees$^2$, respectively (Table~\ref{tbl-11obs}). All of our observations were submitted to the MPC using the standard ``OBS80'' formatting at that time. Although it is now possible to include detailed error analysis in ADES-format submissions, it was not standard at the time, and we did not re-evaluate this information given the level of effort that would have been required post-facto. Astrometry for objects that we later observed with \nh\ were, however, re-analyzed with those uncertainties and that process is reported in detail in \citet{Porter2018,Porter2022}.

Table~\ref{tbl-11obj} provides a summary of the objects discovered during this three year search effort, sorted by increasing \deltav. The information provided also includes the object orbital elements (\textit{a,e,i}), the estimated $H_V$-magnitude, measured R-magnitude, accumulated object arc, discovery date, number of observations, MPC designation if one exists, the dynamical classification for the object, the encounter date and spacecraft encounter distance as well as the delta velocity required for such an encounter.  The dynamical classification shown is meant to be a rough guide where ``C'' is for some type of classical KBO, ``S'' is a scattered object, and those with pairs of numbers denote the type of mean-motion resonance with Neptune.  A close-approach date prior to the planned date of the post-Pluto-encounter trajectory correction maneuver (TCM) has no \deltav\ listed.  One object had a close approach just days after the TCM planning date and would have required an extremely large \deltav.  In this case only ``$>$'' is shown to denote a large value that is of no interest.  The final object in the table is believed to be a real detection of a KBO, but was only seen twice and had insufficient information to constrain its orbit well enough to compute useful close approach circumstances.

The result of the efforts through the end of 2013 made a significant improvement by finding objects that \nh\ would pass close to but the closest, 2011JY31, was not close enough, requiring a factor of two more \deltav\ than available. Without any course adjustment we expected to get to within 0.15 au of 2011JY31 on 7 October 2018.  If we used all of the available fuel, this would still leave us with a flyby distance of $\sim$10 million km and no fuel to support the encounter. As with the earlier search, many of the objects we found became useful
distant encounter targets for \nh.

The results to this point were such that most of the search area was covered to a brightness limit that could be reasonably reached by the best ground-based facilities available.  There was a still a slim chance that we would find something suitable in 2014, but it became clear that another tool was required to ensure finding a target.  The tool of choice to extend the search was the Hubble Space Telescope (HST).  The use of HST was by no means assured so a parallel full search effort was also carried out from ground-based facilities in 2014.  As we began to consider the use of HST, the Hyper Suprime-camera (HSC) was commissioned for use on Subaru (\citet{Miyazaki2018}).  This new camera had a field of view of 1.5 deg$^2$, considerably larger than the search area for that year.  The time allocated for our search on HSC was granted as shared-risk and the weather conditions and instrument problems prevented any new and useful search data.  In the end, the ground-based data from 2014 and the following years were used only for recoveries and orbit refinement work for the distant encounter targets and will not be discussed further in this work. A subsequent new search for additional fly-by targets  begun on Subaru in 2020 and onward will be discussed in other papers such as \citet{Fraser2024}. 

\section{HST search -- 2014}

The immutable properties of the \nh\ flyby of Pluto gave rise to strong constraints on the search efforts.  Re-targeting the spacecraft to the extended mission destination would use the least amount of fuel if it was done as soon as possible after the Pluto encounter. Due to the slow data transmission time from \nh\ to the ground, the project wanted a significant amount of time to downlink critical data before re-targeting.  The time for the trajectory correction maneuver (TCM) was set for October 2015.  The analysis and commanding development for the TCM usually takes a couple of months so when that work began, a good orbit had to be known for the new target.  It was not enough to have just discovered the object prior to the TCM, we had to find it soon enough to get sufficient data to get an accurate orbit for the object as well. Our analysis showed that we needed astrometry spanning at least a year from which an accurate enough orbit estimate could be obtained to support the TCM\null. These constraints meant we absolutely had to find an object in 2014.

The obvious advantages of HST are the stable PSF and the very fine image scale.  The main difficulty with using HST is that the best survey camera for this project, the Wide Field Camera 3 (WFC3)/UVIS (with a image scale of 0.0395 arcsec/pixel) has a rather small FOV at just 0.0021 deg$^2$ (\citet{Dressel2019}). Fortunately, 2014 was the year where the search area reached its minimum size of 0.3 deg$^2$.  Covering 80\% of this area thus would require 83 separate pointings of the telescope.

In the end, we secured two allocations of time on HST -- 40 orbits for a pilot study which if successful would lead to another 154 orbits for the full survey.  The pilot study was supported by Director's Discretionary (DD) time while the full survey came from the Cycle 23 GO allocation (GO-13633, PI Spencer). The agreed upon success criteria for the pilot study was to find two cold-classical KBOs as a rudimentary check on the predicted size distribution down to this small size.  The pilot was successful in finding two objects and we were given clearance to conduct the full survey (83 search fields - 7.29 sq arcmin each, $\sim$605 sq arcmin total). 28 orbits from the full survey allocation were available for additional followup observations on newly discovered objects.

\subsection{Survey design}

The goal of the search was to find an object that \nh\ could reach for a close flyby investigation.  This goal allowed a much narrower scope for the search than a generic search for KBOs.  In this case, we could fine tune the data collection and processing to maximize our ability to find the object we sought.  It was entirely permissible to be blind to objects that \nh\ could not reach.

The area to search was defined by two primary constraints, (1) the region of space that \nh\ could reach after October 2015 and (2) the properties of the Kuiper Belt.  The predicted trajectory of \nh\ defined a time-dependent path through space and the amount of fuel on board set the size of a cone centered on that path.  Any KBO in that cone of 130 m/sec \deltav\ at the right time is encounterable.  The requirement of being in the cone at the right time provided a constraint on the orbit an object would need to have.  Our knowledge of the spatial distribution of the Kuiper Belt and the size distribution of KBOs was used with that positional constraint to construct a virtual population of KBOs that were a subset of all objects having the right orbital properties to be at the right place at the right time. 

There were only two dynamical groups that had a useful probability of being in the accessible cone: cold-classical and hot-classical objects.  Of these, the cold-classicals were strongly favored over hot-classicals with a ratio of 4:1.   We adopted a KBO population model which used the orbital distribution derived from the Canada-France Ecliptic Plane Survey (CFEPS) model of the Kuiper Belt \citet{Petit2011}. The size distribution was constrained by the absolute and apparent magnitude distribution of KBOs from recent models (\citet{Fuentes2010}, \citet{Fraser2014}). At the time of the survey proposal we deemed this ``FEPS/Fraser'' model to be somewhat conservative; the true number to be found might actually be higher.  We then used this orbit population to derive a set of semi-major axis, eccentricity, and inclination values that were consistent with this model of the Kuiper Belt.  From there, we selected random values of the angular elements.  If the randomly drawn elements placed the virtual object in the accessible cone, the elements were saved.  At the end of this process we had a selection of orbital elements representing virtual objects that were all encounterable.

Given the set of encounterable virtual objects, we could then project their positions back to some prior time and thus know the region of sky that needed to be searched to find a real object. Figure~\ref{fig-HST} shows a plot of the sky plane centered on the cloud of virtual objects in our sample.  Each dot represents one virtual object and the cloud is roughly described by a two-dimensional Gaussian distribution.  Each square represents one field of the WFC3 camera on HST and the fields are laid out to cover 80\% of the distribution.  We chose this coverage as a compromise between completeness and a practical request of time for the search. Covering a larger region would have taken much more time with quickly diminishing returns for that time investment.  The properties of this cloud of virtual objects was time-dependent, but we ignored this in designing the search, provided the data could be obtained quickly enough.  However, the orientation of the imaging camera is governed by the roll angle of the spacecraft as a function of time.  The figure depicts the image orientation at the time of the pilot survey, shown in green.  The remainder of the search fields, shown in blue, have the orientation expected for that part of the full search.  In reality, the true search grid could not avoid small gaps between adjacent fields, but the missed search area was small enough to ignore.

A special virtual orbit was chosen from the entire cloud.  This orbit was the closest to the center of the cloud and was used to set the pointing and tracking for HST\null. This strategy works well because in this case, the differential rate of motion of the cloud of virtual objects was 0.5~pixels (20 mas) during a single exposure for fields on the outer edge of the distribution thus avoiding any trailing losses for objects of interest.  Other KBOs, such as resonant or scattered disk objects, would have different rates of motion and were not expected to be detected at all.  All search observations were based on this same guiding orbit throughout the survey.

\begin{center}
\includegraphics[scale=0.5]{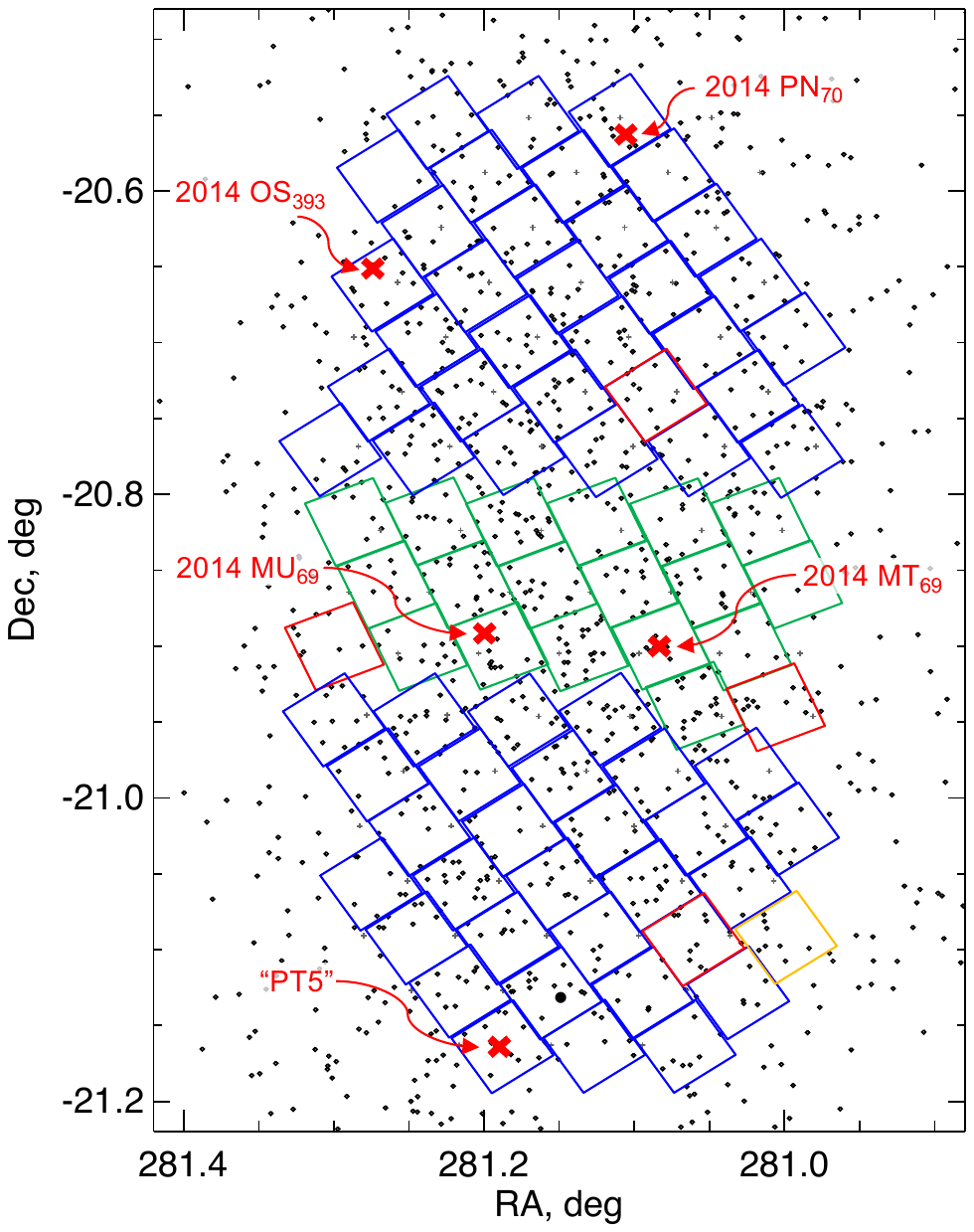}
\figcaption{\label{fig-HST}\scriptsize
A plot of the 83 HST search fields. Each box marks the approximate area of a combined WFC3/UVIS field using both detectors. 5 fields ended up not being useful due to guide star failures (red) or moonlight (orange). The green boxes represent the 40 orbit HST pilot fields and the blue show the additional full program fields excluding the repeated fields to follow up on the detected objects. The 5 discoveries from the search are marked with a red X and labeled with the eventual object designation. The black dots are a simulated KBO population from the CFEPS model with hot and cold Classical KBOs included. }
\end{center}

The search observations were designed as a pair of orbits looking at a given search field from the set shown in Fig.~\ref{fig-HST}.  There was a small amount of motion during one visit and sufficient motion between two visits to get an initial estimate of the target distance and orbital inclination.  This two orbit time span was sufficient to produce an initial orbit constraint that reduced the list of objects requiring followup and further orbit refinement, but was not in itself enough to know if an object was encounterable.  However, by the very nature of the survey design, all detected objects were highly likely to be interesting.

The optimal time spacing was for back-to-back orbits.  In some cases, this was impossible so 4.2\% of the fields had a one-orbit gap between the two epochs.  The increased time base was marginally useful for constraining the orbit of a detected object, but came at the price of a much larger search effort.  Even longer time gaps were categorically excluded from consideration in scheduling observations.  No other constraints were placed on scheduling the search observations other than to pick a set of fields near the center of the distribution for the pilot project.  In retrospect, we should have increased the lunar avoidance distance constraint rather than using the default settings.  Two of our search fields (marked in orange) were taken close enough to a full moon that the extra scattered light made the images useless for the survey. Table \ref{tbl-hst-short} gives a list of each HST field and the results of our searches and Fig.~\ref{fig-HSTsearch} shows a sample image pair. 


\begin{deluxetable}{cccccccccc}
\tablecaption{Summary of New Horizons Search Fields\label{tbl-hst-short}}
\tablewidth{0pt}
\tablehead{
\colhead{VisitID}&
\colhead{Target Name}&
\colhead{R.A.}&
\colhead{Dec.}&
\colhead{Starting Julian Date}&
\colhead{Calendar Date}&
\colhead{Visit}&
\colhead{Span}&
\colhead{Notes}\\
& & J2000& J2000& & & (hrs)& (hrs)& 
}
\startdata
icii01& KBO1-R6C3& 281.17843328& -20.93119929& 2456830.82368& 2014-06-22& 1.83& 4.66& guide star failure\\
icii03& KBO1-R6C4& 281.22944104& -20.93139568& 2456830.95131& 2014-06-22& 0.69& 3.19& & \\
icii02& KBO1-R6C3& 281.17459231& -20.93146782& 2456831.01766& 2014-06-22& 0.69& 4.66& guide star failure\\
icii04& KBO1-R6C4& 281.22681176& -20.93157966& 2456831.08404& 2014-06-22& 0.69& 3.19& & \\
icii05& KBO1-R7C3& 281.15852089& -20.89350538& 2456832.87601& 2014-06-24& 0.69& 3.19& & \\
icii07& KBO1-R7C4& 281.21071802& -20.89361286& 2456832.94238& 2014-06-24& 0.69& 3.19& 0720090F = 2014MT69\\
icii06& KBO1-R7C3& 281.15587505& -20.89369358& 2456833.00874& 2014-06-24& 0.69& 3.19& & \\
icii08& KBO1-R7C4& 281.20807172& -20.89380121& 2456833.07513& 2014-06-24& 0.69& 3.19& 0720090F = 2014MT69\\
icii09& KBO1-R7C5& 281.24569450& -20.89493518& 2456833.87156& 2014-06-25& 0.69& 3.19& & \\
icii13& KBO1-R7C7& 281.35140830& -20.89497445& 2456833.93794& 2014-06-25& 0.69& 3.19& & \\
icii14& KBO1-R7C7& 281.34875510& -20.89516486& 2456834.07068& 2014-06-25& 0.69& 3.19& & \\
icii11& KBO1-R7C6& 281.27929818& -20.89634889& 2456834.86709& 2014-06-26& 0.69& 3.19& 1110113Y = 2014MU69\\
icii15& KBO1-R7C8& 281.38501070& -20.89635644& 2456834.93348& 2014-06-26& 0.69& 3.19& guide star failure\\
icii12& KBO1-R7C6& 281.27663921& -20.89654118& 2456834.99983& 2014-06-26& 0.69& 3.19& 1110113Y = 2014MU69\\
icii16& KBO1-R7C8& 281.38235169& -20.89654884& 2456835.06620& 2014-06-26& 0.69& 3.19& guide star failure\\
\enddata
\tablecomments{\scriptsize
(This table is available in its entirety in machine-readable form in the online journal. A portion is shown here for guidance regarding its form and content.)
}
\end{deluxetable}

\begin{center}
\includegraphics[scale=0.80]{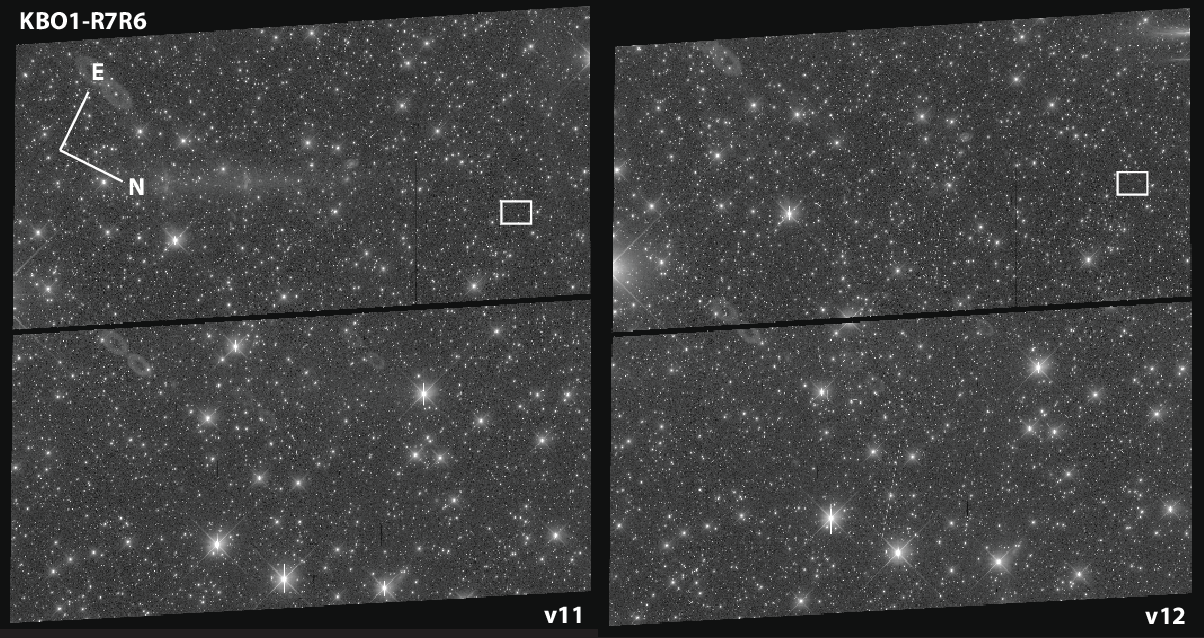}
\figcaption{\label{fig-HSTsearch}\scriptsize
Sample image from field KBO1-R7R6 in each of 2 visits separated by 1.05 hours. This field was part of the pilot search and 2014 MU69 is in the top chip on the right side in the area marked by the white box.  The visit numbers of HST program 13633 are shown on the lower right hand corner of each image.}
\end{center}

We used the WFC3/UVIS instrument with the F350LP filter for all observations.  The data volume for the full FOV readout led to a particular exposure time choice.  Five 370-second exposures fully fill a single visit without losing time to data management overhead.  Some visits needed a small tweak (no more than a few seconds) to get the sequence of observations to fit in a single visibility window.  The pair of visits was further constrained to be taken with the same guide star pair and same roll angle.  Due to the guiding orbit used, all background stars were trailed in the image.  This factor in the images required the development of new techniques to obtain accurate astrometry of the discovered KBOs.

\subsection{Data analysis}

Each set of five images from a single visit was processed to generate a stellar template image. Referenced and stacked on the star positions, the stack is a median of the images.  This template had only stars with cosmic-ray strikes and KBOs naturally filtered out.  Each image was then processed to subtract the template image.  This step suppressed the stars, including the wings of the stellar images.  In the cores, there is too much residual noise to permit searching those areas for KBOs.  In the wings, the KBOs can be completely hidden until the faint outer part of the star image is subtracted.  For most visits, the trailing of the stellar images is not exactly the same in all the orbits.  As a result, the stacked template does not completely remove the core of the star trails and leaves a characteristic residual image.

The next step was stacking the set of five images.  This process was done twice, once with and once without subtracting the stars.  Two things happen at the same time: the images are re-sampled to eliminate the distortion and register on the same linear 40 mas/pixel grid while shifting the image based on a given velocity vector.  Since the images are not Nyquist sampled nor cleaned of cosmic ray strikes and other blemishes, we chose to use bilinear interpolation for the re-sampling.  This choice was cosmetically much better than using a proper sinc interpolation.

\subsection{Rate Corrections}

In order to sensitize the search to KBOs on targetable orbits which differed from the nominal guiding orbit, we conducted a “shift-and-stack” digital tracking approach. Frames were shifted in software to counter the differential sky-plane motion between the nominal guide orbit and other potential targetable KBO orbits. As mentioned previously, the source population of targetable KBO orbits was drawn from the "CFEPS/Fraser" model and they were down-selected to a small set of guide orbits using the methods of \citet{Parker2010}. Unlike previous KBO searches using the digital tracking approach, frame-by-frame shifts were not parameterized as a grid of linear rates of motion in the sky plane; instead, the frame-by-frame motion of each guide orbit was computed and a frame-wise sky-plane offset was computed from the exact motion of a KBO on that orbit as seen from the HST reference frame. The down-selected orbit grid was chosen with a density such that no KBO in the large source sample of synthetic targetable KBOs would have a single frame-wise correction that differed by more than a single HST WFC3 pixel ($\sim$0.039 arcsec).

The sample of virtual objects used to design the search provided a means to calculate the bounds for the search in velocity space.  The size for the velocity bins was set such that the worst-case smear within a bin was no more than one pixel across the five images in a single visit.  There are two important consequences of this design.  First, an object is unlikely to be seen in more than one velocity bin, especially at the detection limit of the survey.  Second, the number of discrete stacks to search varied with the time-span of the set.  For observation pairs of the same field that occurred on back-to-back HST orbits (1.05 hours), typically of order 20 guide orbits were required to achieve the required sampling density leading to 20 image stacks to search. For observation pairs that spanned larger periods of time, a rapidly-increasing density ($\propto t^2$) of guide orbits was required and we ended up with as many as 80 stacks to search.  Due to the increased workload and need for rapid turn-around, we adjusted our observation constraints as the program progressed to strongly prefer the shorter time spacing.

Figure  \ref{fig-offsets} illustrates one resulting grid of linear sky-plane offsets — for the entire population of targetable orbits as well as the chosen guide orbits — from the first image in a single orbit to the last, computed for visit icii11 (where (486958) Arrokoth was discovered). 

\begin{center}
\includegraphics[scale=1.0,trim=3cm 9cm 4cm 6cm, clip=true]{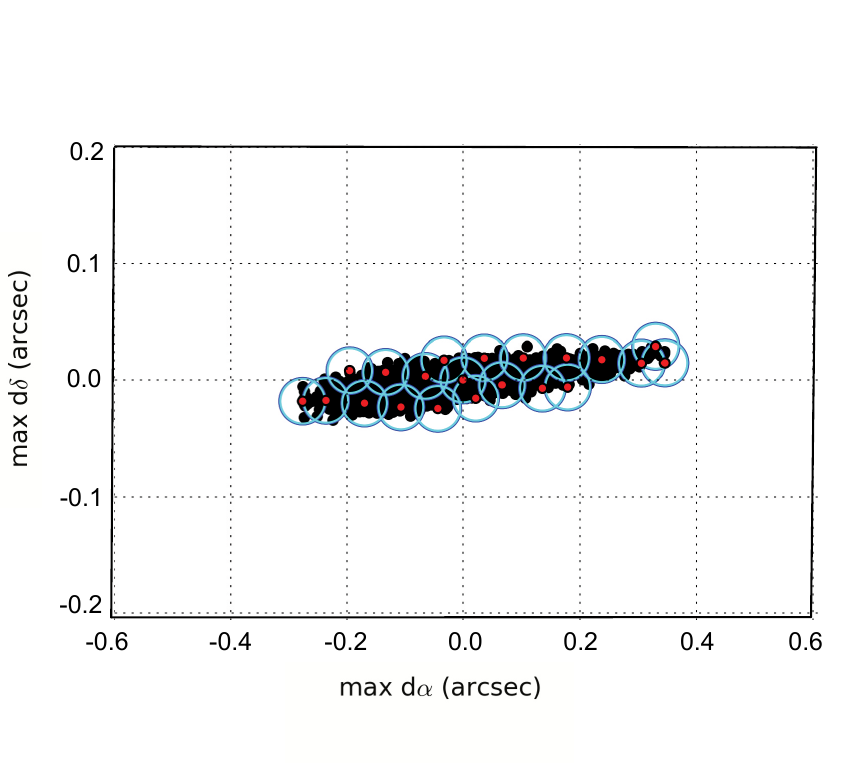}

\figcaption{\label{fig-offsets}\scriptsize
Sky-plane offsets from the nominal guide orbit over the course of a single HST orbit for visit icii11. Black points are the motion of all targetable KBOs from the synthetic orbit sample. Red points are the sky plane motions of the chosen guide orbit grid, and blue circles are the buffers of sky motion each covers to within a single HST WFC3 pixel. The point at (0,0) is the nominal guide orbit.}
\end{center}

We had five nearly independent teams pursuing their own search tools all built more or less around the same concepts of the shift-stacking.  This discussion largely concentrates on the pipeline written by Buie since that was responsible for all five of the discoveries within the survey project. We will touch on aspects of the other systems as needed.

In one of the final stages of development for the software, we implanted fake objects into our data.  This was designed primarily to measure the sensitivity limits of our data.  In the end, it was far more important as a training dataset to test the software.  During the pilot phase, significant portions of the data were scanned multiple times with no results. The test data convinced us that we had the expected sensitivity and that the software was only finding objects when the shift was zero.  This was quickly determined to indicate a sign error in the velocity stacking and Buie's detections in the pilot survey quickly followed after correcting the error.

\subsubsection{Searching for Candidates}

The search methodology employed a mix of automated and manual steps.  Each of the shift-stacked images was scanned for sources.  The scanning was done with {\tt findsrc} with a background window (WINDOW) of 20 pixels and a separation between the window and the target pixel (GAP) of 6 pixels.  The detection threshold was set at 3$\sigma$.  With this setting, the number of false detections per stacked images was quite large -- typically in the range of 3000 to 6000 per velocity stacked image and well beyond the ability to visually scan all candidates. To reduce the number of false positives, the pairs of visits were combined.  A candidate worth examining was required to be seen on both visits with the same velocity and at a relative position within the stacks that was consistent with that velocity.  A threshold on this test was set to 1.7 pixels multiplied by the ratio of the largest velocity to the mean  velocity in the bin.  The value of 1.7 was empirically tuned to set the number of  false positives to a manageable level.  No threshold on consistent brightness was imposed at this step.  For good data, there were typically $\sim$30 candidates per image per velocity, but the worst visit pairs had 10 times more.

The visit-combined detection lists fed a tool that built a composite graphic image summarizing the detection candidate.  This graphic contained small thumbnail images of the detections in each stack as well as the relevant portion of the original image.  In all cases, the thumbnails were centered to the nearest pixel of the position of the candidate object.  Also included was an aperture based magnitude estimate.  Figure \ref{fig-good} shows the vetting graphic for (486958) Arrokoth.  In this figure, the circle in the top left image is drawn at the measured location for the first visit stack.  The top right image shows the stack for the second visit and the circle is drawn at the location that is predicted based on the stacking velocity.  The size of the circles are comparable to the size of the positional coincidence imposed to find the candidates.  The bottom of the figure contains two rows of five thumbnail images each. The top set of five is from the first visit and the bottom set is from the second visit.  Each set of five is ordered in increasing time from left to right.  The only processing done on the bottom images is the suppression of background stars.  In this example, stars are seen as white streaks with some black pixels around the edge.  Other than the object itself, the rest of the bright points are cosmic ray strikes that do not correspond to any object.  This example shows the success of our spurious noise suppression algorithm.  In the stacks there are a few low-level artifacts that are easily dismissed.  For example, in the first stack there is a dark smudge to the lower right of the object.  This is due to the residual of a star that was not perfectly subtracted.  In both visits, there are other faint bright smudges but these are far enough away and at uncorrelated velocities so they are never considered.  This graphic was designed to allow quick examination of all candidates for a determination of being real or not.  In this case, there is clear evidence in all ten images for a source that leads to the two stacks.  One of these ten (first image in second row) shows the object as being very close to a cosmic ray strike. The estimated apparent magnitude in both stacks is superimposed on the graphic below each detection.

\begin{center}
\includegraphics[scale=0.60]{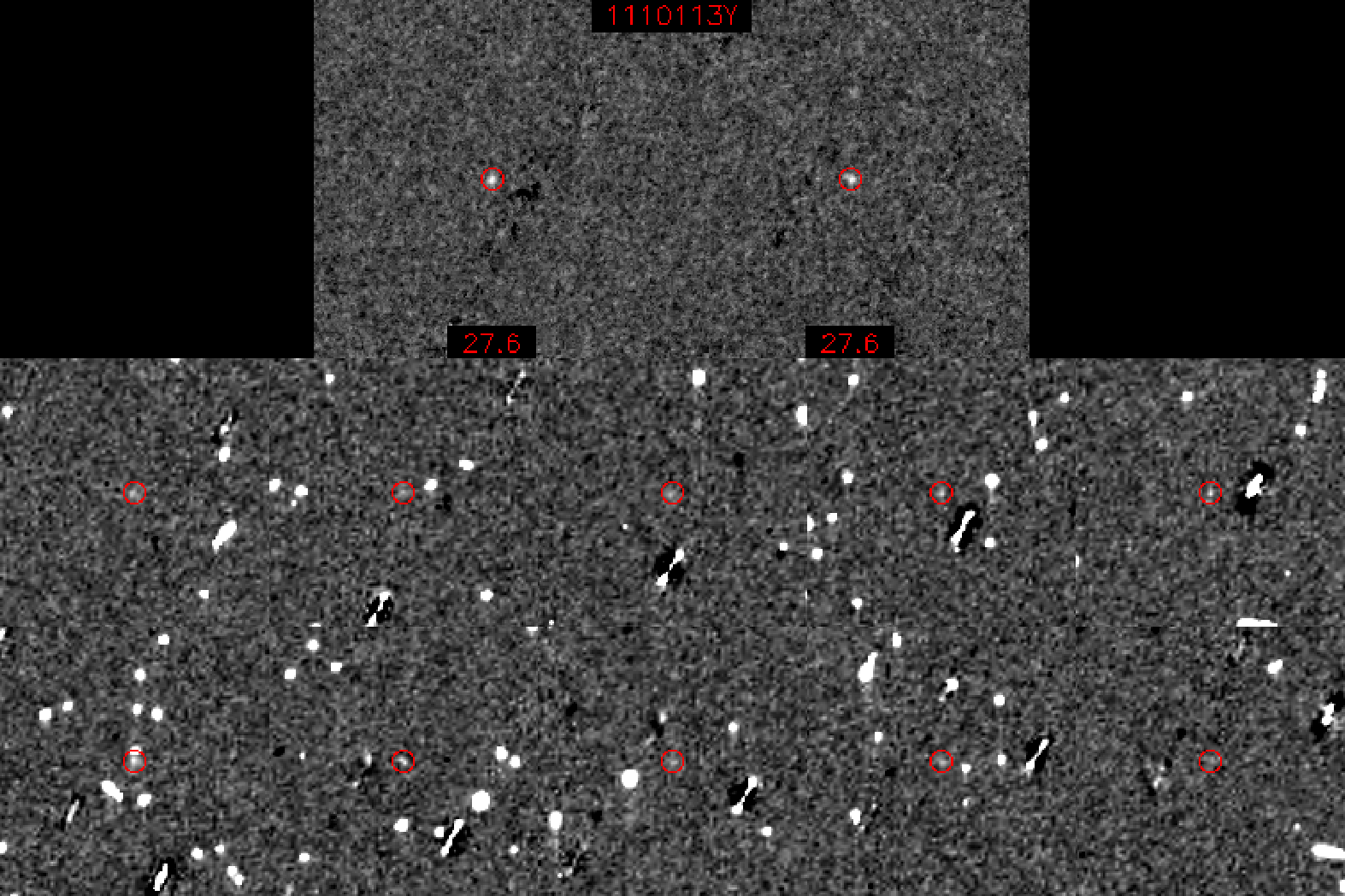}
\figcaption{\label{fig-good}\scriptsize
Confirmation graphic for discovery of (486958) Arrokoth.
}
\end{center}

Figure \ref{fig-false} shows an example of a false positive that was discarded during the vetting process.  The image stacks show an apparent source and is seen to agree with the velocity.  The first clue against it being real is the lack of agreement in the magnitudes.  This is apparent both in the numbers and visual appearance.  The individual images provide the rest of the evidence against.  In this case, the first visit alone is enough to discard the candidate. Image 1 has no source.  Image 2 and 3 are on the edge of a subtracted star.  Image 4 has a bright cosmic ray strike.  Image 5 looks more like a background noise fluctuation.  Similar logic applies to the second visit, Image 1 and 5 show nothing.  In this case, there was clearly a confluence of spurious detections in each set of 5 that triggered a false positive detection. This combined graphic made it very easy to make a decision on the candidate.

\begin{center}
\includegraphics[scale=0.60]{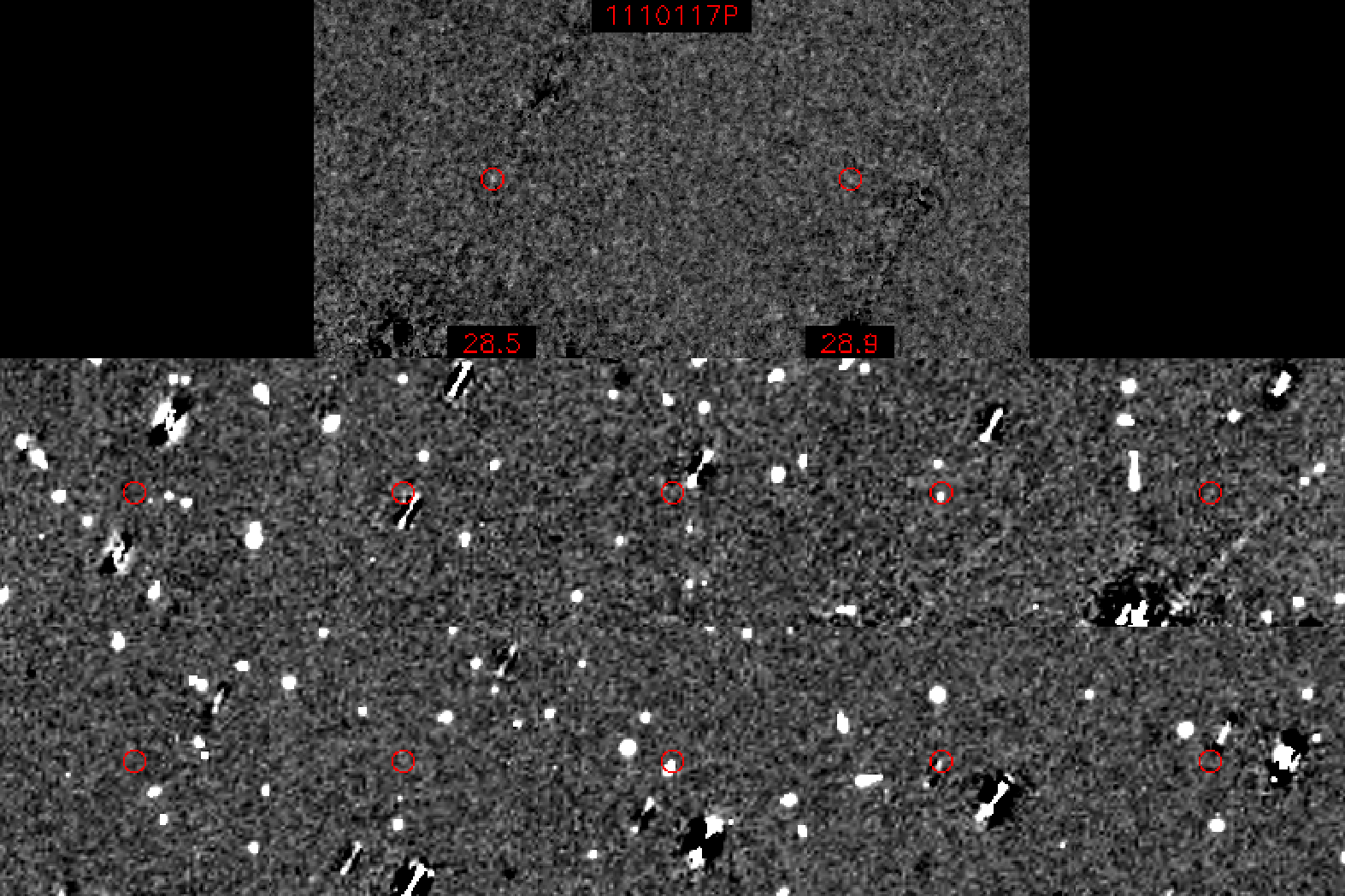}
\figcaption{\label{fig-false}\scriptsize
Confirmation graphics for a typical false positive detection.
}
\end{center}

It is perhaps possible to encode all of the logic used to grade candidate sources to avoid the manual step.  No attempt was made to do this.  The vetting process was already very fast and took much less than 10\% of the total time to process an entire visit pair.  Also, the time required to develop a fully tested automated tool was completely impractical on the timescale in which we needed to complete our search efforts.

\subsubsection{Cross-Checking Searches}

In order to cross-check our discoveries and to make sure no potentially real object was missed, our team compared the results of two independent search algorithms. The second algorithm worked with the HST ``drizzled'' images which are both flat-fielded and distortion-corrected through the HST calibration pipeline. The intensity unit for these images is in electrons/s. Figure \ref{fig-hal} records the steps of the second pipeline. Basically the full visit list of images for a pair of visits was read into a buffer. A star template was created by co-adding the images weighted by their exposure times, excluding high sigma pixels, and stacking on the positions of a selection of 6 user identified stars, ideally spread across the array. This process could be iterative if necessary. Once the star template for each visit was created it was subtracted from each individual image and a composite image for each visit was created for the family of KBO orbits provided by the rate correction previously discussed (i.e. if the family of orbit rates had 10 entries one would have 10 composite images for each HST visit to evaluate). The composite images for each pair of HST visits to the same position and at the same shift rate was then automatically searched for KBOs with image stamps of the discovered objects saved as .png files for visual confirmation. A display tool allowed one to blink-compare the results and the individual shifted and differenced images could also be brought up for searching or confirmation. 

\begin{center}
\includegraphics[scale=0.60]{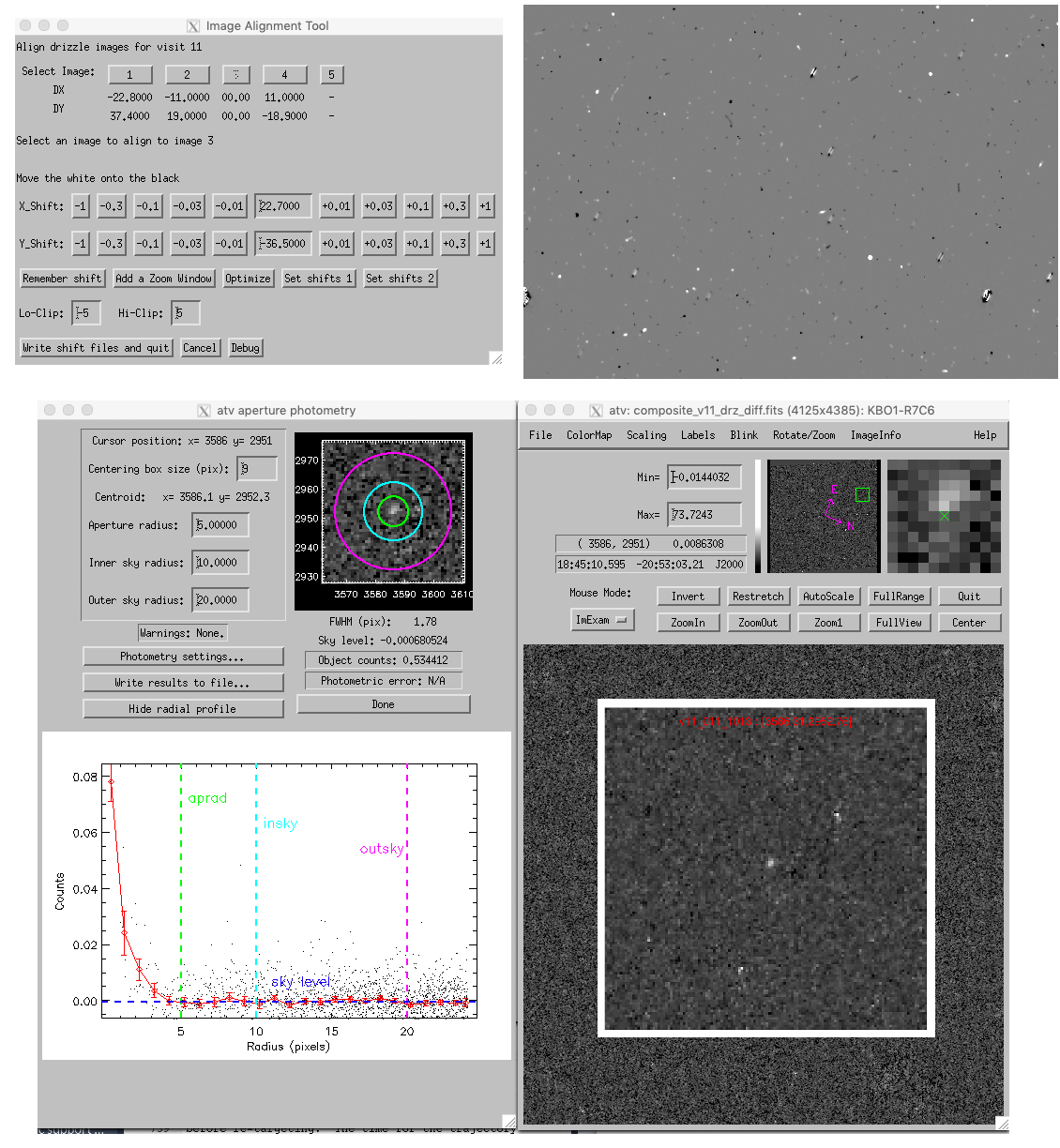}
\figcaption{\label{fig-hal}\scriptsize
Sample images from our second reduction pipeline for the 2014 MU69 discovery image. (Top) Images from a single visit are shifted both by eye, then through an optimization routine to remove the stars. (Bottom) The star subtracted images are then searchable for KBOs in the field. The inlay on the bottom right is an extracted .png used for evaluation of reality.}
\end{center}

In addition to identifying potential targets via an algorithm, an independent manual-visual search was conducted in parallel.  To systematically search each HST visit, a custom IDL program (``HSTlooker'')  was created to present side by side views of stacked and original images from corresponding visits, matching the predicted locations of an object from one visit to the next, and connecting the pixels of the stacked image to the data that created them.  This tool drew from the prior work on searching the ground-based data.  The \textit{HSTlooker} program allowed users to cycle through the stacked and original images for each velocity shift and  systematically cover the full frame to ensure no part of the image was missed.  Each portion of the shifted and stacked images was visually scanned to identify any artifacts that persisted through median combination.  Artifacts found in the stacks were verified via visual examination of the five original images to identify the source of the bright median pixel, be it a KBO or incidental alignment of cosmic rays, star subtraction remnants, scattered light or bad column. All five KBOs were independently ``discovered'' via this process, albeit well after the objects were first located with the computer driven detection processes.

\subsection{Measuring KBO Positions}

The next step in the processing was to obtain accurate positions for the validated objects.  These measurements were always based on the original images, not the stacks.  To start, a manually supervised aperture-based centroid was used.  At this step, if the centroid looked reasonable for the position of the object, it was kept.  If not, a manually derived position based on judgement of the position of the object was used. This aperture measurement is called a navigation position and fixes the region of interest for the next phase.

Given a navigation location, a region of interest with a half-width of 9 pixels was defined.  Within the region, an PSF-fit was performed on the object with the navigation position as its starting point. The PSF used was generated by TinyTim \citep{Krist2011}.  This location is saved separately from the navigation location.  We used bad-pixel masks for the PSF fitting step to avoid contamination from cosmic ray strikes or residual blemishes from imperfectly subtract background stars.  Most of the time the measurements were derived from the star-subtracted images but occasionally a better result was possible on the untouched data.

\subsection{Astrometric support catalog}

The support catalog developed for the ground-based search turned out to be ideally suited for the HST data as well.  There was useful overlap between the catalog and well-measured stars in the field such that there were generally $\sim$200 stars for a single detector in WFC3. The much improved image scale and smaller PSF enabled even more precise astrometry and the CFHT-based catalog held up amazingly well.  Despite the quality of the catalog, the HST data clearly showed the limitations of the underlying astrometric catalogs.  This led to the development of more and more intricate corrections to minimize the obvious systematic errors. None of these difficulties affected our ability to determine initial orbits or predict future positions for followup but it became clear that final targeting of an object would benefit from a better catalog. When the Gaia catalog was released in 2016 \citep{Gaia2016a,Gaia2016b,Lindegren2016}, we switched to using that as the reference catalog for the CFHT data.  We will return to a more in-depth discussion of our support catalogs in section 4.7.

\subsection{Measuring Star Positions and Updating the WCS}

The world-coordinate system (WCS) header information \citep{Gennaro2018} provides the means to transform pixel coordinates from the image to sky-plane coordinates.  The WFC3 camera suffers from significant distortions and must be accurately known for precision astrometry.  \citet{Kozhurina2016} did extensive work to map out the image scale and focal plane distortion. This map provides relative calibration down to the level of 0.01 pix (0.15 $\micron$) or better between any pair of pixels in the image.  The absolute pointing information comes from the guide stars observed with the fine-guidance sensors (FGS) that are providing pointing and guiding control.  These stars are known to have typical errors of 0.1-0.2 arcsec but sometimes are significantly worse than that.  As a consequence, the absolute pointing information is of a much lower quality and must be improved.

To start with we measured the stars that are in the support catalog with PSF fitting.  Again, we used the TinyTim PSF appropriate for the location of each star on the detector.  Still, the star fitting poses one additional challenge due to tracking of the KBO cloud.  All star images show significant trailing which changes on every exposure.  The dominant apparent motion of the stars is due to parallax as HST moves around the Earth in its orbit. Even though the images look like linear trails most of the time, the trailing is actually a time-dependent quantity and the residence time of a star at any given point on the trail varies throughout the exposure.

To address this time-variable trail, we modified the PSF fitting process by introducing a convolution kernel into the model image production. The routine {\tt jitterk.pro} takes the time-dependent position and computes the necessary kernel.  More details about this tool can be found in Appendix~\ref{app-idl}.  The time-dependent position is derived from the HST-centric position of the guiding orbit that represents the center of the cloud of candidate objects.  Also required was the definitive post-observation HST orbit provided by JPL\footnote{\url{ http://naif.jpl.nasa.gov/pub/naif/HST/kernels/spk/hst.bsp}}.

One minor limitation of our tools is that the image distortion was not taken into account for the star smear kernel.  The kernel is computed for the center of the array and then used over the entire image.  There are very subtle shifting features in the residuals from the fit across the field that look like the model image is trailed at a slightly different angle from the image. We looked for patterns in the astrometric analysis that could be traced back to this approximation but no convincing signatures were found. Once the new Gaia catalog \citep{Gaia2018} is included the pipeline WCS can be dramatically improved but this will only ever be constrained by the position of two stars.  By using more stars in the field, the solutions can always be improved.

Our final WCS update consisted of solving for a translation of the reference position in the WCS stored as CRVAL1 and CRVAL2 \citep{Gonzaga2012} and a small angle rotation to the CD matrix.  All other components of the WCS remain untouched.  In practice, the rotation corrections are extremely small and amount to worst-case corrections in the image corners that are much smaller than a pixel (a few milliarcsecs at most).  The apparent offsets in our data are larger than one would expect from just the errors in the guide star catalog but this is due to the bookkeeping of the pointing information for data taken with moving target tracking.  Our pointing solution update removes all of these effects.

We received a report of errors in the reconstructed trajectory of HST deduced from astrometric observations of C/2013 A1 (Siding Springs) relevant to its close passage to Mars \citep{Farnocchia2016}.  In that work, the internal errors from the HST observations did not agree with the post-fit errors when combined with other sources.  In that work the HST data were de-weighted to 0.3 arcsec due to presumed errors in the HST position at the level of 10 km.  If this interpretation is correct, we should presume this same error is present in our data. However, our targets are nearly 30 times further away and perhaps that reduces the problem.  Taken at face value, this issue indicates HST positional errors could contribute as much as 10 mas to the total error budget.  However, in light of our successful occultation work on Arrokoth \citep{Buie2020}, this level of error in our astrometry is just not seen and we have no evidence in our data and reductions for any problems with the reconstructed positions of HST that supported our astrometric work.

\subsection{Final Astrometry on KBOs}

The KBO position for each image is derived using the methodology described in \citet{Porter2018} (sections 2.3-2.4) and further discussed in \citet{Porter2022}. In short, a MCMC algorithim is used in combination with Tiny Tim PSFs to determine the optimum KBO pixel position PDF which then has the WCS applied to extract the R.A./Dec sky-plane position. All measurements are provided in J2000 (EME200).  Table \ref{tbl-hstast-full2} lists the positions for all objects measured in the HST survey.   


\startlongtable
\begin{deluxetable}{ccccccccccc}
\tablecaption{HST Astrometry\label{tbl-hstast-full2}}
\tablewidth{0pt}
\tablehead{
\colhead{Object}&
\colhead{UT Date}&
\colhead{RA}&
\colhead{Dec}&
\colhead{R Mag.}&
\colhead{}&
\colhead{Object}&
\colhead{UT Date}&
\colhead{RA}&
\colhead{Dec}&
\colhead{R Mag.}
}
\startdata
486958  &2014 06  26.36924 &18 45 10.671&-20 53 03.06 & 26.8 &&486958  &2016 07  03.64376 &18 55 39.182&-20 43 53.57 & 27.4 \\     
486958  &2014 06  26.37539 &18 45 10.636&-20 53 03.10 & 27.0 &&486958  &2016 07  03.64965 &18 55 39.142&-20 43 53.65 & 27.4 \\      
486958  &2014 06  26.38153 &18 45 10.595&-20 53 03.18 & 27.0 &&486958  &2016 07  03.65595 &18 55 39.107&-20 43 53.67 & 27.3 \\      
486958  &2014 06  26.38768 &18 45 10.557&-20 53 03.23 & 26.5 &&486958  &2016 07  03.66214 &18 55 39.070&-20 43 53.61 & 28.2 \\      
486958  &2014 06  26.39382 &18 45 10.521&-20 53 03.31 & 26.7 &&486958  &2016 07  03.66832 &18 55 39.036&-20 43 53.48 & 28.4 \\      
486958  &2014 06  26.50197 &18 45 10.019&-20 53 03.77 & 26.2 &&486958  &2016 07  03.71002 &18 55 38.855&-20 43 54.02 & 27.8 \\      
486958  &2014 06  26.50812 &18 45 09.983&-20 53 03.80 & 27.1 &&486958  &2016 07  03.71591 &18 55 38.819&-20 43 54.09 & 28.4 \\      
486958  &2014 06  26.51426 &18 45 09.943&-20 53 03.88 & 27.2 &&486958  &2016 07  03.72222 &18 55 38.780&-20 43 54.12 & 28.7 \\      
486958  &2014 06  26.52041 &18 45 09.905&-20 53 03.95 & 27.3 &&486958  &2016 07  03.72840 &18 55 38.742&-20 43 54.05 & 27.9 \\      
486958  &2014 06  26.52656 &18 45 09.870&-20 53 04.01 & 27.4 &&486958  &2016 07  03.73458 &18 55 38.708&-20 43 53.99 & 27.7 \\     
486958  &2014 08  02.54538 &18 42 15.467&-20 56 43.22 & 27.5 &&486958  &2016 07  03.77627 &18 55 38.528&-20 43 54.50 & 28.0 \\     
486958  &2014 08  02.55152 &18 42 15.433&-20 56 43.22 & 27.1 &&486958  &2016 07  03.78216 &18 55 38.493&-20 43 54.55 & 27.8 \\     
486958  &2014 08  02.55767 &18 42 15.402&-20 56 43.19 & 25.1 &&486958  &2016 07  03.78847 &18 55 38.454&-20 43 54.53 & 28.1 \\      
486958  &2014 08  02.56381 &18 42 15.373&-20 56 43.16 & 27.0 &&486958  &2016 07  03.79465 &18 55 38.413&-20 43 54.53 & 29.2 \\      
486958  &2014 08  03.73983 &18 42 10.581&-20 56 50.45 & 26.5 &&486958  &2016 07  03.80083 &18 55 38.381&-20 43 54.45 & 28.2 \\      
486958  &2014 08  03.74598 &18 42 10.547&-20 56 50.41 & 27.0 &&486958  &2016 07  25.70106 &18 53 53.037&-20 46 32.22 & 27.4 \\      
486958  &2014 08  03.75212 &18 42 10.517&-20 56 50.40 & 28.1 &&486958  &2016 07  25.70717 &18 53 53.003&-20 46 32.26 & 27.1 \\      
486958  &2014 08  03.75827 &18 42 10.499&-20 56 50.37 & 27.6 &&486958  &2016 07  25.71328 &18 53 52.967&-20 46 32.32 & 26.8 \\      
486958  &2014 08  21.31900 &18 41 08.704&-20 58 32.35 & 26.2 &&486958  &2016 07  25.71939 &18 53 52.930&-20 46 32.37 & 27.3 \\      
486958  &2014 08  21.32514 &18 41 08.680&-20 58 32.41 & 25.8 &&486958  &2016 07  25.72550 &18 53 52.897&-20 46 32.43 & 27.5 \\     
486958  &2014 08  21.33129 &18 41 08.655&-20 58 32.52 & 26.7 &&486958  &2016 10  21.20737 &18 51 48.303&-20 53 07.58 & 28.3 \\ 
486958  &2014 08  21.33744 &18 41 08.627&-20 58 32.59 & 27.2 &&486958  &2016 10  21.21348 &18 51 48.318&-20 53 07.64 & 26.4 \\ 
486958  &2014 08  21.34358 &18 41 08.606&-20 58 32.63 & 26.6 &&486958  &2016 10  21.22570 &18 51 48.335&-20 53 07.52 & 26.8 \\ 
486958  &2014 08  21.38535 &18 41 08.512&-20 58 32.73 & 27.4 &&486958  &2016 10  21.23181 &18 51 48.345&-20 53 07.45 & 27.7 \\ 
486958  &2014 08  21.39150 &18 41 08.486&-20 58 32.79 & 27.0 &&486958  &2016 10  24.25448 &18 51 56.623&-20 53 06.39 & 26.3 \\ 
486958  &2014 08  21.39764 &18 41 08.462&-20 58 32.88 & 26.8 &&486958  &2016 10  24.26059 &18 51 56.630&-20 53 06.39 & 26.5 \\ 
486958  &2014 08  21.40379 &18 41 08.436&-20 58 32.94 & 26.5 &&486958  &2016 10  24.26670 &18 51 56.644&-20 53 06.31 & 26.5 \\     
486958  &2014 08  21.40994 &18 41 08.412&-20 58 33.00 & 26.7 &&486958  &2016 10  24.27281 &18 51 56.654&-20 53 06.21 & 27.1 \\     
486958  &2014 08  23.17695 &18 41 03.420&-20 58 42.47 & 27.5 &&486958  &2016 10  24.27892 &18 51 56.667&-20 53 06.16 & 26.2 \\     
486958  &2014 08  23.18310 &18 41 03.397&-20 58 42.52 & 27.0 &&14OS393 &2014 06  25.40498 &18 45 34.850&-20 37 37.23 &       \\    
486958  &2014 08  23.18924 &18 41 03.373&-20 58 42.64 & 27.2 &&14OS393 &2014 06  25.42497 &18 45 34.740&-20 37 37.33 &       \\
486958  &2014 08  23.19539 &18 41 03.350&-20 58 42.70 & 27.7 &&14OS393 &2014 06  25.44258 &18 45 34.668&-20 37 37.35 &       \\
486958  &2014 08  23.20153 &18 41 03.324&-20 58 42.77 & 26.9 &&14OS393 &2014 06  25.53807 &18 45 34.190&-20 37 37.94 &       \\     
486958  &2014 08  23.24332 &18 41 03.236&-20 58 42.85 & 26.1 &&14OS393 &2014 06  27.36456 &18 45 25.216&-20 37 48.44 &       \\     
486958  &2014 08  23.24946 &18 41 03.213&-20 58 42.90 & 26.1 &&14OS393 &2014 06  27.37556 &18 45 25.161&-20 37 48.47 &       \\      
486958  &2014 08  23.25561 &18 41 03.189&-20 58 42.98 & 26.3 &&14OS393 &2014 07  30.75208 &18 42 46.475&-20 41 23.18 & 26.6 \\      
486958  &2014 08  23.26175 &18 41 03.163&-20 58 43.05 & 26.9 &&14OS393 &2014 07  30.77051 &18 42 46.374&-20 41 23.15 & 26.1 \\      
486958  &2014 08  23.26790 &18 41 03.140&-20 58 43.13 & 26.1 &&14OS393 &2014 07  30.96961 &18 42 45.524&-20 41 24.49 & 26.1 \\      
486958  &2014 10  15.06699 &18 40 43.948&-21 01 45.51 & 26.5 &&14OS393 &2014 07  30.97576 &18 42 45.495&-20 41 24.45 & 26.0 \\      
486958  &2014 10  15.07313 &18 40 43.955&-21 01 45.55 & 26.2 &&14OS393 &2014 08  21.47018 &18 41 27.395&-20 43 44.28 & 26.1 \\     
486958  &2014 10  15.07928 &18 40 43.962&-21 01 45.58 & 25.8 &&14OS393 &2014 08  21.54268 &18 41 27.179&-20 43 44.74 & 25.7 \\
486958  &2014 10  15.08542 &18 40 43.966&-21 01 45.64 & 25.4 &&14OS393 &2014 08  23.30972 &18 41 22.194&-20 43 55.43 & 26.3 \\
486958  &2014 10  15.09157 &18 40 43.977&-21 01 45.65 & 25.5 &&14OS393 &2014 08  23.37606 &18 41 22.010&-20 43 55.84 & 26.2 \\
486958  &2014 10  15.13334 &18 40 44.093&-21 01 45.60 & 26.7 &&14OS393 &2014 08  23.38835 &18 41 21.963&-20 43 55.99 & 26.2 \\     
486958  &2014 10  15.13949 &18 40 44.100&-21 01 45.61 & 26.9 &&14OS393 &2014 08  23.39450 &18 41 21.938&-20 43 56.08 & 26.3 \\     
486958  &2014 10  15.14563 &18 40 44.107&-21 01 45.65 & 27.1 &&14OS393 &2014 08  23.40064 &18 41 21.916&-20 43 56.16 & 25.9 \\     
486958  &2014 10  15.15178 &18 40 44.113&-21 01 45.71 & 27.4 &&14OS393 &2014 10  16.06833 &18 41 05.096&-20 47 37.17 & 26.4 \\      
486958  &2014 10  15.15792 &18 40 44.121&-21 01 45.73 & 26.8 &&14OS393 &2014 10  24.10764 &18 41 26.554&-20 47 46.18 & 26.0 \\      
486958  &2014 10  16.19482 &18 40 46.483&-21 01 46.56 & 26.9 &&14PN70  &2014 08  06.45430 &18 41 40.006&-20 37 42.58 & 26.6 \\      
486958  &2014 10  16.20097 &18 40 46.508&-21 01 46.94 & 29.0 &&14PN70  &2014 08  06.46659 &18 41 39.945&-20 37 42.56 & 26.3 \\      
486958  &2014 10  16.20711 &18 40 46.497&-21 01 46.67 & 27.7 &&14PN70  &2014 08  06.47274 &18 41 39.915&-20 37 42.57 & 26.5 \\ 
486958  &2014 10  16.21326 &18 40 46.506&-21 01 46.71 & 26.9 &&14PN70  &2014 08  06.65952 &18 41 39.214&-20 37 43.87 & 26.0 \\      
486958  &2014 10  16.21941 &18 40 46.513&-21 01 46.73 & 28.4 &&14PN70  &2014 08  25.25857 &18 40 39.013&-20 39 41.71 & 26.6 \\      
486958  &2014 10  22.36438 &18 41 02.502&-21 01 49.71 & 26.0 &&14PN70  &2014 09  02.06975 &18 40 21.481&-20 40 26.87 & 26.7 \\      
486958  &2014 10  22.37053 &18 41 02.513&-21 01 49.82 & 26.9 &&14PN70  &2014 09  02.07590 &18 40 21.462&-20 40 26.96 & 27.0 \\      
486958  &2014 10  22.37667 &18 41 02.524&-21 01 49.87 & 26.9 &&14PN70  &2014 09  02.08204 &18 40 21.443&-20 40 27.05 & 26.7 \\      
486958  &2014 10  22.38282 &18 41 02.533&-21 01 49.94 & 27.0 &&14PN70  &2014 09  02.08819 &18 40 21.424&-20 40 27.08 & 26.6 \\      
486958  &2014 10  22.38897 &18 41 02.545&-21 01 49.95 & 27.2 &&14PN70  &2014 09  02.12996 &18 40 21.369&-20 40 27.11 & 26.3 \\      
486958  &2015 05  04.69178 &18 54 04.856&-20 45 17.13 & 28.3 &&14PN70  &2014 09  02.13611 &18 40 21.354&-20 40 27.24 & 26.7 \\      
486958  &2015 05  04.69791 &18 54 04.834&-20 45 17.09 & 27.0 &&14PN70  &2014 09  02.14225 &18 40 21.335&-20 40 27.35 & 26.5 \\     
486958  &2015 05  04.70403 &18 54 04.815&-20 45 17.03 & 27.1 &&14PN70  &2014 09  02.14840 &18 40 21.315&-20 40 27.42 & 26.2 \\ 
486958  &2015 05  04.71015 &18 54 04.796&-20 45 16.99 & 26.7 &&14PN70  &2014 09  02.15454 &18 40 21.299&-20 40 27.47 & 25.5 \\     
486958  &2015 07  04.60444 &18 50 04.474&-20 49 11.58 & 27.0 &&14PN70  &2014 10  15.27217 &18 40 25.447&-20 43 14.81 & 26.4 \\     
486958  &2015 07  04.61057 &18 50 04.437&-20 49 11.60 & 26.8 &&14PN70  &2014 10  22.90739 &18 40 44.701&-20 43 25.60 & 26.1 \\     
486958  &2015 07  04.61670 &18 50 04.399&-20 49 11.61 & 27.4 &&14PN70  &2014 10  22.91968 &18 40 44.723&-20 43 25.68 & 24.5 \\     
486958  &2015 07  04.62284 &18 50 04.360&-20 49 11.62 & 26.8 &&14PN70  &2015 05  04.75190 &18 53 27.895&-20 28 23.23 & 26.7 \\     
486958  &2015 07  04.62897 &18 50 04.324&-20 49 11.64 & 26.8 &&14PN70  &2015 05  04.75802 &18 53 27.878&-20 28 23.21 & 26.4 \\     
486958  &2016 03  15.00829 &18 59 12.343&-20 42 13.66 & 27.0 &&14PN70  &2015 05  04.76414 &18 53 27.858&-20 28 23.16 & 26.3 \\     
486958  &2016 03  15.01440 &18 59 12.355&-20 42 13.66 & 27.1 &&14PN70  &2015 05  04.77027 &18 53 27.836&-20 28 23.11 & 26.7 \\     
486958  &2016 03  15.02051 &18 59 12.363&-20 42 13.62 & 27.3 &&14PN70  &2015 05  04.77639 &18 53 27.821&-20 28 23.06 & 26.3 \\     
486958  &2016 03  15.02662 &18 59 12.373&-20 42 13.52 & 26.7 &&14MT69  &2014 06  24.44453 &18 44 46.035&-20 52 50.73 & 26.9 \\      
486958  &2016 03  15.03273 &18 59 12.386&-20 42 13.41 & 27.1 &&14MT69  &2014 06  24.45068 &18 44 45.998&-20 52 50.81 & 27.8 \\      
486958  &2016 05  15.09107 &18 59 07.565&-20 39 50.46 & 26.6 &&14MT69  &2014 06  24.46297 &18 44 45.919&-20 52 50.92 & 29.0 \\      
486958  &2016 05  15.09718 &18 59 07.541&-20 39 50.50 & 26.5 &&14MT69  &2014 06  24.59571 &18 44 45.266&-20 52 51.65 & 26.8 \\      
486958  &2016 05  15.10329 &18 59 07.518&-20 39 50.47 & 26.6 &&14MT69  &2014 08  02.61786 &18 41 40.228&-20 56 50.25 & 27.5 \\      
486958  &2016 05  15.10940 &18 59 07.491&-20 39 50.45 & 27.0 &&14MT69  &2014 08  02.63016 &18 41 40.162&-20 56 50.20 & 27.1 \\      
486958  &2016 05  15.11551 &18 59 07.468&-20 39 50.38 & 26.8 &&14MT69  &2014 08  03.80002 &18 41 35.393&-20 56 57.75 & 27.1 \\      
486958  &2016 07  03.57750 &18 55 39.511&-20 43 53.11 & 27.5 &&14MT69  &2014 08  03.80616 &18 41 35.362&-20 56 57.74 & 27.5 \\      
486958  &2016 07  03.58339 &18 55 39.474&-20 43 53.17 & 26.6 &&14MT69  &2014 08  03.81231 &18 41 35.328&-20 56 57.75 & 28.1 \\      
486958  &2016 07  03.59587 &18 55 39.400&-20 43 53.14 & 28.8 &&14MT69  &2014 08  03.81845 &18 41 35.297&-20 56 57.72 & 27.0 \\      
486958  &2016 07  03.60205 &18 55 39.361&-20 43 53.07 & 28.0 &&14MT69  &2014 08  03.82460 &18 41 35.266&-20 56 57.72 & 27.0 \\ 
\enddata
\end{deluxetable}

\subsection{Targetability Assessment}

\begin{deluxetable}{cccccccccc}
\tabletypesize{\scriptsize}
\tablecaption{HST List of Potential Targets\label{tbl-hst-targets}}
\tablewidth{0pt}
\tablehead{
\colhead{Name}&
\colhead{Mag}&
\colhead{Est. Diameter}&
\colhead{Target}&
\colhead{First}&
\colhead{Arc}&
\colhead{Encounter}&
\colhead{\deltav}&
\colhead{Enc. Helio}&
\colhead{Notes}\\
& & Range (km)& Prob & Seen & (days) & & (m/sec) & Distance &
}
\startdata
\multicolumn{1}{m{2cm}}{PT1, 1110113Y (2014MU69)} &	26.8&	28-53&1.00	 	&6/26/14	& 118		& Dec 2018&59&43.4&	\multicolumn{1}{m{4cm}}{100\% targetable with multiple models and assumptions, selected for flyby}\\
\multicolumn{1}{m{2cm}}{PT2, E31007AI (2014OS393)} &	26.3&	34-66&0.25-0.55& 7/30/14&	83		& Feb 2019&181&43.3&	 \multicolumn{1}{m{4cm}}{Observed as a DKBO} \\
\multicolumn{1}{m{2cm}}{PT3, G12000JZ (2014PN70)} 	&	26.4&	34-66&0.04-0.90& 8/6/14	& 76		& Apr 2019&118&44.1&	\multicolumn{1}{m{4cm}}{Observed as a DKBO}\\
\multicolumn{1}{m{2cm}}{PT4, 0720090F (2014MT69)} & 27.4&	21-40&		0.&	 	6/24/14&	0.05& \nodata&	\nodata		&\nodata&\multicolumn{1}{m{4cm}}{Unlikely to be targetable based on first follow-up - no further follow-up requested}\\
\multicolumn{1}{m{2cm}}{PT5, 4510067S} 	& 26.9&	29-55&		0.&	 	7/8/14&	0.00&	\nodata	&		\nodata&\nodata&\multicolumn{1}{m{4cm}}{Inaccessible based on discovery images- no follow-up requested}\\
\enddata
\end{deluxetable}

Our HST discoveries are listed in Table~\ref{tbl-hst-targets}.  The apparent brightness measured is shown along with the positional data.  The photometry tabulated has a very large uncertainty and the variations in the measurements is a consequence of the low signal-to-noise ratio and not any intrinsic variability in the objects.  In the case of Arrokoth, a much more extensive dataset was collected and reported by \citet{Benecchi2019} but no rotational modulation could be identified.  The post-flyby results from \nh\ showed that our vantage point from the Earth is essentially looking down the rotation pole thus explaining the lack of lightcurve.  In fact, we suspect that this geometric circumstance helped us find Arrokoth since it maintained a consistent brightness just above the detection limit of HST\null.  A strong lightcurve would have likely led to missing followup observations and complicating the orbit determination work.

The astrometric measurements in Table~\ref{tbl-hstast-full2} were used as time progressed to produce orbit fits and to evaluate the likely \deltav\ needed to re-target \nh\ for a close encounter. On this basis alone, the candidate object 4510067S (PT5) was immediately discarded as out of reach and no further followup observations were attempted. For the remaining candidates, we obtained additional observations to improve the orbit estimates (GO-13663, GO-13311).  Objects 0720090F (PT4, 2014 MT69) and E31007AI (PT2, 2014 OS393) were eventually revealed to also be beyond reach, the latter became a distant encounter object.  The objects G12000JZ (PT3, 2014 PN70) and 1110113Y (PT1, 2014 MU69) survived as potential candidates into 2015.  Of the two objects 2014PN70 was slightly brighter, implying a larger object, but the required \deltav\ was at the very limit if the remaining resources and would leave very limited margins. In the end, our choice was (486958) Arrokoth (formerly 2014 MU69) which required the smallest trajectory maneuver and left ample fuel reserves for other contingencies. Table~\ref{tbl-hst-orbits} summarizes the final orbital elements for the three objects that were carried forward from this search based on all data collected through the end of 2016 with estimates of their uncertainties.


\begin{deluxetable}{ccccc}
\tablecaption{Orbital Elements for HST Discoveries \label{tbl-hst-orbits}}
\tablewidth{0pt}
\tablehead{
\colhead{Object}&
\colhead{a (au)}&
\colhead{e}&
\colhead{i (deg)}&
\colhead{Hr Mag.}
}
\startdata
486958    & 44.085$\pm$0.0001& 0.0355$\pm$0.0001& 2.4513$\pm$0.0001& 11.01\\
2014OS393 & 44.013$\pm$0.1956& 0.060$\pm$0.051&   3.818$\pm$0.007&   10.07\\
2014PN70  & 44.071$\pm$0.006&  0.058$\pm$0.003&   4.119$\pm$0.004&   10.12\\
2014MT69  & 47$\pm$10&         0.15$\pm$1.04&     3.10$\pm$0.33&     11.46\\
{\em 4510067S}  & {\em 91}&                {\em 0.7}&               {\em 4}&                 {\em 11}\\
\enddata
\tablecomments{\scriptsize
The values in italics are just estimates based on limited data.  No additional data were ever acquired to improve these estimates.
}
\end{deluxetable}

\section{Discussion/Lessons Learned}

In summary, our combined ground-based efforts from Subaru and Magellan resulted in the discovery of 80 new KBOs. An additional 5 KBOs came out of our HST search including the fly-by target (486958) Arrokoth. Eleven of these objects (excluding Arrokoth) became \nh\ long-range observation targets which received measurements at phase angles far greater than can ever be observed from earth-orbit locations. These observations are allowing us to map surface scattering properties of small KBOs for the first time (\citet{Verbiscer2019},\citet{Verbiscer2022}). The tools we developed using the excellent Gaia astrometric catalog have also allowed us to make occultation predictions for many of these objects at high precision (\citet{Porter2022}).  The successful Arrokoth occultations were an excellent validation of this approach \citep{Buie2020}.

We also gained significant insight into searching for KBOs in general, but in the galactic plane specifically. When we started we expected that a target search would be a relatively straight forward process resulting in multiple potential mission objects to select from or prioritize among. Many on the search team had previously been part of dedicated KBO search projects from the time of the Kuiper Belt's observational discovery and we thought we had a pretty good handle on the Kuiper Belt population density and distribution. 

Our initial 2004-2005 survey didn't find a target, but it served a critical role in learning just how challenging it was to work in extremely crowded galactic plane fields. It also provided sample datasets for developing tools and techniques to deal with identifying faint objects in high density star fields and was thus very successful in moving us forward in the search. These lessons learned were carried forward into the second 2011-2014 survey epoch and helped us to optimize both data collection and level of effort or image analysis. In particular we learned that image quality, both observing conditions and seeing stability, was critical. Observations at high ($>$1.6 airmass) or under non-photometric conditions were insufficient to meet the search goals. We also quickly learned the importance of having an astrometic star catalog that went deeper than the standard databases of the time (USNO B1.0). Our solution was to use MegaCam on the CFHT from 2012-2018 to build a deeper (fainter magnitude) high precision astrometric star sub-catalog specific to our search fields. Stephen Gwyn built these catalogues from a variety of references, building off various Gaia (DR2) and 2MASS references sets for the global frame and then internally building up to measure high precision relative positions. When the official Gaia DR2 version became available we re-calibrated the final catalogue. This took a tremendous amount of effort, but using this secondary reference sub-net made a huge difference in the earlier efforts and was essential for the second and third epoch searches. 

Our search effort was also fairly early in the timeline of large data volumes among astronomical projects. Due to the volume of information, we quickly learned the critical importance of having effective pipelines that could process the data inclusive of data quality metrics on image characteristics, performance etc. so that one could quickly identify if quality data were being collected and also how prioritize data analysis. These lessons have directly informed searches for additional objects following the \nh\ fly-by of Arrokoth. The new search epoch, 2019 and following, as a result, are far better designed and characterized \citep{Fraser2024}. Likewise, it became clear by the end of 2013 that the likelihood of the data we were collecting at the time for identifying a fly-by object was minimal at best and a new strategy needed to be employed, hence a search with HST. 

Design of the HST search was itself a critical point in the discovery process as it required a full review of the state of knowledge of the Kuiper Belt at the time, in 2014. A 10-year improvement was made in our understanding of the absolute density and extent of the region in addition to search technique and analysis gains. We also accurately estimated what we would be able to find using HST. Finding 2 objects that we could consider targeting out of the sample of 5 detected was in fact an expected outcome rather than an under or overestimate. It is safe to say that by the time the survey was completed we knew the types of targets we were looking for, and had a good handle on the systematics, characteristcs and processes required for success.

Looking to the future, searching for additional KBOs, this work has led us to push our techniques even further into building deeper images stacks, applying machine learning, and considering additional (non-classical) KBO populations. These details are left to a future publication. However, it can't be overstated that the work presented here was a critical step along the pathway for identifying a fly-by object for \nh\ and in designing, characterizing and executing these sorts of search surveys while also sharing the methods and tools along the way. 

\section{Acknowledgments}

This work is dedicated to the memory of Peter Collins (1949-2017), instrumental to much of the software for the 2004 search that began this entire project.

Based on observations made with the NASA/ESA Hubble Space Telescope, obtained at the Space Telescope Science Institute, which is operated by the Association of Universities for Research in Astronomy, Inc., under NASA contract NAS5-26555. These observations are associated with programs 13633, 14053, 14092, 14485, and 14629. We are greatly indebted to Mike Ryschkewitsch (John's Hopkins Applied Physics Lab), Jim Green (Director of NASA's Solar System Exploration Division) and Ken Sembach (Space Telescope Science Institute) for their efforts in helping us to acquire the time for these observations. 

This paper includes data gathered with the 6.5 meter Magellan Telescopes located at Las Campanas Observatory, Chile. It is also based on data collected at the Subaru Telescope and retrieved from the HSC data archive system, which is operated by Subaru Telescope and  Astronomy Data Center at National Astronomical Observatory of Japan.  Access to the Subaru telescope was supported through NOAO Keck-Subaru exchange time, NASA HQ support and direct support from Taiwan through the Academia Sinica Institute of Astronomy and Astrophysics. The Subaru Telescope is operated by the National Astronomical Observatory of Japan. Likewise, the authors wish to recognize and acknowledge the very significant cultural role and reverence that the summit of Maunakea has always had within the indigenous Hawaiian community.  We are most fortunate to have the opportunity to conduct observations from this mountain.

This research used the Canadian Astronomy Data Centre facilities operated by the National Research Council of Canada with the support of the Canadian Space Agency. This work also made use of the Gemini Observatory Archive, NASA’s Astrophysics Data System Bibliographic Services, and the JPL HORIZONS web interface (\url{https://ssd.jpl.nasa.gov/horizons.cgi}). Some of the follow-up astrometry to refine orbits of newly discovered objects to access object targetability by \nh{}  were obtained at the international Gemini Observatory, a program of NSF’s NOIRLab, which is managed by the Association of Universities for Research in Astronomy (AURA) under a cooperative agreement with the National Science Foundation on behalf of the Gemini Observatory partnership: the National Science Foundation (United States), National Research Council (Canada), Agencia Nacional de Investigaci\'{o}n y Desarrollo (Chile), Ministerio de Ciencia, Tecnolog\'{i}a e Innovaci\'{o}n (Argentina), Minist\'{e}rio da Ci\^{e}ncia, Tecnologia, Inova\c{c}\~{o}es e Comunica\c{c}\~{o}es (Brazil), and Korea Astronomy and Space Science Institute (Republic of Korea).

The Zooniverse.org platform, development of which is funded by generous support, including a Global Impact Award from Google, and by a grant from the Alfred P. Sloan Foundation, hosted and promoted the ``Ice Hunters'' project with all of the data handling and analysis being prepared by the authors of this paper.

This work was also supported by a Harvey Mudd College Clinic Program team in 2009, consisting of faculty member Patrick Little, and team members Austin Lee, Claire Robinson, Florian Scheulen, Steven Berry, Chris Sauro, and Cullen McMahon.  We also thank James Miles of Boulder, Colorado for his volunteer work in support of software development, observing support, and data processing.

\appendix

\section{Astrometric description of an image\label{app-ast}}

The astrometric tools for this project pre-dated any development of a FITS standard for the astrometric coordinate description of an image that required a non-linear transformation.  The current standard system is described in \citet{Greisen2002} and \citet{Calabretta2002}. As a result, Buie developed a system for this purpose.  This appendix describes the system and how it is encoded into a FITS header.  As the FITS standard was developed, the system described here was adjusted to be complementary to any existing system by using a set of non-overlapping keywords so that both can co-exist in the same header.  This alternate system was designed to be used with software that can modify an existing FITS header and place these keywords in the header (an operation that is now very commonplace).

The task at hand is to describe a transformation from raw coordinates internal to some image or digital detector to coordinates on the celestial sphere, that is, right ascension and declination. This process involves the four coordinate systems indicated below,

\begin{equation}
(x,y) \Leftrightarrow (x_m,y_m) \Leftrightarrow (\xi,\eta) \Leftrightarrow (\alpha,\delta)
\end{equation}

\noindent where $(x,y)$ denote the native coordinate system implied by the image array or detector pixels, $(x_m,y_m)$ are the monolith coordinates, $(\xi,\eta)$ are the tangent plane coordinates, and $(\alpha,\beta)$ are the right ascension and declination on the sky. For this problem one needs to be able to convert between any coordinate system as indicated. Note that these are considered to be linked in the order shown.  Thus, to compute tangent plane coordinates from raw pixel positions you first convert to monolith coordinates and then to the tangent plane.  In the case of a single detector focal plane, the monolith and native coordinate systems are one and the same.  A monolith coordinate system is a virtual pixel scale that mimics the native coordinate system of a chip but this single system extends to include the area of all of the detectors.  Typically, detector coordinates are never negative and the implementation of this system uses (0,0) for the coordinates of the first pixel in the image and the coefficients and IDL software reflect this design choice.

\subsection{Tangent plane and celestial sphere}

The transformation between the tangent plane $(\xi,\eta)$ and celestial sphere $(\alpha,\delta)$ can be found on p. 283 from \citet{Smart1977} and is reproduced below:
\begin{eqnarray}
\eta & = & {{\tan \delta - \tan \delta_0 \cos(\alpha - \alpha_0)} \over
       {\cos(\alpha - \alpha_0)}+\tan\delta\tan\delta_0}\hskip10pt {\rm and}\label{eqn-eta} \\
\xi & = & {{\sin(\alpha - \alpha_0)} \over
          {\cos \delta (\cos(\alpha-\alpha_0)+\tan\delta \tan\delta_0 )}\label{eqn-xi}},
\end{eqnarray}
where $\alpha_0$ and $\delta_0$ are the coordinates of the tangent point. These expressions have been re-written from those found in the reference to minimize the impact of mathematical singularities in the equations and to speed their execution on a computer.  Despite the re-write, take note that equation~\ref{eqn-xi} still contains a singularity at the celestial poles ($\pm\pi/2$). The inverse transformation is written,
\begin{eqnarray}
\tan(\alpha - \alpha_0) &=& {{\xi} \over {\cos\delta_0 - \eta\sin\delta_0}}\label{eqn-invra},~{\rm and} \\
\tan\delta &=& {{\cos(\alpha-\alpha_0)(\eta+\tan\delta_0)} \over {1-\eta\tan\delta_0}}\label{eqn-invdec}~.
\end{eqnarray}
Note that care must be taken when evaluating equation~\ref{eqn-invra} so that all 4 quadrants of the arctangent are preserved.  Equations \ref{eqn-eta} and \ref{eqn-xi} are evaluated by the IDL procedure {\tt astrd2sn.pro}. Equations \ref{eqn-invra} and \ref{eqn-invdec} are evaluated by IDL procedure {\tt astsn2rd.pro}.  Both transformations are included in {\tt astcvt.pro}.  By virtue of these transformations, one can treat tangent plane and sky coordinates as completely interchangeable and the discussion below will often make little distinction between the two.  Tangent plane coordinates are also referred to as ``standard coordinates.''  The natural units of calculating the tangent plane coordinates would be radians.  However, it is usually more convenient to work in units of arcseconds instead.

\subsection{Image and tangent plane}

A plane tangent to a sphere at a given point is also known as the tangent plane.  This plane is very similar to the physical mapping that takes place when a camera takes a picture of the sky.  For small enough angular range images the mapping is linear.  Departures from non-linearity are usually a consequence of optical distortions in the camera optics but also inevitable with a large field-of-view.  The mapping from detector (image) pixels to the tangent plane is thus an approximate process that needs to be adjusted to match the characteristics of the image so that a sufficiently accurate mapping is achieved.

Digital image data collected with a single device is a common type of data and represent a simple initial case for consideration.  The mapping between these two coordinate systems can be accomplished by any set of basis functions capable of reaching the desired accuracy.  A common and sufficient basis set is an $n$-degree polynomial in x and y.  In its simplest, linear form the transformation would look like this:
\begin{eqnarray}
\xi &=& a_0 + a_1 x + a_2 y \\
\eta &=& b_0 + b_1 x + b_2 y
\end{eqnarray}
where the $a_n$ and $b_n$ coefficients determine the transformation.  This approach is just a power series expansion in $x$ and $y$. Typically, the coefficients are determined from some least-squares fit between the image and a stellar catalog.  For the sake of this discussion the method of determination is not important.  A linear expansion works well for sufficiently small detectors but the systems we used for this search required higher order terms and the IDL-based tools support up to a fifth order expansion.  Some of the original tools that were used as a starting point for this development used an $r$ term ($r=\sqrt{x^2+y^2}$), but this term is not part of the polynomial basis set and is incorrect to use in a cartesian framework.  Support for this term is included in the IDL tools, but is present for historical reasons and should not be used.  Other basis sets are possible, such as Chebyshev and Legendre polynomials.  Thus, to specify this transformation it is necessary to indicate which basis set is to be used and then which terms are provided along with the coefficients. The IDL routine {\tt astcvt.pro} supports all three of these basis sets. Note that this methodology does not require or enforce orthogonality between the $\xi$ and $\eta$ coordinates.

If the transformation is non-linear it becomes important to specify the location of the optical axis.  This point is special in that it will represent a point of symmetry in the mapping.  If the device is well fit by a linear transformation it is common to use the center of the device as the optical center.  In this case the center $(x_c,y_c)$ is defined to be $(n_x/2,n_y/2)$ where $n_x$ and $n_y$ are the $x$ and $y$ size of the image, in pixels.  In non-linear cases it works best to use the optical axis position, wherever that might be.  While it is important to know the optical axis position, the determination of this via a least-squares fit often leads to a poorly constrained answer. Knowledge of this position will usually give a better answer than relying on a fit but a precise determination is not needed.

It is often necessary to rescale x and y prior to solving for coefficients.  For a linear polynomial fit this rescaling is optional. For non-linear polynomials, the rescaling also helps make the solutions more robust even for a simple polynomial basis set.  For Chebyshev and Legendre polynomials, the rescaling is required since the input range of values must be between -1 and 1. Normally this is done by dividing by a renormalization factor, $d$, such that $x'=x/d$ and $y'=y/d$. Renormalization is usually done by using the size of the image array or detector.  One form of normalization is $d = \sqrt{n_x^2 + n_y^2}$.  However, this can lead to an incomplete filling of the allowed range of $x'$ and $y'$ between $\pm1$.  A better choice is $d = \rm{max}(n_x/2,n_y/2)$ if the center is used for the optical axis position.  Otherwise, use the maximum of the absolute value of the corners of the array after subtracting the optical axis position.  In the IDL software, the renormalization is carried only in the monolith coordinates.

\subsection{Image mosaic conversions}

A closely related problem to a single image is dealing with a mosaic of images from an array of detectors within a single camera.  They are also related though a presumed constant physical bond between detectors. Processing and fitting individual images from a mosaic can, in principle, be handled as if they were each independent images subject to the treatment in the previous section.  Unfortunately, such handling is subject to numerical instabilities in the fitting process leading to often erroneous coefficients.  For the case of a mosaic it is still true that there is a native raw coordinate system for each device that is identical to the $(x,y)$ system in the previous section. This case benefits from a new coordinate system intermediate between $(x,y)$ and $(\xi,\eta)$ that we will call the monolith coordinates denoted by the subscript $m$.  A well-constructed mosaic camera is one where all of the imaging devices lie in the focal plane of the instrument.  In such cases, the raw coordinates of each device is related to the other devices by a translation and a rotation.  The following treatment explicitly ignores any out-of-plane tilt a detector might have.  The choice of the monolithic coordinate system is arbitrary, so for convenience we choose one of the devices to be co-joined with the monolithic system. It is also useful to choose this one device to be the one that contains or is at least closest to the optical axis.  In this coordinate system the position of the optical axis is defined to coincide with the origin of the monolithic system. The transformation from the $i^{\rm{th}}$ raw device to monolith coordinates is thus,
\begin{eqnarray}
x_m^i &=& ((A^i_{1,1}(x-x_c^i) + A^i_{1,2}(y-y_c^i))/d \\
y_m^i &=& ((A^i_{2,1}(x-x_c^i) + A^i_{2,2}(y-y_c^i))/d
\end{eqnarray}
where $A$ is a rotation matrix related to a rotation angle $\theta$ by
\begin{eqnarray}
A = \left ( {\begin{array}{cc}
cos\theta& \sin\theta\\
-\sin\theta& \cos\theta
\end{array}}
\right )
\end{eqnarray}
Note that the elements of $A$ are often referred to as the ``CD matrix'' found in FITS headers for WCS (world coordinate system) conversions \citep{Greisen2002,Calabretta2002}. The values for $(x_0,y_0)$ are often found in the {\tt CRPIXn} keywords.  The choice of the reference chip leads naturally to $\theta=0$ for that image. With this construction, the $x_0^i$ values are the raw coordinates of the optical axis for each image.  While it is useful and convenient to impose no rotation between the monolith system and the anchor chip, there are cases were a global rotation is useful. The rotation for the anchor chip will impart that global rotation if applied at the right step.  In our IDL software, {\tt astcvt.pro}, this global rotation is supported.

\subsection{Inverting an astrometric solution}

The equations above make it very easy to go from position $(x,y)$ to sky-plane coordinates, either $(\alpha,\delta)$ or $(\xi,\eta)$. However, it is often very handy to be able to go in the other direction. That is what {\tt astcvt.pro} does.  Note that the convertion between $(\alpha,\delta)$ and $(\xi,\eta)$ in either direction is very simple and is not discussed here.  If the solution is linear the conversion is deterministic and exact via an analytic expression.  If the solution is non-linear the conversion requires an iterative numerical solution. The routine {\tt astcvt} will do this inversion for any input solution.

\subsection{FITS keywords}

All of the keywords in the Buie system start with {\tt AST}.  The variables expressions shown refer back to the quantities defined above.  The units of the variables are provided in brackets.

{\tt ASTINFO} (boolean) -- Flag to indicate this header contains the astrometric information in this format.

{\tt ASTXCEN} (float,$x_0$,[pixels]) -- X pixel location of the optical axis.

{\tt ASTYCEN} (float,$x_0$,[pixels]) -- Y pixel location of the optical axis.

{\tt ASTPROT} (float,[radians]) -- Chip pre-rotation angle.

{\tt ASTRNORM} (double,$d$,[pixels]) -- Normalization factor.

{\tt ASTTRA} (double,$\alpha_0$,[HMS string]) -- Right ascension of the tangent point.  It is not required, but highly desireable that this point be coincident with ($x_0,y_0$).

{\tt ASTTDEC} (double,$\delta_0$,[DMS string]) -- Declination of the tangent point.  It is not required, but highly desireable that this point be coincident with ($x_0,y_0$).

The following three-keyword set is given for every term in the basis set. The end of the keyword is a numeric digit, starting at 1 and increasing for as many terms as are required for the mapping. The order of the terms is arbitrary.

{\tt ASTTNi} (string) -- Name of the term.  The names indicate the term and the basis set to be used.  There are three basis sets supported by {\tt astterms.pro}: simple polynomial, Chebyshev degree 1 or degree 2 polynomials, and Legendre polynomials (M=0).  The recognized term names are {\tt CONST}, {\tt X}, {\tt Y}, {\tt XX}, {\tt YY}, {\tt XY}, {\tt XXX}, {\tt YYY}, {\tt XYY}, {\tt XXY}, {\tt XXXX}, {\tt YYYY}, {\tt XYYY}, {\tt XXYY}, {\tt XXXY}, {\tt XXXXX}, {\tt YYYYY}, {\tt XYYYY}, {\tt XXYYY}, {\tt XXXYY}, {\tt XXXXY}, or all of the preceeding with a {\tt T} prepended for degree 1 Chebychev polynomials, a {\tt U} prepended for degree 2 Chebychev polynomials, or {\tt P} prepended for Legendre polynomials.  The order of the term is indicated by the repetition of {\tt X} or {\tt Y}.  For example, the $x^3$ term is denoted with {\tt XXX}.

{\tt ASTCXi} (double, $a_i$, arcsec) -- The transformation coefficients from $(x,y)$ to $\xi$.  The units are in arcsec raised to the degree of the term.

{\tt ASTCEi} (double, $b_i$, arcsec) -- The transformation coefficients from $(x,y)$ to $\eta$.  The units are in arcsec raised to the degree of the term.

\subsection{Useful Buie IDL library tools}

The preceeding sections describe the mathematical basis and algorithmic approach needed to extract astrometry from an image.  This approach can, of course, be implemented in most programming languages.  The rest of this section describes our specific implementation of tools in the IDL programming language.  The following software tools are in direct support of handling the operations related to astrometry and can be used directly or to guide development and testing of other versions. The top level tools are the ones most often used to build applications to extract or use astrometry.  The low-level routines are usually just used internally in the library.  The mid-level routines are somewhere in between the other two categories.

\subsubsection{Top-level routines}

{\tt astcvt} -- Convert from one astrometric coordinate system to another given an astrometric description of the image.  The coordinate systems supported are raw pixel positions in the image (xy), detector mosaic monolithic coordinates (XY), tangent plane coordinates in radians (sn), tangent plane coordinates in arcsec (SN), and equatorial celestial coordinates (rd).

{\tt astinfo} -- Read (or write) astrometric information from (to) the header.

{\tt mkastinfo} -- Create a synthetic astrometric description.  This is intended only for creating linear descriptions.

\subsubsection{Mid-level routines}

{\tt astchi1} -- Compute the goodness of fit for a solution based on some rotation and offset.  This is used internally to help solve for inter-chip rotations in a multi-detector camera.

\subsubsection{Low-level routines}

{\tt asteval} -- Evaluate an astrometric polynomial function given the independent variables $(x,y)$, a list of terms, and the polynomial coefficients.

{\tt astrd2sn} -- Convert from $(\alpha,\delta)$ to $(\xi,\eta)$.  This routine is called by {\tt astcvt} but is useful on its own.

{\tt astrd2xy} -- Convert from $(\alpha,\delta)$ to $(x,y)$.  This routine works only for linear transformations.  {\tt astcvt} is almost always
preferred.

{\tt astsn2rd} -- Convert from $(\xi,\eta)$ to $(\alpha,\delta)$.  This routine is called by {\tt astcvt} but is useful on its own.

{\tt astsn2xy} -- Convert from $(\xi,\eta)$ to $(x,y)$.  This routine is called by {\tt astcvt} and is rarely called directly.

{\tt astsolve.pro} -- This routine performs the least-squares fit for an astrometric  solution. The inputs are correlated lists of pixel coordinates $(x,y)$ and tangent plane coordinates $(\xi,\eta)$ along with a list of fitting terms to use.  This tool performs a robust fit, automatically culling discrepant positions based on the noise statistics in the fit.  Considerable work is required before-hand to set up the correlated lists but this tool does the basic work of the fit.  One example of an interactive tool for solving for astrometry that feeds this solver is {\tt astrom.pro}.  This one example is a heavily used tool for other projects but the scale of the data reduction process for the \nh\ search was such that it was more efficient to write dedicated and very non-general solving tools.

{\tt astterms} -- Evaluate the independent vectors to match an astrometric polynomial function.

{\tt astxy2rd} -- Convert from $(x,y)$ to $(\alpha,\delta)$. {\tt astcvt} is usually used rather than this tool.

{\tt astxy2sn} -- Convert from $(x,y)$ to $(\xi,\eta)$.  This routine is called by {\tt astcvt} and is rarely called on its own.

\section{Details of key Buie IDL library routines\label{app-idl}}

The following routines were particularly useful in the data processing and results presented in this paper.  Many more were also used but these contain either the essential computational engines or interesting and perhaps non-obvious algorithmic concepts.  This discussion is provided to highlight the algorithms used rather than to be a guide for the use of this specific implementation.

\subsection{basphote.pro}

Aperture photometry summation and centroiding is provided by {\tt basphote.pro}.  This routine and its use has been discussed in \citet{Buie1992} and \citet{Buie1996}.  Alternate versions are available in C and were also ported to IRAF many decades ago.  The object aperture integral is the sum of all pixels weighted by the exact area of overlap between the pixel and the circular aperture.  Sky signal is determined from a sample of pixels whose centers lie between the inner and outer radii of the sky annulus.  The robust mean of the sample ({\tt robomean.pro}) is computed and the sky mean and the sky noise is used to remove sky signal from the object aperture summation and to propagate into the final noise estimation.  Detector gain, exposure time, and detector read-noise can also be provided for more accurate and consistent flux and noise computations.  During the measurement, the input position is treated as a rough location.  The image is scanned for the maximum within the distance of the object aperture radius.  From this position, an initial centroid is calculated without subtracting sky. This first centroid is then used to place the sky annulus, determine sky, then compute the object aperture flux and final centroid.

\subsection{dewarp.pro}

This tool was created in a collaborative project with a Harvey Mudd College clinic team.  The basic function of this tool is to map image data from one grid to another.  The input is an astronomical image and an image descriptor that contains an astrometric solution for the image. Another input is a second astrometric solution for an image to be constructed. The two solutions are used to map flux in the input image to the correct location in the output image.  The images are resampled, guided by the astrometric solution, using an interpolator that is equivalent to a sinc interpolation.

This tool had many uses in our project.  It can be used to map one image onto another eliminating any distortion, sampling, or rotation from the resampled image and then replicating any distortion in the destination image.  This type of operation is particularly valuable when building a deep stack of the sky that matches a single image in preparation for subtracting that template without sampling the original image.

For the ground-based searches it was more advantageous to define a synthetic, linear image that all search data mapped to for a particular field.  In this case, all images are resampled but the result is strictly linear and registered.  Given that the image transformation is always done with full astrometric knowledge, the resampled images are just as valid for astrometric measurements of discovered objects as the originals.

\subsection{findsrc.pro}

Automatic source detection is important element of getting astrometry and photometry from a digital sky image.  This tool is similar in function to the DAOPHOT tool, SExtractor \citep{Stetson1987,Bertin1996}.  However, the internal algorithmic workings of this program are quite a bit different and its output can have very different characteristics.  In particular, the number of false positives is usually much lower.

The first step in this tool is a scan of the image to find pixels that exceed a given threshold {\tt SIGTHRESH} relative to the noise in the background of the image.  For each pixel, four windows on the background are computed.  These windows extend away from the pixel in the cardinal directions of the image ($\pm x$ and $\pm y$).  The length of each window is controlled by the {\tt WINDOW} keyword and defaults to 6 pixels. The windows are displaced by {\tt GAP} pixels from the target pixel with a default of 2 pixels.  This value is normally set to a value similar to the FWHM of the PSF in the image.  From these windows, four sky signal values are computed with an associated standard deviation.  If the pixel signal exceeds the threshold above the sky in 3 or 4 directions, that pixel is flagged.  The result of this step is a detection array with a binary value for the pixel exceeding the threshold or not.  Note that an additional check is made to restrict the detected pixels to have a signal level less than the saturation limit imposed by the {\tt MAXPHOTSIG} keyword.

The next step is to collapse the detection array.  This is needed since many pixels from the same source can be flagged as detections.  This step is performed by an internal rountine, {\tt collapse}.  The detection image is eroded so that only one pixel is left flagged per source and that pixel is at the local maximum in the image.  The local maximum are further constrained to be at least {\tt GAP} pixels apart.  The erosion logic is designed to also collapse large regions from saturated stars and minimize the number of pixels flagged as detected.  This step is reasonably effective but local variations along bleed trail from a saturated source can have local maxima that will confuse this tool and lead to a string of detections along the trail.

The eroded detection image is then a map of source locations.  Aperture photometry is computed with {\tt basphote.pro} for each flagged pixel in this map.  The radius of the object aperture defaults to {\tt GAP} but can be altered with the {\tt OBJRAD} keyword if desired.  The sky annulus for a local sky computation is set to objrad+1 and objrad+5.  Those sources with sensible values for FWHM and instrumental magnitude are kept.  Also, the SNR of the source must also be above the value of {\tt SIGTHRESH}.  During the photometry calculation, a centroid location is computed as well.

At the end, the centroid position, FWHM, flux, magnitude, magnitude error, SNR, sky and sky noise are returned to the caller or possible saved to a file for later processing.  This routine is reasonably fast with the most expensive part being the computation of the aperture photometry and centroids.

\subsection{frmdxdy.pro}

One of the most important tasks during the determination of an astrometric solution is the correlation of two lists of positions.  In this case, we need to take a list of star positions, usually in $(x,y)$ coordinates and determine the correspondence into a list of star catalog positions, usually in $(\alpha,\delta)$.  The algorithm used here was inspired by a method developed by Bruce Koehn at Lowell Observatory for the LONEOS project.  Approximate information about the image scale and orientation is used to convert the catalog positions to estimated pixels positions.  This tool is extremely resilient against differences in photometric depth between the two lists. Rough knowledge of the plate scale is needed.  The position shift between the two lists can be large with no penalty on speed of execution or quality of result.  The rotation of the field, however, must be known very well.  The requirement on this knowledge gets more stringent as the field-of-view increases.  This limitation can be ameliorated somewhat by restricting the initial determination of shift to a subset of a large image.  The presence of uncharacterized non-linearities in the image also limits the ability to find a good shift but images subsets help here too.

\subsection{frmdxyr.pro}

This routine is an attempt to fix the problem of needing to know an exact rotation angle for {\tt frmdxdy}.  Rotation is scanned over a range and a binary search is run to find the best rotation angle that gives a sensible list correlations.  This tool can be very time-intensive to run if the rotation range is large and is really only a stop-gap measure when there is small rotation error in the datasets.  This tool is completely useless for the class of problem where the rotation is not known at all.  For most astronomical data, one usually has a pretty good idea of the information for these routines.

\subsection{jitterk.pro}

PSF-fitting to an image to get the position and flux of a source requires a good understanding or characterization of the PSF.  For stellar data taken with stellar guiding, knowing the PSF is sufficient and is also easy to determine from the data (see {\tt psfstack}).  For solar system observations, the sources of interest are moving.  The situation is even more complex for Hubble Space Telescope data where parallax in introduced.

This tool was specifically written to support HST data reductions though it can be usefully used on ground-based data as well.  The input is a time-history of the motion of an object relative to the pixel coordinates during an exposure.  In the case of the HST search data for this project, we track the KBO.  As a consequence, the stars are trailed by the target motion and parallax induced by the telescope's motion around the Earth.  In principle, one can also add in the measured jitter of HST pointing as measured with its fine-guidance sensors but this level of detail was never needed.

At first glance, most of the trailed HST images look like linear trails. In practice, this visual appearance is mis-leading.  The parallactic motion can still be along a line but sometimes, the rate of motion is slower at the start of the exposure than at the end or vice versa.  If the images are fitted as lines and the line center used, there can be an effective temporal shift in the derived astrometry.  This tool builds a time-history of the pointing that is used to create a convolution kernel that can be applied to the PSF prior to the fitting process.  This ensures that the temporal registration between smeared and un-smeared targets is preserved.

\subsection{lowess.pro}

This tool is a LOcally-WEighted Scatterplot Smoother.  It provides a means to draw a smooth curve through data to follow the trend in the data even in the presence of noise without imposing a smoothing or fitting function.

\subsection{ois.pro}

This tool was originally written by \citet{Miller2008} as a port of the \citet{Alard1998} optimal image subtraction (OIS) software into IDL\null.  That tool was very slow and not very practical to use.  Our version was optimized during a Harvey Mudd College clinic project and runs at essentially the same speed as the original C-code.  An additional enhancement was added by \citet{Miller2008} to encode a $\delta$-function basis set for the PSF manipulations rather than the nested gaussian approach from \citet{Alard1998}. During our extensive work on the ground-based data for this project, we found the $\delta$-function basis set, coupled with very precise astrometric registration eliminated all need for spatially variable convolution kernels in the OIS process and that feature of the software was never used.

\subsection{robomean.pro}

Numerous operations in our data require the computation of a mean from a sample of data.  Most real data unfortunately are affected by samples that do not fit the underlying noise distribution of the quantity being measured.  This tool is designed to provide a robust mean from a sample is is probably the most heavily used routine from the Buie library. The goal is to iteratively process the data until all of the surviving points are consistent with the noise distribution in the surviving points.  A threshold can be set on the filtering and is normally set to 3$\sigma$. Convergence depends on getting convergence in the value of the mean, standard deviation, skew, and kurtosis consistent with the errors on these four quantities as points are (possibly) removed.  This is used for measuring sky brightness around a source for position and flux measurements as well as just about any other place where a mean is desired.

\subsection{psfstack.pro}

Many operations require a PSF for an image.  This routine determines a numerical PSF by stacking many sources from the images to get an average PSF.  An input list of sources is provided and they are stacked in descending order of SNR down to a limit that defaults to 25.  Sources that are not good PSFs (such as a double-star blend) are excluded.

\subsection{skyfit.pro}

The degree of image crowding in the image made it difficult to measure the signal level due to the light from the sky.  A typical method is to collect a large sample of pixels across the image and then compute a robust mean, removing unusual outliers caused by such things as pixels illuminated by stars.  This method requires the sample to be dominated by actual sky measurements.  Levels of contamination up to 10-20\% are easily removed but when that fraction rises too high our usual tool failed and returned an estimate of the sky that is systematically and significantly too bright.

This new routine includes robust estimation and simultaneous fitting of a low-order 2-D polynomial to the sky background with one important enhancement.  An examination of a histogram of the random image pixels shows an asymmetric distribution.  The appearance of this histogram in Fig.~\ref{fig-sky} reveals what looks like a normal distribution for signal at or below the mode of the distribution.  Signals higher than the mode are contaminated by faint star signals that are just barely brighter than the sky and thus add flux to the apparent distribution of signal from the sky.  If there were no stars, the distribution will always look like a symmetric Gaussian but as the number of star-illuminated pixels increases the distribution becomes more and more distended on the upper half.  Pixels below the mode are minimally affected by background star light.

\begin{center}
\includegraphics[scale=0.5]{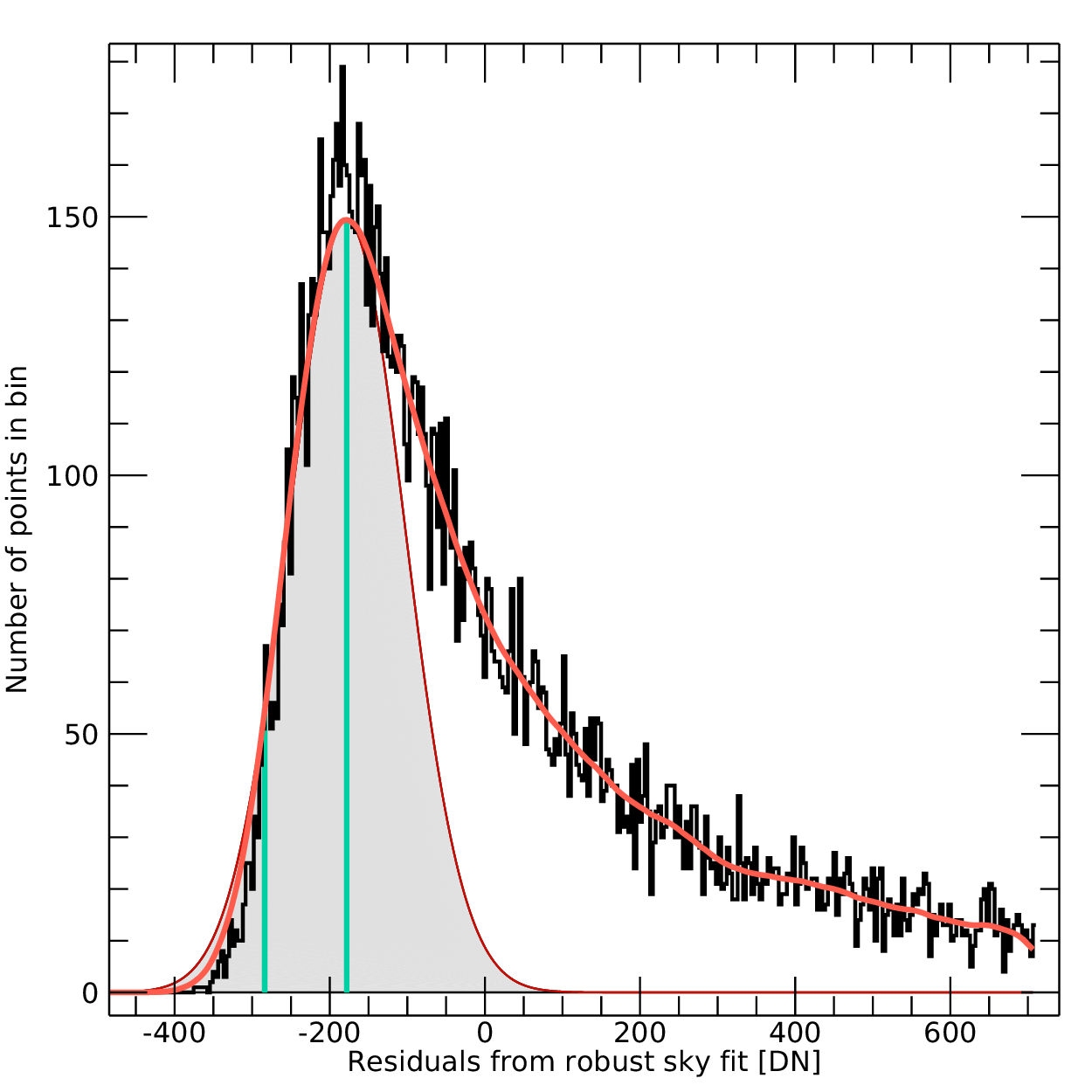}
\figcaption{\label{fig-sky}
Sky signal example.  The plotted histogram shows the residuals from a 2-d fit to the sky in a single image.  The thick orange curve is a lowess curve of the histogram and the peak is used as an estimate of the mode.  The vertical green lines mark the mode and the $-1\sigma$ point on the low side of the distribution.  The gray shaded region shows the approximate Gaussian representation of the sky signal and noise.
}
\end{center}

In detail, the program first selects 20,000 random pixels from the image (2k by 4k array size) representing 0.2\% of the pixels.  A first-order 2-D polynomial is then fitted to these pixels.  Those pixels whose residuals are more than 3$\sigma$ from the mean residual are rejected.  If there are new rejected pixels, the fitting process is repeated.  When there are no additional pixels removed with up to 10 passes, the 2-D fitting is deemed complete.  The mode of the surviving pixels is computed and then the 1/e half-width is measured from a smoothed curve based on the data using {\tt lowess.pro} applied to data falling below the mode.  The true sky signal is assumed to be normally distributed and we use the mode and 1/e half-width to approximate the sky signal and its uncertainty.  In the example demonstrated in Fig.~\ref{fig-sky}, 23\% of the samples were excluded by the initial robust sky fit and the plotted histogram is of the surviving samples.  The mode from this histogram is then added to the constant term of the 2-d polynomial fit and the width of the grey shaded region is taken to represent the sky noise in the image.  In the histogram shown, the portion above the gaussian represents pixels that are contaminated by faint starlight.  The fraction thus contaminated are too numerous to be excluded by typical outlier rejection algorithms.  Without this correction to the sky fit, we would systematically fail to detect the faintest sources which happen to be the most likely to be of interest in our search. This approach still contains a small systematic bias to a higher sky level but this bias is significantly smaller than the sky noise level and was thus neglected. Note that zero in this plot is effectively the naive mean of the sky computed without sufficient regard to the residual star contamination. In this case, the correction to the sky level is nearly 200 DN which is roughly a 3-$\sigma$ shift.

\subsection{stacker.pro}

Building image templates or stacks is done with this program.  A set of images can be added up and a robust mean is estimated from the stack at each pixel. This tool also supports shifting the images to get a coherent stack on a object moving with respect to the pixel coordinates.

\section{Ice Hunters -- Citizen Science Participation}\label{ap-icehunt}

The rise of citizen science projects has been a consequence of attempts to harness the interest and attention of a large number of initially untrained people to tackle a large and either tedious or difficult to code task, generally in the form of sorting, classification, or searching through data.  The search for KBOs in these crowded stellar fields was seen to be an interesting case.

For this project, all the data were pre-processed as described in the main section of this paper.  The basic product of these calculations were difference images where the stars are (mostly) subtracted and moving objects are seen as white and black pairs.  We provided annotated training images  to acquaint the participants with what to look for.  After successful training, access to live data was granted.

The images being scanned were much larger (2k by 4k) than could be displayed given the technology at the time.  Rather than rely on image navigation driven by the user, we chose to present sub-images from the larger images.  The size of the sub-array was dictated by the least capable graphical system to be used -- in this case we were limited by the capabilities of the first version of the Apple iPad and a sub-array size of 424 x 424.  To eliminate missing objects on the edge, we built-in a 10\% overlap between adjacent sub-arrays.  At the time of the sub-array creation, a fixed stretch of the image was used -- scaled to $-3$ to $+8\sigma$ of the sky signal and the result was saved as a PNG-format image.

Breaking up the images resulted in a very large number of files.  These files had to be loaded on a cloud server which was hosted on Amazon Web Services (AWS).  The tool for uploading data to the server was GUI-based and designed for small batches of files.  The transfer process was extremely time-consuming and would often fail if too many files were attempted at once.  This problem proved to be a nearly fatal bottleneck for this project.

The system was designed to present images to a participant for examination. Performing a complete search meant that each sub-image was viewed and processed by up to 16 different people with a small fraction seen as many as 40 times. The list of images was stepped through as each load request came in.  Our design allowed for very fast processing of the data given enough participants.  Once an image was viewed enough times, it was removed from the candidate list for processing.  The goal during processing was to ensure that there were always images in the queue waiting for processing.  In practice it was difficult to load images to AWS faster than they could be processed by the citizen scientists.

This method of processing created a very large number of false positive detections.  The total number of flagged transients was 728,954 and we had to apply some automated filtering steps to reduce the work load.  Any transient seen $\ge$3 times was considered worthy of a confirmation effort.  This threshold excluded 61\% of the flagged transients.  An additional 13\% were excluded on the basis of visual examination of the candidate. The surviving candidates were visually graded into 3 different classes: a likely KBO (124, 0.02\%), a likely main-belt asteroid (16,011, 2.2\%), and a likely variable star (79,383, 11\%). An additional 93,734 (13\%) candidates could not be excluded or placed into any of these three classes and simply left marked as ``unknown'' and not considered further.

The final performance of this system versus our normal processing methods was the same for KBO detections.  The citizen science approach was far more effective at finding and recording all of the other category of time variable elements contained in the data for essentially the same amount of effort.  The citizen science approach would also be very effective if the data were populated with simulated objects for the purposes of characterizing the observational biases.

The following is a list of all those that participated as an ``Ice Hunter:''
T.  Abbate,
T.  Abrahamson,
M.  Adam,
J.  Adamson,
A.  Agbedor,
B.  Alexander,
U.  Allan,
X.  Alldredge,
A. M. H.  Allen,
D.  Alton
% ...
and
I.  Zormpa.

\vspace{5mm}
\facility{Subaru, Magellan, HST, CFHT, Gemini}

\bibliographystyle{aasjournal}
\bibliography{references.bib}

\begin{thebibliography}{}
\expandafter\ifx\csname natexlab\endcsname\relax\def\natexlab#1{#1}\fi
\providecommand{\url}[1]{\href{#1}{#1}}
\providecommand{\dodoi}[1]{doi:~\href{http://doi.org/#1}{\nolinkurl{#1}}}
\providecommand{\doeprint}[1]{\href{http://ascl.net/#1}{\nolinkurl{http://ascl.net/#1}}}
\providecommand{\doarXiv}[1]{\href{https://arxiv.org/abs/#1}{\nolinkurl{https://arxiv.org/abs/#1}}}

\bibitem[{{Alard} \& {Lupton}(1998)}]{Alard1998}
{Alard}, C., \& {Lupton}, R.~H. 1998, \apj, 503, 325, \dodoi{10.1086/305984}

\bibitem[{{Benecchi} {et~al.}(2019){Benecchi}, {Porter}, {Buie}, {Zangari},
  {Verbiscer}, {Noll}, {Stern}, {Spencer}, \& {Parker}}]{Benecchi2019}
{Benecchi}, S.~D., {Porter}, S.~B., {Buie}, M.~W., {et~al.} 2019, \icarus, 334,
  11, \dodoi{10.1016/j.icarus.2019.01.023}

\bibitem[{{Bernstein} \& {Khushalani}(2000)}]{Bernstein2000}
{Bernstein}, G., \& {Khushalani}, B. 2000, \aj, 120, 3323,
  \dodoi{10.1086/316868}

\bibitem[{Bernstein {et~al.}(2004)Bernstein, Trilling, Allen, Brown, Holman, \&
  Malhotra}]{Bernstein2004}
Bernstein, G.~M., Trilling, D.~E., Allen, R.~L., {et~al.} 2004, The
  Astronomical Journal, 128, 1364, \dodoi{10.1086/422919}

\bibitem[{{Bertin} \& {Arnouts}(1996)}]{Bertin1996}
{Bertin}, E., \& {Arnouts}, S. 1996, \aaps, 117, 393,
  \dodoi{10.1051/aas:1996164}

\bibitem[{{Boulade} {et~al.}(2003){Boulade}, {Charlot}, {Abbon}, {Aune},
  {Borgeaud}, {Carton}, {Carty}, {Da Costa}, {Deschamps}, {Desforge},
  {Eppell{\'e}}, {Gallais}, {Gosset}, {Granelli}, {Gros}, {de Kat}, {Loiseau},
  {Ritou}, {Rouss{\'e}}, {Starzynski}, {Vignal}, \& {Vigroux}}]{Boulade2003}
{Boulade}, O., {Charlot}, X., {Abbon}, P., {et~al.} 2003, in Society of
  Photo-Optical Instrumentation Engineers (SPIE) Conference Series, Vol. 4841,
  \procspie, ed. M.~{Iye} \& A.~F.~M. {Moorwood}, 72--81,
  \dodoi{10.1117/12.459890}

\bibitem[{Brown {et~al.}(2018)Brown, Vallenari, Prusti, de~Bruijne, Babusiaux,
  Bailer-Jones, Biermann, Evans, Eyer, Jansen, Jordi, Klioner, Lammers,
  Lindegren, Luri, Mignard, Panem, Pourbaix, Randich, Sartoretti, Siddiqui,
  Soubiran, van Leeuwen, Walton, Arenou, Bastian, Cropper, Drimmel, Katz,
  Lattanzi, Bakker, Cacciari, Casta{\~{n}}eda, Chaoul, Cheek, {De Angeli},
  Fabricius, Guerra, Holl, Masana, Messineo, Mowlavi, Nienartowicz, Panuzzo,
  Portell, Riello, Seabroke, Tanga, Th{\'{e}}venin, Gracia-Abril, Comoretto,
  Garcia-Reinaldos, Teyssier, Altmann, Andrae, Audard, Bellas-Velidis, Benson,
  Berthier, Blomme, Burgess, Busso, Carry, Cellino, Clementini, Clotet,
  Creevey, Davidson, {De Ridder}, Delchambre, Dell'Oro, Ducourant,
  Fern{\'{a}}ndez-Hern{\'{a}}ndez, Fouesneau, Fr{\'{e}}mat, Galluccio,
  Garc{\'{i}}a-Torres, Gonz{\'{a}}lez-N{\'{u}}{\~{n}}ez, Gonz{\'{a}}lez-Vidal,
  Gosset, Guy, Halbwachs, Hambly, Harrison, Hern{\'{a}}ndez, Hestroffer,
  Hodgkin, Hutton, Jasniewicz, Jean-Antoine-Piccolo, Jordan, Korn,
  Krone-Martins, Lanzafame, Lebzelter, L{\"{o}}ffler, Manteiga, Marrese,
  Mart{\'{i}}n-Fleitas, Moitinho, Mora, Muinonen, Osinde, Pancino, Pauwels,
  Petit, Recio-Blanco, Richards, Rimoldini, Robin, Sarro, Siopis, Smith,
  Sozzetti, S{\"{u}}veges, Torra, van Reeven, Abbas, {Abreu Aramburu}, Accart,
  Aerts, Altavilla, {\'{A}}lvarez, Alvarez, Alves, Anderson, Andrei, {Anglada
  Varela}, Antiche, Antoja, Arcay, Astraatmadja, Bach, Baker,
  Balaguer-N{\'{u}}{\~{n}}ez, Balm, Barache, Barata, Barbato, Barblan, Barklem,
  Barrado, Barros, Barstow, {Bartholom{\'{e}} Mu{\~{n}}oz}, Bassilana,
  Becciani, Bellazzini, Berihuete, Bertone, Bianchi, Bienaym{\'{e}},
  Blanco-Cuaresma, Boch, Boeche, Bombrun, Borrachero, Bossini, Bouquillon,
  Bourda, Bragaglia, Bramante, Breddels, Bressan, Brouillet,
  Br{\"{u}}semeister, Brugaletta, Bucciarelli, Burlacu, Busonero, Butkevich,
  Buzzi, Caffau, Cancelliere, Cannizzaro, Cantat-Gaudin, Carballo, Carlucci,
  Carrasco, Casamiquela, Castellani, Castro-Ginard, Charlot, Chemin, Chiavassa,
  Cocozza, Costigan, Cowell, Crifo, Crosta, Crowley, Cuypers†, Dafonte,
  Damerdji, Dapergolas, David, David, de~Laverny, {De Luise}, {De March},
  de~Martino, de~Souza, de~Torres, Debosscher, del Pozo, Delbo, Delgado,
  Delgado, {Di Matteo}, Diakite, Diener, Distefano, Dolding, Drazinos,
  Dur{\'{a}}n, Edvardsson, Enke, Eriksson, Esquej, {Eynard Bontemps}, Fabre,
  Fabrizio, Faigler, Falc{\~{a}}o, {Farr{\`{a}}s Casas}, Federici, Fedorets,
  Fernique, Figueras, Filippi, Findeisen, Fonti, Fraile, Fraser,
  Fr{\'{e}}zouls, Gai, Galleti, Garabato, Garc{\'{i}}a-Sedano, Garofalo,
  Garralda, Gavel, Gavras, Gerssen, Geyer, Giacobbe, Gilmore, Girona,
  Giuffrida, Glass, Gomes, Granvik, Gueguen, Guerrier, Guiraud,
  Guti{\'{e}}rrez-S{\'{a}}nchez, Haigron, Hatzidimitriou, Hauser, Haywood,
  Heiter, Helmi, Heu, Hilger, Hobbs, Hofmann, Holland, Huckle, Hypki, Icardi,
  Jan{\ss}en, {Jevardat de Fombelle}, Jonker, Juh{\'{a}}sz, Julbe, Karampelas,
  Kewley, Klar, Kochoska, Kohley, Kolenberg, Kontizas, Kontizas, Koposov,
  Kordopatis, Kostrzewa-Rutkowska, Koubsky, Lambert, Lanza, Lasne, Lavigne, {Le
  Fustec}, {Le Poncin-Lafitte}, Lebreton, Leccia, Leclerc, Lecoeur-Taibi,
  Lenhardt, Leroux, Liao, Licata, Lindstr{\o}m, Lister, Livanou, Lobel,
  L{\'{o}}pez, Managau, Mann, Mantelet, Marchal, Marchant, Marconi, Marinoni,
  Marschalk{\'{o}}, Marshall, Martino, Marton, Mary, Massari, Matijevi{\v{c}},
  Mazeh, McMillan, Messina, Michalik, Millar, Molina, Molinaro, Moln{\'{a}}r,
  Montegriffo, Mor, Morbidelli, Morel, Morris, Mulone, Muraveva, Musella,
  Nelemans, Nicastro, Noval, O'Mullane, Ord{\'{e}}novic,
  Ord{\'{o}}{\~{n}}ez-Blanco, Osborne, Pagani, Pagano, Pailler, Palacin,
  Palaversa, Panahi, Pawlak, Piersimoni, Pineau, Plachy, Plum, Poggio,
  Poujoulet, Pr{\v{s}}a, Pulone, Racero, Ragaini, Rambaux, Ramos-Lerate,
  Regibo, Reyl{\'{e}}, Riclet, Ripepi, Riva, Rivard, Rixon, Roegiers, Roelens,
  Romero-G{\'{o}}mez, Rowell, Royer, Ruiz-Dern, Sadowski, {Sagrist{\`{a}}
  Sell{\'{e}}s}, Sahlmann, Salgado, Salguero, Sanna, Santana-Ros, Sarasso,
  Savietto, Schultheis, Sciacca, Segol, Segovia, S{\'{e}}gransan, Shih,
  Siltala, Silva, Smart, Smith, Solano, Solitro, Sordo, {Soria Nieto}, Souchay,
  Spagna, Spoto, Stampa, Steele, Steidelm{\"{u}}ller, Stephenson, Stoev, Suess,
  Surdej, Szabados, Szegedi-Elek, Tapiador, Taris, Tauran, Taylor, Teixeira,
  Terrett, Teyssandier, Thuillot, Titarenko, {Torra Clotet}, Turon, Ulla,
  Utrilla, Uzzi, Vaillant, Valentini, Valette, van Elteren, {Van Hemelryck},
  van Leeuwen, Vaschetto, Vecchiato, Veljanoski, Viala, Vicente, Vogt, von
  Essen, Voss, Votruba, Voutsinas, Walmsley, Weiler, Wertz, Wevers,
  Wyrzykowski, Yoldas, {\v{Z}}erjal, Ziaeepour, Zorec, Zschocke, Zucker,
  Zurbach, \& Zwitter}]{Gaia2018}
Brown, A. G.~A., Vallenari, A., Prusti, T., {et~al.} 2018, Astronomy {\&}
  Astrophysics, 616, A1, \dodoi{10.1051/0004-6361/201833051}

\bibitem[{{Buie}(1996)}]{Buie1996}
{Buie}, M.~W. 1996, in Astronomical Society of the Pacific Conference Series,
  Vol. 101, Astronomical Data Analysis Software and Systems V, ed. G.~H.
  {Jacoby} \& J.~{Barnes}, 135

\bibitem[{{Buie} \& {Bus}(1992)}]{Buie1992}
{Buie}, M.~W., \& {Bus}, S.~J. 1992, \icarus, 100, 288,
  \dodoi{10.1016/0019-1035(92)90101-C}

\bibitem[{{Buie} {et~al.}(2020){Buie}, {Porter}, {Tamblyn}, {Terrell},
  {Parker}, {Baratoux}, {Kaire}, {Leiva}, {Verbiscer}, {Zangari}, {Colas},
  {Diop}, {Samaniego}, {Wasserman}, {Benecchi}, {Caspi}, {Gwyn}, {Kavelaars},
  {Ocampo Ur{\'\i}a}, {Rabassa}, {Skrutskie}, {Soto}, {Tanga}, {Young},
  {Stern}, {Andersen}, {Arango P{\'e}rez}, {Arredondo}, {Artola}, {B{\^a}},
  {Ballet}, {Blank}, {Bop}, {Bosh}, {Camino L{\'o}pez}, {Carter},
  {Castro-Chac{\'o}n}, {Caycedo Desprez}, {Caycedo Guerra}, {Conard},
  {Dauvergne}, {Dean}, {Dean}, {Desmars}, {Dieng}, {Bousso Dieng}, {Diouf},
  {Dorego}, {Dunham}, {Dunham}, {Durantini Luca}, {Edwards}, {Erasmus}, {Faye},
  {Faye}, {Ferrario}, {Ferrell}, {Finley}, {Fraser}, {Friedli}, {Galvez Serna},
  {Garcia-Migani}, {Genade}, {Getrost}, {Gil-Hutton}, {Gimeno}, {Golub},
  {Gonz{\'a}lez Murillo}, {Grusin}, {Gurovich}, {Hanna}, {Henn}, {Hinton},
  {Hughes}, {Josephs}, {Joya}, {Kammer}, {Keeney}, {Keller}, {Kramer},
  {Levine}, {Lisse}, {Lovell}, {Mackie}, {Makarchuk}, {Manzano}, {Mbaye},
  {Mbaye}, {Melia}, {Moreno}, {Moss}, {Ndaiye}, {Ndiaye}, {Nelson}, {Olkin},
  {Olsen}, {Ospina Moreno}, {Pasachoff}, {Pereyra}, {Person}, {Pinz{\'o}n},
  {Pulver}, {Quintero}, {Regester}, {Resnick}, {Reyes-Ruiz}, {Rolfsmeier},
  {Ruhland}, {Salmon}, {Santos-Sanz}, {Santucho}, {Sep{\'u}lveda Ni{\~n}o},
  {Sickafoose}, {Silva}, {Singer}, {Skipper}, {Slivan}, {Smith}, {Spagnotto},
  {Stephens}, {Strabala}, {Tamayo}, {Throop}, {Torres Ca{\~n}as}, {Toure},
  {Traore}, {Tsang}, {Turner}, {Vanegas}, {Venable}, {Wilson}, {Zuluaga}, \&
  {Zuluaga}}]{Buie2020}
{Buie}, M.~W., {Porter}, S.~B., {Tamblyn}, P., {et~al.} 2020, \aj, 159, 130,
  \dodoi{10.3847/1538-3881/ab6ced}

\bibitem[{{Calabretta} \& {Greisen}(2002)}]{Calabretta2002}
{Calabretta}, M.~R., \& {Greisen}, E.~W. 2002, \aap, 395, 1077,
  \dodoi{10.1051/0004-6361:20021327}

\bibitem[{{Dressel}(2019)}]{Dressel2019}
{Dressel}, L. 2019, {Wide Field Camera 3 Instrument Handbook, Version 12.0}
  (Space Telescope Science Institute)

\bibitem[{{Dressler} {et~al.}(2006){Dressler}, {Hare}, {Bigelow}, \&
  {Osip}}]{Dressler2006}
{Dressler}, A., {Hare}, T., {Bigelow}, B.~C., \& {Osip}, D.~J. 2006, in Society
  of Photo-Optical Instrumentation Engineers (SPIE) Conference Series, Vol.
  6269, \procspie, 62690F, \dodoi{10.1117/12.670573}

\bibitem[{{Elliot} {et~al.}(2005){Elliot}, {Kern}, {Clancy}, {Gulbis},
  {Millis}, {Buie}, {Wasserman}, {Chiang}, {Jordan}, {Trilling}, \&
  {Meech}}]{Elliot2005}
{Elliot}, J.~L., {Kern}, S.~D., {Clancy}, K.~B., {et~al.} 2005, \aj, 129, 1117,
  \dodoi{10.1086/427395}

\bibitem[{{Farnocchia} {et~al.}(2016){Farnocchia}, {Chesley}, {Micheli},
  {Delamere}, {Heyd}, {Tholen}, {Giorgini}, {Owen}, \&
  {Tamppari}}]{Farnocchia2016}
{Farnocchia}, D., {Chesley}, S.~R., {Micheli}, M., {et~al.} 2016, \icarus, 266,
  279, \dodoi{10.1016/j.icarus.2015.10.035}

\bibitem[{{Fraser} {et~al.}(2014){Fraser}, {Brown}, {Morbidelli}, {Parker}, \&
  {Batygin}}]{Fraser2014}
{Fraser}, W.~C., {Brown}, M.~E., {Morbidelli}, A., {Parker}, A., \& {Batygin},
  K. 2014, \apj, 782, 100, \dodoi{10.1088/0004-637X/782/2/100}

\bibitem[{{Fraser} {et~al.}(2008){Fraser}, {Kavelaars}, {Holman}, {Pritchet},
  {Gladman}, {Grav}, {Jones}, {MacWilliams}, \& {Petit}}]{Fraser2008}
{Fraser}, W.~C., {Kavelaars}, J.~J., {Holman}, M.~J., {et~al.} 2008, \icarus,
  195, 827, \dodoi{10.1016/j.icarus.2008.01.014}

\bibitem[{{Fraser} {et~al.}(2024){Fraser}, {Porter}, {Peltier}, {Kavelaars},
  {Verbiscer}, {Buie}, {Stern}, {Spencer}, {Benecchi}, {Terai}, {Takashi},
  {Yoshida}, {Gerdes}, J., {Lin}, {Gwyn}, {Smotherman}, {Fabbro}, {Singer},
  {Alexander}, {Arimatsu}, {Banks}, {Bray}, {El-Maarr}, {Ferrell}, {Fuse},
  {Glass}, {Holt}, {Hong}, {Ishimaru}, {Johnson}, {Lauer}, {Leiva}, {Lykawka},
  {Marschall}, {Nu\'n\~ez}, {Postman}, {Quirico}, {Rhoden}, {Simpson},
  {Schenk}, {Skrutskie}, J., \& {Throop}}]{Fraser2024}
{Fraser}, W.~C., {Porter}, S.~B., {Peltier}, L., {et~al.} 2024, \psj\ submitted

\bibitem[{{Fuentes} {et~al.}(2010){Fuentes}, {Holman}, {Trilling}, \&
  {Protopapas}}]{Fuentes2010}
{Fuentes}, C.~I., {Holman}, M.~J., {Trilling}, D.~E., \& {Protopapas}, P. 2010,
  \apj, 722, 1290, \dodoi{10.1088/0004-637X/722/2/1290}

\bibitem[{{Gaia Collaboration} {et~al.}(2016{\natexlab{a}}){Gaia
  Collaboration}, {Prusti}, {de Bruijne}, {Brown}, {Vallenari}, {Babusiaux},
  {Bailer-Jones}, {Bastian}, {Biermann}, {Evans}, {Eyer}, {Jansen}, {Jordi},
  {Klioner}, {Lammers}, {Lindegren}, {Luri}, {Mignard}, {Milligan}, {Panem},
  {Poinsignon}, {Pourbaix}, {Randich}, {Sarri}, {Sartoretti}, {Siddiqui},
  {Soubiran}, {Valette}, {van Leeuwen}, {Walton}, {Aerts}, {Arenou}, {Cropper},
  {Drimmel}, {H{\o}g}, {Katz}, {Lattanzi}, {O'Mullane}, {Grebel}, {Holland},
  {Huc}, {Passot}, {Bramante}, {Cacciari}, {Casta{\~n}eda}, {Chaoul}, {Cheek},
  {De Angeli}, {Fabricius}, {Guerra}, {Hern{\'a}ndez}, {Jean-Antoine-Piccolo},
  {Masana}, {Messineo}, {Mowlavi}, {Nienartowicz}, {Ord{\'o}{\~n}ez-Blanco},
  {Panuzzo}, {Portell}, {Richards}, {Riello}, {Seabroke}, {Tanga},
  {Th{\'e}venin}, {Torra}, {Els}, {Gracia-Abril}, {Comoretto},
  {Garcia-Reinaldos}, {Lock}, {Mercier}, {Altmann}, {Andrae}, {Astraatmadja},
  {Bellas-Velidis}, {Benson}, {Berthier}, {Blomme}, {Busso}, {Carry},
  {Cellino}, {Clementini}, {Cowell}, {Creevey}, {Cuypers}, {Davidson}, {De
  Ridder}, {de Torres}, {Delchambre}, {Dell'Oro}, {Ducourant}, {Fr{\'e}mat},
  {Garc{\'\i}a-Torres}, {Gosset}, {Halbwachs}, {Hambly}, {Harrison}, {Hauser},
  {Hestroffer}, {Hodgkin}, {Huckle}, {Hutton}, {Jasniewicz}, {Jordan},
  {Kontizas}, {Korn}, {Lanzafame}, {Manteiga}, {Moitinho}, {Muinonen},
  {Osinde}, {Pancino}, {Pauwels}, {Petit}, {Recio-Blanco}, {Robin}, {Sarro},
  {Siopis}, {Smith}, {Smith}, {Sozzetti}, {Thuillot}, {van Reeven}, {Viala},
  {Abbas}, {Abreu Aramburu}, {Accart}, {Aguado}, {Allan}, {Allasia},
  {Altavilla}, {{\'A}lvarez}, {Alves}, {Anderson}, {Andrei}, {Anglada Varela},
  {Antiche}, {Antoja}, {Ant{\'o}n}, {Arcay}, {Atzei}, {Ayache}, {Bach},
  {Baker}, {Balaguer-N{\'u}{\~n}ez}, {Barache}, {Barata}, {Barbier}, {Barblan},
  {Baroni}, {Barrado y Navascu{\'e}s}, {Barros}, {Barstow}, {Becciani},
  {Bellazzini}, {Bellei}, {Bello Garc{\'\i}a}, {Belokurov}, {Bendjoya},
  {Berihuete}, {Bianchi}, {Bienaym{\'e}}, {Billebaud}, {Blagorodnova},
  {Blanco-Cuaresma}, {Boch}, {Bombrun}, {Borrachero}, {Bouquillon}, {Bourda},
  {Bouy}, {Bragaglia}, {Breddels}, {Brouillet}, {Br{\"u}semeister},
  {Bucciarelli}, {Budnik}, {Burgess}, {Burgon}, {Burlacu}, {Busonero}, {Buzzi},
  {Caffau}, {Cambras}, {Campbell}, {Cancelliere}, {Cantat-Gaudin}, {Carlucci},
  {Carrasco}, {Castellani}, {Charlot}, {Charnas}, {Charvet}, {Chassat},
  {Chiavassa}, {Clotet}, {Cocozza}, {Collins}, {Collins}, {Costigan}, {Crifo},
  {Cross}, {Crosta}, {Crowley}, {Dafonte}, {Damerdji}, {Dapergolas}, {David},
  {David}, {De Cat}, {de Felice}, {de Laverny}, {De Luise}, {De March}, {de
  Martino}, {de Souza}, {Debosscher}, {del Pozo}, {Delbo}, {Delgado},
  {Delgado}, {di Marco}, {Di Matteo}, {Diakite}, {Distefano}, {Dolding}, {Dos
  Anjos}, {Drazinos}, {Dur{\'a}n}, {Dzigan}, {Ecale}, {Edvardsson}, {Enke},
  {Erdmann}, {Escolar}, {Espina}, {Evans}, {Eynard Bontemps}, {Fabre},
  {Fabrizio}, {Faigler}, {Falc{\~a}o}, {Farr{\`a}s Casas}, {Faye}, {Federici},
  {Fedorets}, {Fern{\'a}ndez-Hern{\'a}ndez}, {Fernique}, {Fienga}, {Figueras},
  {Filippi}, {Findeisen}, {Fonti}, {Fouesneau}, {Fraile}, {Fraser}, {Fuchs},
  {Furnell}, {Gai}, {Galleti}, {Galluccio}, {Garabato}, {Garc{\'\i}a-Sedano},
  {Gar{\'e}}, {Garofalo}, {Garralda}, {Gavras}, {Gerssen}, {Geyer}, {Gilmore},
  {Girona}, {Giuffrida}, {Gomes}, {Gonz{\'a}lez-Marcos},
  {Gonz{\'a}lez-N{\'u}{\~n}ez}, {Gonz{\'a}lez-Vidal}, {Granvik}, {Guerrier},
  {Guillout}, {Guiraud}, {G{\'u}rpide}, {Guti{\'e}rrez-S{\'a}nchez}, {Guy},
  {Haigron}, {Hatzidimitriou}, {Haywood}, {Heiter}, {Helmi}, {Hobbs},
  {Hofmann}, {Holl}, {Holland}, {Hunt}, {Hypki}, {Icardi}, {Irwin}, {Jevardat
  de Fombelle}, {Jofr{\'e}}, {Jonker}, {Jorissen}, {Julbe}, {Karampelas},
  {Kochoska}, {Kohley}, {Kolenberg}, {Kontizas}, {Koposov}, {Kordopatis},
  {Koubsky}, {Kowalczyk}, {Krone-Martins}, {Kudryashova}, {Kull}, {Bachchan},
  {Lacoste-Seris}, {Lanza}, {Lavigne}, {Le Poncin-Lafitte}, {Lebreton},
  {Lebzelter}, {Leccia}, {Leclerc}, {Lecoeur-Taibi}, {Lemaitre}, {Lenhardt},
  {Leroux}, {Liao}, {Licata}, {Lindstr{\o}m}, {Lister}, {Livanou}, {Lobel},
  {L{\"o}ffler}, {L{\'o}pez}, {Lopez-Lozano}, {Lorenz}, {Loureiro},
  {MacDonald}, {Magalh{\~a}es Fernandes}, {Managau}, {Mann}, {Mantelet},
  {Marchal}, {Marchant}, {Marconi}, {Marie}, {Marinoni}, {Marrese},
  {Marschalk{\'o}}, {Marshall}, {Mart{\'\i}n-Fleitas}, {Martino}, {Mary},
  {Matijevi{\v{c}}}, {Mazeh}, {McMillan}, {Messina}, {Mestre}, {Michalik},
  {Millar}, {Miranda}, {Molina}, {Molinaro}, {Molinaro}, {Moln{\'a}r},
  {Moniez}, {Montegriffo}, {Monteiro}, {Mor}, {Mora}, {Morbidelli}, {Morel},
  {Morgenthaler}, {Morley}, {Morris}, {Mulone}, {Muraveva}, {Musella},
  {Narbonne}, {Nelemans}, {Nicastro}, {Noval}, {Ord{\'e}novic},
  {Ordieres-Mer{\'e}}, {Osborne}, {Pagani}, {Pagano}, {Pailler}, {Palacin},
  {Palaversa}, {Parsons}, {Paulsen}, {Pecoraro}, {Pedrosa}, {Pentik{\"a}inen},
  {Pereira}, {Pichon}, {Piersimoni}, {Pineau}, {Plachy}, {Plum}, {Poujoulet},
  {Pr{\v{s}}a}, {Pulone}, {Ragaini}, {Rago}, {Rambaux}, {Ramos-Lerate},
  {Ranalli}, {Rauw}, {Read}, {Regibo}, {Renk}, {Reyl{\'e}}, {Ribeiro},
  {Rimoldini}, {Ripepi}, {Riva}, {Rixon}, {Roelens}, {Romero-G{\'o}mez},
  {Rowell}, {Royer}, {Rudolph}, {Ruiz-Dern}, {Sadowski}, {Sagrist{\`a}
  Sell{\'e}s}, {Sahlmann}, {Salgado}, {Salguero}, {Sarasso}, {Savietto},
  {Schnorhk}, {Schultheis}, {Sciacca}, {Segol}, {Segovia}, {Segransan},
  {Serpell}, {Shih}, {Smareglia}, {Smart}, {Smith}, {Solano}, {Solitro},
  {Sordo}, {Soria Nieto}, {Souchay}, {Spagna}, {Spoto}, {Stampa}, {Steele},
  {Steidelm{\"u}ller}, {Stephenson}, {Stoev}, {Suess}, {S{\"u}veges}, {Surdej},
  {Szabados}, {Szegedi-Elek}, {Tapiador}, {Taris}, {Tauran}, {Taylor},
  {Teixeira}, {Terrett}, {Tingley}, {Trager}, {Turon}, {Ulla}, {Utrilla},
  {Valentini}, {van Elteren}, {Van Hemelryck}, {van Leeuwen}, {Varadi},
  {Vecchiato}, {Veljanoski}, {Via}, {Vicente}, {Vogt}, {Voss}, {Votruba},
  {Voutsinas}, {Walmsley}, {Weiler}, {Weingrill}, {Werner}, {Wevers},
  {Whitehead}, {Wyrzykowski}, {Yoldas}, {{\v{Z}}erjal}, {Zucker}, {Zurbach},
  {Zwitter}, {Alecu}, {Allen}, {Allende Prieto}, {Amorim},
  {Anglada-Escud{\'e}}, {Arsenijevic}, {Azaz}, {Balm}, {Beck}, {Bernstein},
  {Bigot}, {Bijaoui}, {Blasco}, {Bonfigli}, {Bono}, {Boudreault}, {Bressan},
  {Brown}, {Brunet}, {Bunclark}, {Buonanno}, {Butkevich}, {Carret}, {Carrion},
  {Chemin}, {Ch{\'e}reau}, {Corcione}, {Darmigny}, {de Boer}, {de Teodoro}, {de
  Zeeuw}, {Delle Luche}, {Domingues}, {Dubath}, {Fodor}, {Fr{\'e}zouls},
  {Fries}, {Fustes}, {Fyfe}, {Gallardo}, {Gallegos}, {Gardiol}, {Gebran},
  {Gomboc}, {G{\'o}mez}, {Grux}, {Gueguen}, {Heyrovsky}, {Hoar}, {Iannicola},
  {Isasi Parache}, {Janotto}, {Joliet}, {Jonckheere}, {Keil}, {Kim},
  {Klagyivik}, {Klar}, {Knude}, {Kochukhov}, {Kolka}, {Kos}, {Kutka}, {Lainey},
  {LeBouquin}, {Liu}, {Loreggia}, {Makarov}, {Marseille}, {Martayan},
  {Martinez-Rubi}, {Massart}, {Meynadier}, {Mignot}, {Munari}, {Nguyen},
  {Nordlander}, {Ocvirk}, {O'Flaherty}, {Olias Sanz}, {Ortiz}, {Osorio},
  {Oszkiewicz}, {Ouzounis}, {Palmer}, {Park}, {Pasquato}, {Peltzer}, {Peralta},
  {P{\'e}turaud}, {Pieniluoma}, {Pigozzi}, {Poels}, {Prat}, {Prod'homme},
  {Raison}, {Rebordao}, {Risquez}, {Rocca-Volmerange}, {Rosen}, {Ruiz-Fuertes},
  {Russo}, {Sembay}, {Serraller Vizcaino}, {Short}, {Siebert}, {Silva},
  {Sinachopoulos}, {Slezak}, {Soffel}, {Sosnowska}, {Strai{\v{z}}ys}, {ter
  Linden}, {Terrell}, {Theil}, {Tiede}, {Troisi}, {Tsalmantza}, {Tur},
  {Vaccari}, {Vachier}, {Valles}, {Van Hamme}, {Veltz}, {Virtanen}, {Wallut},
  {Wichmann}, {Wilkinson}, {Ziaeepour}, \& {Zschocke}}]{Gaia2016a}
{Gaia Collaboration}, {Prusti}, T., {de Bruijne}, J.~H.~J., {et~al.}
  2016{\natexlab{a}}, \aap, 595, A1, \dodoi{10.1051/0004-6361/201629272}

\bibitem[{{Gaia Collaboration} {et~al.}(2016{\natexlab{b}}){Gaia
  Collaboration}, {Brown}, {Vallenari}, {Prusti}, {de Bruijne}, {Mignard},
  {Drimmel}, {Babusiaux}, {Bailer-Jones}, {Bastian}, {Biermann}, {Evans},
  {Eyer}, {Jansen}, {Jordi}, {Katz}, {Klioner}, {Lammers}, {Lindegren}, {Luri},
  {O'Mullane}, {Panem}, {Pourbaix}, {Randich}, {Sartoretti}, {Siddiqui},
  {Soubiran}, {Valette}, {van Leeuwen}, {Walton}, {Aerts}, {Arenou}, {Cropper},
  {H{\o}g}, {Lattanzi}, {Grebel}, {Holland}, {Huc}, {Passot}, {Perryman},
  {Bramante}, {Cacciari}, {Casta{\~n}eda}, {Chaoul}, {Cheek}, {De Angeli},
  {Fabricius}, {Guerra}, {Hern{\'a}ndez}, {Jean-Antoine-Piccolo}, {Masana},
  {Messineo}, {Mowlavi}, {Nienartowicz}, {Ord{\'o}{\~n}ez-Blanco}, {Panuzzo},
  {Portell}, {Richards}, {Riello}, {Seabroke}, {Tanga}, {Th{\'e}venin},
  {Torra}, {Els}, {Gracia-Abril}, {Comoretto}, {Garcia-Reinaldos}, {Lock},
  {Mercier}, {Altmann}, {Andrae}, {Astraatmadja}, {Bellas-Velidis}, {Benson},
  {Berthier}, {Blomme}, {Busso}, {Carry}, {Cellino}, {Clementini}, {Cowell},
  {Creevey}, {Cuypers}, {Davidson}, {De Ridder}, {de Torres}, {Delchambre},
  {Dell'Oro}, {Ducourant}, {Fr{\'e}mat}, {Garc{\'\i}a-Torres}, {Gosset},
  {Halbwachs}, {Hambly}, {Harrison}, {Hauser}, {Hestroffer}, {Hodgkin},
  {Huckle}, {Hutton}, {Jasniewicz}, {Jordan}, {Kontizas}, {Korn}, {Lanzafame},
  {Manteiga}, {Moitinho}, {Muinonen}, {Osinde}, {Pancino}, {Pauwels}, {Petit},
  {Recio-Blanco}, {Robin}, {Sarro}, {Siopis}, {Smith}, {Smith}, {Sozzetti},
  {Thuillot}, {van Reeven}, {Viala}, {Abbas}, {Abreu Aramburu}, {Accart},
  {Aguado}, {Allan}, {Allasia}, {Altavilla}, {{\'A}lvarez}, {Alves},
  {Anderson}, {Andrei}, {Anglada Varela}, {Antiche}, {Antoja}, {Ant{\'o}n},
  {Arcay}, {Bach}, {Baker}, {Balaguer-N{\'u}{\~n}ez}, {Barache}, {Barata},
  {Barbier}, {Barblan}, {Barrado y Navascu{\'e}s}, {Barros}, {Barstow},
  {Becciani}, {Bellazzini}, {Bello Garc{\'\i}a}, {Belokurov}, {Bendjoya},
  {Berihuete}, {Bianchi}, {Bienaym{\'e}}, {Billebaud}, {Blagorodnova},
  {Blanco-Cuaresma}, {Boch}, {Bombrun}, {Borrachero}, {Bouquillon}, {Bourda},
  {Bouy}, {Bragaglia}, {Breddels}, {Brouillet}, {Br{\"u}semeister},
  {Bucciarelli}, {Burgess}, {Burgon}, {Burlacu}, {Busonero}, {Buzzi}, {Caffau},
  {Cambras}, {Campbell}, {Cancelliere}, {Cantat-Gaudin}, {Carlucci},
  {Carrasco}, {Castellani}, {Charlot}, {Charnas}, {Chiavassa}, {Clotet},
  {Cocozza}, {Collins}, {Costigan}, {Crifo}, {Cross}, {Crosta}, {Crowley},
  {Dafonte}, {Damerdji}, {Dapergolas}, {David}, {David}, {De Cat}, {de Felice},
  {de Laverny}, {De Luise}, {De March}, {de Martino}, {de Souza}, {Debosscher},
  {del Pozo}, {Delbo}, {Delgado}, {Delgado}, {Di Matteo}, {Diakite},
  {Distefano}, {Dolding}, {Dos Anjos}, {Drazinos}, {Duran}, {Dzigan},
  {Edvardsson}, {Enke}, {Evans}, {Eynard Bontemps}, {Fabre}, {Fabrizio},
  {Faigler}, {Falc{\~a}o}, {Farr{\`a}s Casas}, {Federici}, {Fedorets},
  {Fern{\'a}ndez-Hern{\'a}ndez}, {Fernique}, {Fienga}, {Figueras}, {Filippi},
  {Findeisen}, {Fonti}, {Fouesneau}, {Fraile}, {Fraser}, {Fuchs}, {Gai},
  {Galleti}, {Galluccio}, {Garabato}, {Garc{\'\i}a-Sedano}, {Garofalo},
  {Garralda}, {Gavras}, {Gerssen}, {Geyer}, {Gilmore}, {Girona}, {Giuffrida},
  {Gomes}, {Gonz{\'a}lez-Marcos}, {Gonz{\'a}lez-N{\'u}{\~n}ez},
  {Gonz{\'a}lez-Vidal}, {Granvik}, {Guerrier}, {Guillout}, {Guiraud},
  {G{\'u}rpide}, {Guti{\'e}rrez-S{\'a}nchez}, {Guy}, {Haigron},
  {Hatzidimitriou}, {Haywood}, {Heiter}, {Helmi}, {Hobbs}, {Hofmann}, {Holl},
  {Holland}, {Hunt}, {Hypki}, {Icardi}, {Irwin}, {Jevardat de Fombelle},
  {Jofr{\'e}}, {Jonker}, {Jorissen}, {Julbe}, {Karampelas}, {Kochoska},
  {Kohley}, {Kolenberg}, {Kontizas}, {Koposov}, {Kordopatis}, {Koubsky},
  {Krone-Martins}, {Kudryashova}, {Kull}, {Bachchan}, {Lacoste-Seris}, {Lanza},
  {Lavigne}, {Le Poncin-Lafitte}, {Lebreton}, {Lebzelter}, {Leccia}, {Leclerc},
  {Lecoeur-Taibi}, {Lemaitre}, {Lenhardt}, {Leroux}, {Liao}, {Licata},
  {Lindstr{\o}m}, {Lister}, {Livanou}, {Lobel}, {L{\"o}ffler}, {L{\'o}pez},
  {Lorenz}, {MacDonald}, {Magalh{\~a}es Fernandes}, {Managau}, {Mann},
  {Mantelet}, {Marchal}, {Marchant}, {Marconi}, {Marinoni}, {Marrese},
  {Marschalk{\'o}}, {Marshall}, {Mart{\'\i}n-Fleitas}, {Martino}, {Mary},
  {Matijevi{\v{c}}}, {Mazeh}, {McMillan}, {Messina}, {Michalik}, {Millar},
  {Miranda}, {Molina}, {Molinaro}, {Molinaro}, {Moln{\'a}r}, {Moniez},
  {Montegriffo}, {Mor}, {Mora}, {Morbidelli}, {Morel}, {Morgenthaler},
  {Morris}, {Mulone}, {Muraveva}, {Musella}, {Narbonne}, {Nelemans},
  {Nicastro}, {Noval}, {Ord{\'e}novic}, {Ordieres-Mer{\'e}}, {Osborne},
  {Pagani}, {Pagano}, {Pailler}, {Palacin}, {Palaversa}, {Parsons}, {Pecoraro},
  {Pedrosa}, {Pentik{\"a}inen}, {Pichon}, {Piersimoni}, {Pineau}, {Plachy},
  {Plum}, {Poujoulet}, {Pr{\v{s}}a}, {Pulone}, {Ragaini}, {Rago}, {Rambaux},
  {Ramos-Lerate}, {Ranalli}, {Rauw}, {Read}, {Regibo}, {Reyl{\'e}}, {Ribeiro},
  {Rimoldini}, {Ripepi}, {Riva}, {Rixon}, {Roelens}, {Romero-G{\'o}mez},
  {Rowell}, {Royer}, {Ruiz-Dern}, {Sadowski}, {Sagrist{\`a} Sell{\'e}s},
  {Sahlmann}, {Salgado}, {Salguero}, {Sarasso}, {Savietto}, {Schultheis},
  {Sciacca}, {Segol}, {Segovia}, {Segransan}, {Shih}, {Smareglia}, {Smart},
  {Solano}, {Solitro}, {Sordo}, {Soria Nieto}, {Souchay}, {Spagna}, {Spoto},
  {Stampa}, {Steele}, {Steidelm{\"u}ller}, {Stephenson}, {Stoev}, {Suess},
  {S{\"u}veges}, {Surdej}, {Szabados}, {Szegedi-Elek}, {Tapiador}, {Taris},
  {Tauran}, {Taylor}, {Teixeira}, {Terrett}, {Tingley}, {Trager}, {Turon},
  {Ulla}, {Utrilla}, {Valentini}, {van Elteren}, {Van Hemelryck}, {van
  Leeuwen}, {Varadi}, {Vecchiato}, {Veljanoski}, {Via}, {Vicente}, {Vogt},
  {Voss}, {Votruba}, {Voutsinas}, {Walmsley}, {Weiler}, {Weingrill}, {Wevers},
  {Wyrzykowski}, {Yoldas}, {{\v{Z}}erjal}, {Zucker}, {Zurbach}, {Zwitter},
  {Alecu}, {Allen}, {Allende Prieto}, {Amorim}, {Anglada-Escud{\'e}},
  {Arsenijevic}, {Azaz}, {Balm}, {Beck}, {Bernstein}, {Bigot}, {Bijaoui},
  {Blasco}, {Bonfigli}, {Bono}, {Boudreault}, {Bressan}, {Brown}, {Brunet},
  {Bunclark}, {Buonanno}, {Butkevich}, {Carret}, {Carrion}, {Chemin},
  {Ch{\'e}reau}, {Corcione}, {Darmigny}, {de Boer}, {de Teodoro}, {de Zeeuw},
  {Delle Luche}, {Domingues}, {Dubath}, {Fodor}, {Fr{\'e}zouls}, {Fries},
  {Fustes}, {Fyfe}, {Gallardo}, {Gallegos}, {Gardiol}, {Gebran}, {Gomboc},
  {G{\'o}mez}, {Grux}, {Gueguen}, {Heyrovsky}, {Hoar}, {Iannicola}, {Isasi
  Parache}, {Janotto}, {Joliet}, {Jonckheere}, {Keil}, {Kim}, {Klagyivik},
  {Klar}, {Knude}, {Kochukhov}, {Kolka}, {Kos}, {Kutka}, {Lainey}, {LeBouquin},
  {Liu}, {Loreggia}, {Makarov}, {Marseille}, {Martayan}, {Martinez-Rubi},
  {Massart}, {Meynadier}, {Mignot}, {Munari}, {Nguyen}, {Nordlander}, {Ocvirk},
  {O'Flaherty}, {Olias Sanz}, {Ortiz}, {Osorio}, {Oszkiewicz}, {Ouzounis},
  {Palmer}, {Park}, {Pasquato}, {Peltzer}, {Peralta}, {P{\'e}turaud},
  {Pieniluoma}, {Pigozzi}, {Poels}, {Prat}, {Prod'homme}, {Raison}, {Rebordao},
  {Risquez}, {Rocca-Volmerange}, {Rosen}, {Ruiz-Fuertes}, {Russo}, {Sembay},
  {Serraller Vizcaino}, {Short}, {Siebert}, {Silva}, {Sinachopoulos}, {Slezak},
  {Soffel}, {Sosnowska}, {Strai{\v{z}}ys}, {ter Linden}, {Terrell}, {Theil},
  {Tiede}, {Troisi}, {Tsalmantza}, {Tur}, {Vaccari}, {Vachier}, {Valles}, {Van
  Hamme}, {Veltz}, {Virtanen}, {Wallut}, {Wichmann}, {Wilkinson}, {Ziaeepour},
  \& {Zschocke}}]{Gaia2016b}
{Gaia Collaboration}, {Brown}, A.~G.~A., {Vallenari}, A., {et~al.}
  2016{\natexlab{b}}, \aap, 595, A2, \dodoi{10.1051/0004-6361/201629512}

\bibitem[{{Gennaro} {et~al.}(2018){Gennaro}, {Anderson}, {Baggett}, {Bajaj},
  {Brammer}, {Bourque}, {Calamida}, {Deustua}, {Dressel}, {Fowler},
  {Khandrika}, {Khozurina-Platais}, Heather~{Kurtz}, {Long}, {Mack}, {Martlin},
  {McCullough}, {McKay}, {Medina}, {de la Pena}, {Pirzkal}, {Russell}, {Sahu},
  {Shanahan}, {Sosey}, {Riess}, {Sabbi}, {Stevenson}, \&
  {Sunnquist}}]{Gennaro2018}
{Gennaro}, M., {Anderson}, J., {Baggett}, S., {et~al.} 2018, {WFC3 Data
  Handbook Version 4.0} (Space Telescope Science Institute)

\bibitem[{{Gonzaga} {et~al.}(2012){Gonzaga}, {Hack}, {Fruchter}, \&
  {Mack}}]{Gonzaga2012}
{Gonzaga}, S., {Hack}, W., {Fruchter}, A., \& {Mack}, J. 2012, {The DrizzlePac
  Handbook} (Space Telescope Science Institute)

\bibitem[{{Greisen} \& {Calabretta}(2002)}]{Greisen2002}
{Greisen}, E.~W., \& {Calabretta}, M.~R. 2002, \aap, 395, 1061,
  \dodoi{10.1051/0004-6361:20021326}

\bibitem[{{Gwyn}(2014)}]{Gwyn2014}
{Gwyn}, S.~D.~J. 2014, Journal of Instrumentation, 9, C04003,
  \dodoi{10.1088/1748-0221/9/04/C04003}

\bibitem[{{Kavelaars} {et~al.}(2021){Kavelaars}, {Petit}, {Gladman},
  {Bannister}, {Alexandersen}, {Chen}, {Gwyn}, \& {Volk}}]{Kavelaars2021}
{Kavelaars}, J.~J., {Petit}, J.-M., {Gladman}, B., {et~al.} 2021, \apjl, 920,
  L28, \dodoi{10.3847/2041-8213/ac2c72}

\bibitem[{{Kozhurina-Platais} {et~al.}(2016){Kozhurina-Platais}, {Mackenty},
  {Golimovski}, {Sirianni}, {Borncamp}, {Anderson}, \&
  {Grogin}}]{Kozhurina2016}
{Kozhurina-Platais}, V., {Mackenty}, J., {Golimovski}, D., {et~al.} 2016, in
  Society of Photo-Optical Instrumentation Engineers (SPIE) Conference Series,
  Vol. 9904, \procspie, 99046I, \dodoi{10.1117/12.2233793}

\bibitem[{{Krist} {et~al.}(2011){Krist}, {Hook}, \& {Stoehr}}]{Krist2011}
{Krist}, J.~E., {Hook}, R.~N., \& {Stoehr}, F. 2011, in Society of
  Photo-Optical Instrumentation Engineers (SPIE) Conference Series, Vol. 8127,
  \procspie, 81270J, \dodoi{10.1117/12.892762}

\bibitem[{{Lindegren} {et~al.}(2016){Lindegren}, {Lammers}, {Bastian},
  {Hern{\'a}ndez}, {Klioner}, {Hobbs}, {Bombrun}, {Michalik}, {Ramos-Lerate},
  {Butkevich}, {Comoretto}, {Joliet}, {Holl}, {Hutton}, {Parsons},
  {Steidelm{\"u}ller}, {Abbas}, {Altmann}, {Andrei}, {Anton}, {Bach},
  {Barache}, {Becciani}, {Berthier}, {Bianchi}, {Biermann}, {Bouquillon},
  {Bourda}, {Br{\"u}semeister}, {Bucciarelli}, {Busonero}, {Carlucci},
  {Casta{\~n}eda}, {Charlot}, {Clotet}, {Crosta}, {Davidson}, {de Felice},
  {Drimmel}, {Fabricius}, {Fienga}, {Figueras}, {Fraile}, {Gai}, {Garralda},
  {Geyer}, {Gonz{\'a}lez-Vidal}, {Guerra}, {Hambly}, {Hauser}, {Jordan},
  {Lattanzi}, {Lenhardt}, {Liao}, {L{\"o}ffler}, {McMillan}, {Mignard}, {Mora},
  {Morbidelli}, {Portell}, {Riva}, {Sarasso}, {Serraller}, {Siddiqui}, {Smart},
  {Spagna}, {Stampa}, {Steele}, {Taris}, {Torra}, {van Reeven}, {Vecchiato},
  {Zschocke}, {de Bruijne}, {Gracia}, {Raison}, {Lister}, {Marchant},
  {Messineo}, {Soffel}, {Osorio}, {de Torres}, \& {O'Mullane}}]{Lindegren2016}
{Lindegren}, L., {Lammers}, U., {Bastian}, U., {et~al.} 2016, \aap, 595, A4,
  \dodoi{10.1051/0004-6361/201628714}

\bibitem[{Lindegren {et~al.}(2018)Lindegren, Hern{\'{a}}ndez, Bombrun, Klioner,
  Bastian, Ramos-Lerate, de~Torres, Steidelm{\"{u}}ller, Stephenson, Hobbs,
  Lammers, Biermann, Geyer, Hilger, Michalik, Stampa, McMillan,
  Casta{\~{n}}eda, Clotet, Comoretto, Davidson, Fabricius, Gracia, Hambly,
  Hutton, Mora, Portell, van Leeuwen, Abbas, Abreu, Altmann, Andrei, Anglada,
  Balaguer-N{\'{u}}{\~{n}}ez, Barache, Becciani, Bertone, Bianchi, Bouquillon,
  Bourda, Br{\"{u}}semeister, Bucciarelli, Busonero, Buzzi, Cancelliere,
  Carlucci, Charlot, Cheek, Crosta, Crowley, de~Bruijne, de~Felice, Drimmel,
  Esquej, Fienga, Fraile, Gai, Garralda, Gonz{\'{a}}lez-Vidal, Guerra, Hauser,
  Hofmann, Holl, Jordan, Lattanzi, Lenhardt, Liao, Licata, Lister,
  L{\"{o}}ffler, Marchant, Martin-Fleitas, Messineo, Mignard, Morbidelli,
  Poggio, Riva, Rowell, Salguero, Sarasso, Sciacca, Siddiqui, Smart, Spagna,
  Steele, Taris, Torra, van Elteren, van Reeven, \& Vecchiato}]{Lindegren2018}
Lindegren, L., Hern{\'{a}}ndez, J., Bombrun, A., {et~al.} 2018, Astronomy {\&}
  Astrophysics, 616, A2, \dodoi{10.1051/0004-6361/201832727}

\bibitem[{{McLeod} {et~al.}(2015){McLeod}, {Geary}, {Conroy}, {Fabricant},
  {Ordway}, {Szentgyorgyi}, {Amato}, {Ashby}, {Caldwell}, {Curley}, {Gauron},
  {Holman}, {Norton}, {Pieri}, {Roll}, {Weaver}, {Zajac}, {Palunas}, \&
  {Osip}}]{McLeod2015}
{McLeod}, B., {Geary}, J., {Conroy}, M., {et~al.} 2015, \pasp, 127, 366,
  \dodoi{10.1086/680687}

\bibitem[{{Miller} {et~al.}(2008){Miller}, {Pennypacker}, \&
  {White}}]{Miller2008}
{Miller}, J.~P., {Pennypacker}, C.~R., \& {White}, G.~L. 2008, \pasp, 120, 449,
  \dodoi{10.1086/588258}

\bibitem[{{Miyazaki} {et~al.}(2002){Miyazaki}, {Komiyama}, {Sekiguchi},
  {Okamura}, {Doi}, {Furusawa}, {Hamabe}, {Imi}, {Kimura}, {Nakata}, {Okada},
  {Ouchi}, {Shimasaku}, {Yagi}, \& {Yasuda}}]{Miyazaki2002}
{Miyazaki}, S., {Komiyama}, Y., {Sekiguchi}, M., {et~al.} 2002, \pasj, 54, 833,
  \dodoi{10.1093/pasj/54.6.833}

\bibitem[{{Miyazaki} {et~al.}(2018){Miyazaki}, {Komiyama}, {Kawanomoto}, {Doi},
  {Furusawa}, {Hamana}, {Hayashi}, {Ikeda}, {Kamata}, {Karoji}, {Koike},
  {Kurakami}, {Miyama}, {Morokuma}, {Nakata}, {Namikawa}, {Nakaya}, {Nariai},
  {Obuchi}, {Oishi}, {Okada}, {Okura}, {Tait}, {Takata}, {Tanaka}, {Tanaka},
  {Terai}, {Tomono}, {Uraguchi}, {Usuda}, {Utsumi}, {Yamada}, {Yamanoi},
  {Aihara}, {Fujimori}, {Mineo}, {Miyatake}, {Oguri}, {Uchida}, {Tanaka},
  {Yasuda}, {Takada}, {Murayama}, {Nishizawa}, {Sugiyama}, {Chiba}, {Futamase},
  {Wang}, {Chen}, {Ho}, {Liaw}, {Chiu}, {Ho}, {Lai}, {Lee}, {Jeng}, {Iwamura},
  {Armstrong}, {Bickerton}, {Bosch}, {Gunn}, {Lupton}, {Loomis}, {Price},
  {Smith}, {Strauss}, {Turner}, {Suzuki}, {Miyazaki}, {Muramatsu}, {Yamamoto},
  {Endo}, {Ezaki}, {Ito}, {Kawaguchi}, {Sofuku}, {Taniike}, {Akutsu}, {Dojo},
  {Kasumi}, {Matsuda}, {Imoto}, {Miwa}, {Suzuki}, {Takeshi}, \&
  {Yokota}}]{Miyazaki2018}
{Miyazaki}, S., {Komiyama}, Y., {Kawanomoto}, S., {et~al.} 2018, \pasj, 70, S1,
  \dodoi{10.1093/pasj/psx063}

\bibitem[{{Monet} {et~al.}(2003){Monet}, {Levine}, {Canzian}, {Ables}, {Bird},
  {Dahn}, {Guetter}, {Harris}, {Henden}, {Leggett}, {Levison}, {Luginbuhl},
  {Martini}, {Monet}, {Munn}, {Pier}, {Rhodes}, {Riepe}, {Sell}, {Stone},
  {Vrba}, {Walker}, {Westerhout}, {Brucato}, {Reid}, {Schoening}, {Hartley},
  {Read}, \& {Tritton}}]{Monet2003}
{Monet}, D.~G., {Levine}, S.~E., {Canzian}, B., {et~al.} 2003, \aj, 125, 984,
  \dodoi{10.1086/345888}

\bibitem[{{Parker} \& {Kavelaars}(2010)}]{Parker2010}
{Parker}, A.~H., \& {Kavelaars}, J.~J. 2010, \icarus, 209, 766,
  \dodoi{10.1016/j.icarus.2010.04.018}

\bibitem[{{Petit} {et~al.}(2011){Petit}, {Kavelaars}, {Gladman}, {Jones},
  {Parker}, {Van Laerhoven}, {Nicholson}, {Mars}, {Rousselot}, {Mousis},
  {Marsden}, {Bieryla}, {Taylor}, {Ashby}, {Benavidez}, {Campo Bagatin}, \&
  {Bernabeu}}]{Petit2011}
{Petit}, J.~M., {Kavelaars}, J.~J., {Gladman}, B.~J., {et~al.} 2011, \aj, 142,
  131, \dodoi{10.1088/0004-6256/142/4/131}

\bibitem[{{Porter} {et~al.}(2018){Porter}, {Buie}, {Parker}, {Spencer},
  {Benecchi}, {Tanga}, {Verbiscer}, {Kavelaars}, {Gwyn}, {Young}, {Weaver},
  {Olkin}, {Parker}, \& {Stern}}]{Porter2018}
{Porter}, S.~B., {Buie}, M.~W., {Parker}, A.~H., {et~al.} 2018, \aj, 156, 20,
  \dodoi{10.3847/1538-3881/aac2e1}

\bibitem[{{Porter} {et~al.}(2022){Porter}, {Spencer}, {Verbiscer}, {Benecchi},
  {Weaver}, {Wen Lin}, {Kavelaars}, {Fraser}, {Gerdes}, {Buie}, {Singer},
  {Parker}, \& {Stern}}]{Porter2022}
{Porter}, S.~B., {Spencer}, J.~R., {Verbiscer}, A., {et~al.} 2022, \psj, 3, 23,
  \dodoi{10.3847/PSJ/ac3491}

\bibitem[{{Smart} \& {Green}(1977)}]{Smart1977}
{Smart}, W.~M., \& {Green}, E. b. R.~M. 1977, {Textbook on Spherical Astronomy}
  (Cambridge University Press)

\bibitem[{{Spencer} {et~al.}(2003){Spencer}, {Buie}, {Young}, {Guo}, \&
  {Stern}}]{Spencer2003}
{Spencer}, J., {Buie}, M., {Young}, L., {Guo}, Y., \& {Stern}, A. 2003, Earth
  Moon and Planets, 92, 483, \dodoi{10.1023/B:MOON.0000031963.58573.97}

\bibitem[{{Stern}(2008)}]{Stern2008}
{Stern}, S.~A. 2008, \ssr, 140, 3, \dodoi{10.1007/s11214-007-9295-y}

\bibitem[{{Stetson}(1987)}]{Stetson1987}
{Stetson}, P.~B. 1987, \pasp, 99, 191, \dodoi{10.1086/131977}

\bibitem[{{Verbiscer} {et~al.}(2019){Verbiscer}, {Porter}, {Benecchi},
  {Kavelaars}, {Weaver}, {Spencer}, {Buie}, {Tholen}, {Buratti}, {Helfenstein},
  {Parker}, {Olkin}, {Parker}, {Stern}, {Young}, {Ennico-Smith}, {Singer},
  {Cheng}, {Lisse}, \& {New Horizons Science Team}}]{Verbiscer2019}
{Verbiscer}, A.~J., {Porter}, S., {Benecchi}, S.~D., {et~al.} 2019, \aj, 158,
  123, \dodoi{10.3847/1538-3881/ab3211}

\bibitem[{{Verbiscer} {et~al.}(2022){Verbiscer}, {Helfenstein}, {Porter},
  {Benecchi}, {Kavelaars}, {Lauer}, {Peng}, {Protopapa}, {Spencer}, {Stern},
  {Weaver}, {Buie}, {Buratti}, {Olkin}, {Parker}, {Singer}, {Young}, \& {New
  Horizons Science Team}}]{Verbiscer2022}
{Verbiscer}, A.~J., {Helfenstein}, P., {Porter}, S.~B., {et~al.} 2022, \psj, 3,
  95, \dodoi{10.3847/PSJ/ac63a6}

\bibitem[{{Zacharias} {et~al.}(2012){Zacharias}, {Finch}, {Girard}, {Henden},
  {Bartlett}, {Monet}, \& {Zacharias}}]{Zacharias2012}
{Zacharias}, N., {Finch}, C.~T., {Girard}, T.~M., {et~al.} 2012, VizieR Online
  Data Catalog, I/322A

\bibitem[{{Zacharias} {et~al.}(2015){Zacharias}, {Finch}, {Subasavage},
  {Bredthauer}, {Crockett}, {Divittorio}, {Furguson}, {Harris}, {Harris},
  {Henden}, {Kilian}, {Munn}, {Rafferty}, {Rhodes}, {Schultheiss}, {Tilleman},
  \& {Wieder}}]{Zacharias2015}
{Zacharias}, N., {Finch}, C., {Subasavage}, J., {et~al.} 2015, VizieR Online
  Data Catalog, I/329

\end{thebibliography}

\allauthors

\end{document}